\documentclass[reqno,11pt]{amsart}
\usepackage[dvipsnames]{xcolor}
\usepackage{geometry,enumitem,comment,bbold}
\usepackage[normalem]{ulem}
\usepackage[utf8x]{inputenc}
\usepackage{amsmath,float}
\usepackage{graphicx}
\usepackage{rotating}
\usepackage{wasysym}
\DeclareGraphicsExtensions{.pdf,.png,.jpg}
\usepackage{setspace}

\usepackage[colorinlistoftodos]{todonotes}

\usepackage{geometry}
\usepackage[round,comma]{natbib}                

\usepackage{booktabs}
\usepackage{amsmath}
\usepackage{amsthm}
\usepackage{amsfonts}
\usepackage{amssymb}
\usepackage{dsfont}

\usepackage{nicefrac}
\usepackage{booktabs}
\usepackage{mathrsfs}
\usepackage{bm} 
\usepackage{mathtools}   
\usepackage{subfigure}

\newtheorem{theorem}{Theorem}[section]
\newtheorem{corollary}[theorem]{Corollary}      
\newtheorem{lemma}[theorem]{Lemma}              
  
\newtheorem{proposition}[theorem]{Proposition}  
\theoremstyle{definition}
\newtheorem{example}[theorem]{Example} 
\newtheorem{definition}[theorem]{Definition} 
\newtheorem{remark}[theorem]{Remark}
\newtheorem{assumption}[theorem]{Assumption}

\numberwithin{equation}{section}

\DeclareMathOperator{\es}{\mathrm{ES}}
\DeclareMathOperator{\LVaR}{\mathrm{LVaR}}

\newcommand{\R}{\mathbb{R}} 
\newcommand{\Q}{\mathbb{Q}}

\newcommand{\E}{\mathbb{E}}

\newcommand{\tn}{\textnormal}

\newcommand{\ind}{\mathbf{1}}
\renewcommand{\P}{\mathbb{P}}

\newcommand{\N}{\mathbb{N}}
\newcommand{\arcsinh}{\mathrm{arcsinh}}

\newcommand{\eps}{\varepsilon}
\newcommand{\supp}{\textnormal{supp}}

\newcommand{\ssd}{\leq_{\textnormal{SSD}}}

\newcommand{\CF}{\mathcal F}
\newcommand{\CA}{\mathcal A}
\newcommand{\CG}{\mathcal G}
\newcommand{\CE}{\mathcal E}
\newcommand{\CX}{\mathcal X}

\newcommand{\CC}{\mathcal C}

\newcommand{\CL}{\mathcal L}
\renewcommand{\Q}{{\mathbb Q}}
\newcommand{\CB}{\mathcal B}

\newcommand{\ph}{\varphi}
\newcommand{\peq}{\preceq}

\newcommand{\vsd}{\leq_{v}}

\newcommand{\worst}{\rho^{\mathrm w}}
\newcommand{\mf}{\mathfrak}

\newcommand{\vinverse}{h}

\newcommand*\diff{\mathop{}\!\mathrm{d}}
\newcommand{\id}{\mathrm{id}}

\renewcommand{\and}{\quad\text{and}\quad}

\usepackage{natbib}
\usepackage[colorlinks,urlcolor=red,citecolor=blue,linkcolor=red]{hyperref}

\begin{document}

\title[Meyer risk measures]{Meyer risk measures}

\author{Christian Laudag\'e}
		\address{Department of Mathematics, RPTU University Kaiserslautern-Landau, Germany.}
    \email{christian.laudage@rptu.de}

    \author{Felix-Benedikt Liebrich}
		\address{Amsterdam School of Economics, University of Amsterdam, Netherlands.}
		\email{f.b.liebrich@uva.nl}

\date{September 1, 2026}

\flushbottom

\begin{abstract}
\setstretch{1.1}

Risk measures summarize the risk profile of financial positions in a single metric, which allows their comparison and supports investment decisions.
When they respect common stochastic orders such as second-order stochastic dominance (SSD), they rest on sound conceptual grounds. Since SSD can be overly conservative and rigid, we study {\em Meyer risk measures}. 
Those are monetary risk measures that are consistent with stochastic orders induced by a threshold utility function~$v$, where the defining test utilities are at least as risk averse as~$v$. 
The flexibility in choosing~$v$ allows to encompass a wide range of stochastic orders, including SSD. 
Our contribution is threefold. First, we show that Meyer risk measures whose threshold utility~$v$ is CARA admit a lower envelope representation via  adjusted risk measures, and analyse how they extend and compare to classical risk measures. 
Second, we show that monotone additive statistics, recently introduced in decision theory, are Meyer risk measures under mild conditions.
Third, impossibility results for the situation beyond the CARA class establish  that monetary risk measures compatible to the associated stochastic orders either fail to exist or cannot satisfy key properties simultaneously. Throughout, we highlight practical implications, for example to portfolio selection and risk management.

\smallskip


\end{abstract}

\maketitle
\thispagestyle{empty} 

\setstretch{1.4}    
\section{Introduction}\label{section.intro}

How should one choose between two random prospects? This question lies at the heart of decision theory, risk management, and financial investment. 
A standard approach, going back to \cite{Hadar} and \cite{Hanoch}, is to compare the prospects in second-order stochastic dominance (SSD) and verify whether there is unanimous preference for one payoff over the other among all expected utility (EU) agents who share a specific probabilistic view of the future; prefer more to less (nondecreasing utility); and are risk averse (concave utility).
In contrast, \cite{LeshnoLevy} have observed that SSD-based decisions can be hijacked by unrealistically risk-averse EU agents. 
Even when one payoff is intuitively clearly preferred, SSD may not support that ranking.
They illustrate this phenomenon with two payoff lotteries: one pays \$1 with certainty, the other \$1,000,000 with 99\% chance and nothing otherwise. 
While virtually all decision makers would prefer the second lottery to the first, this preference is not reflected by an SSD ordering because overly risk-averse agents within the SSD collective prefer the certain payoff, thereby breaking unanimity. 

\begin{figure}[ht]
    \centering
    \subfigure{
        \includegraphics[width=0.4\textwidth]{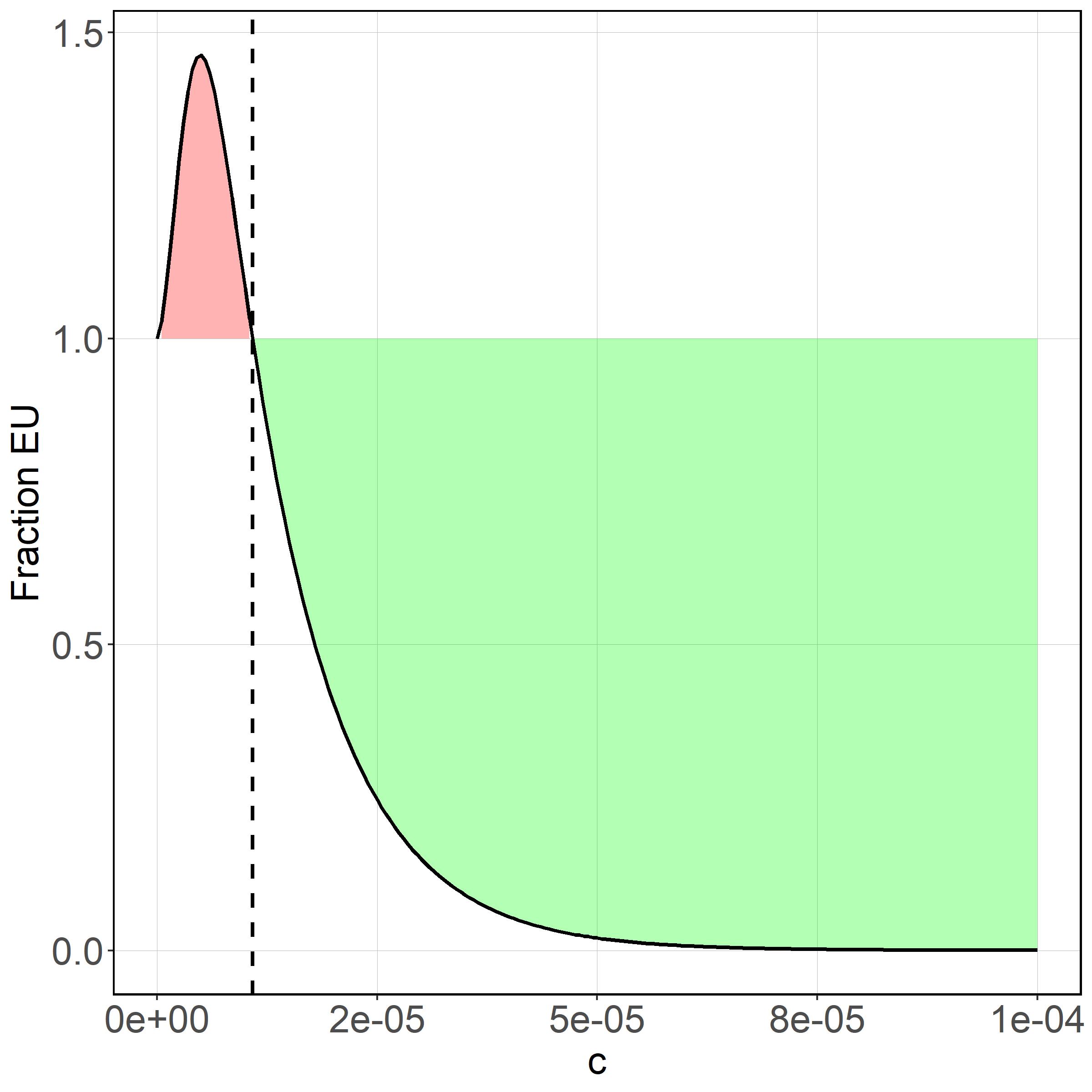}
    }
    \subfigure{
        \includegraphics[width=0.4\textwidth]{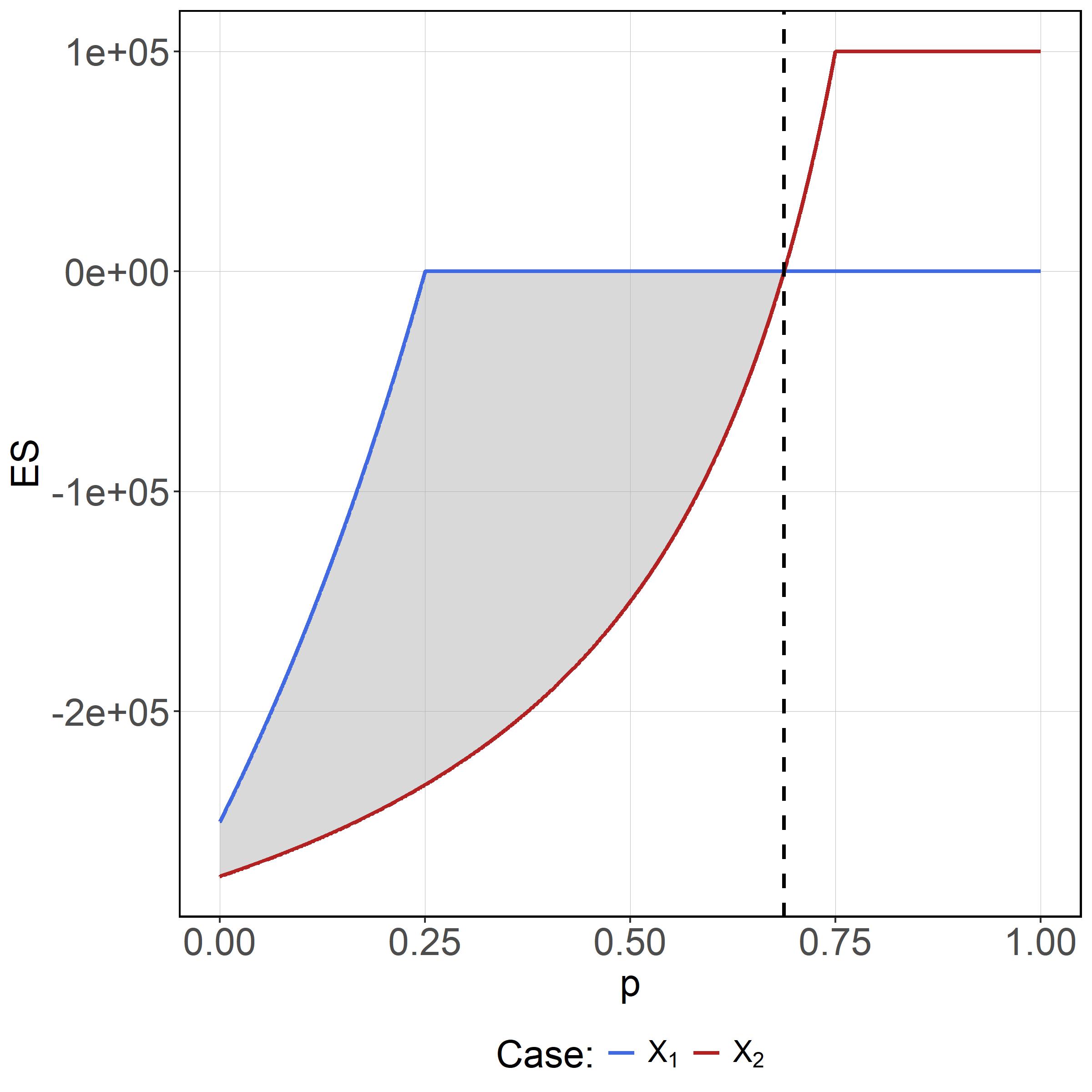}
    }
    \caption{\footnotesize Fraction of expected utilities $\frac{\E[u_c(X_1)]}{\E[u_c(X_2)]}$ (left), and ES-curves for $X_1$ and $X_2$ (right).}
    \label{fig:example_introduction}
\end{figure}

From a risk management perspective, not only overly risk-averse EU agents pose a puzzle; intuitive preferences for specific lottery can also be prevented by the presence of overly risk-loving agents in the consensus collective. 
Inspired by the example of \cite{LeshnoLevy}, consider two lotteries with the following random payoffs: 
\begin{center}
$\begin{array}{c@{\quad\quad\quad}c}
\textbf{Lottery 1} & \textbf{Lottery 2} \\[0.2em]

\begin{array}{l|c}
\text{Outcome} & \text{Probability} \\ \hline
\$1{,}000{,}000 & 25\% \\
\$0             & 75\%
\end{array}
&
\begin{array}{l|c}
\text{Outcome} & \text{Probability} \\ \hline
\$400{,}000    & 75\% \\
-\$100{,}000   & 25\%
\end{array}
\end{array}$
\end{center}
\smallskip
Lottery 2 has a higher winning probability but a smaller potential gain than Lottery 1. 
Moreover, Lottery 2 carries downside risk while Lottery 1 does not.
Preferring Lottery 1 to Lottery 2 therefore seems reasonable. Furthermore, a decision maker choosing Lottery 2 should be required to hold sufficient risk capital to cover potential losses. 
If random variables $X_1$ and $X_2$ represent the risky payoffs of these lotteries, then just like in the aforementioned example of \cite{LeshnoLevy}, the intuitive preference for $X_1$ over $X_2$ will {\em not} be supported by  SSD. 
In contrast to the earlier example, this is because the decision hinges on EU agents that are not sufficiently risk averse.
Consider the exponential utility
$
u_c(x) = -ce^{-cx}
$
with risk aversion $c > 0$ and the EU of $X_1$ relative to the one of $X_2$, 
\[
f(c) :=\frac{\E[u_c(X_1)]}{\E[u_c(X_2)]}.\]
EU agent $u_c$ weakly prefers $X_1$ over $X_2$ if $f(c) \leq 1$.
As can be seen on the left-hand side of Figure~\ref{fig:example_introduction}, the intuitive SSD-preference is prevented because $X_2$ exceeds $X_1$ in EU at low risk aversion levels.
This is also reflected by the Expected Shortfall (ES) curves on the right-hand side of  Figure~\ref{fig:example_introduction}, where $X_1$ being SSD-preferred would mean its ES-curve is not undercut by $X_2$’s.
However, that is precisely what happens at ES confidence levels below 69\%, as indicated by the dashed vertical line. 
In conclusion, excluding EU agents with low risk aversion can ensure that $X_1$ is preferred over $X_2$ by the remaining consensus collective. 

A possibility to solve this issue---originally suggested by~\cite{LeshnoLevy} and revisited later by \cite{Tzeng,Tsetlin,Mueller}, and \cite{Wu}---is to modify the set of relevant utility functions, excluding agents whose risk aversion is deemed extreme, and adding locally risk-loving EU agents. These contributions have been preceded by \cite{MeyerJET,Meyer}, where the risk aversion of the decision-making EU collective is controlled directly. This approach is crucial for the present paper and has been revisited, e.g., by \cite{Huang} and \cite{LiuMeyer2025}. 
Several of these contributions, e.g.,~\cite{Mueller}, lead to fractional-degree interpolation between classical first- and second order stochastic dominance. Building on this observation,~\cite{Huang} call their framework \textit{fractional degree stochastic dominance}.\footnote{~A related perspective, establishing a direct link to fractional integration, is used in~\cite{Fishburn-1976,Fishburn-1980}, where a stochastic order grounded in the so-called Riemann-Liouville fractional integral is proposed.} 

We follow \cite{Meyer} and exclude insufficiently risk-averse agents using a fixed twice-differentiable, strictly increasing {\em threshold utility function} $v$, which draws the line between sufficient and insufficient risk aversion. 
To be precise, the resulting {\em $v$-SD order} modifies SSD by including only test utilities $u$ that are strictly increasing, twice differentiable, and at least as risk averse as $v$:    
$-\frac{u''}{u'}\ge -\frac{v''}{v'}$.
If $v$ is an exponential utility as above with constant absolute risk aversion, we recover the stochastic orders recently studied by \cite{Huang} as interpolations between classical integer-degree stochastic orders.
However, $v$-SD orders differ markedly from approaches controlling marginal utilities $u'$ as in \cite{LeshnoLevy} and \cite{Mueller}.
For a comparison of various fractional stochastic orders, see \cite{Fractional}.
Note that bounding the absolute risk aversion of the decision collective from below is universal and also covers cases in which risk aversion is bounded from above, or assumed to be sandwiched between two bounds; see Appendix~\ref{sec:otherMeyer}.

Nevertheless, SSD- and $v$-SD-orders share the significant drawback that they are incomplete and do not produce a ranking within arbitrary pairs of random prospects.
The solution is to use a {\em risk measure}---see \cite{ADEH} and \cite[Chapter 4]{FoeSch}---to summarise prospects with a single value and choose the one with lower risk. 
In contrast to using solely a $v$-SD-order, one always reaches a decision with this approach.
Risk measures typically comply with the preference of more to less by definition, and many practically relevant examples {\em respect} SSD by weakly ordering payoffs that are already weakly ordered in SSD, like in case of the ES class.\footnote{~For an axiomatisation of ES, see \cite{WangZitikis}, and for discussions of its practical use, see \cite{Embrechts,Koch-Medina}. For the interplay between SSD and portfolio selection based on risk measures, see \cite{Levy-1992,Ogryczak,Gotoh}.} 

To the best of our knowledge, risk measures that respect $v$-SD orders have not been studied. 
We therefore introduce and analyse these risk measures, which we call {\em Meyer risk measures} in honour of Meyer's original contribution.
Our main findings, summarised below, fall into three categories. 

In Section~\ref{sec:representation_vsd}, we focus on the case when threshold utility $v$ exhibits constant absolute risk aversion (is CARA), both at a theoretical and a practical level. 
Our first main result, Theorem~\ref{thm:repEXPSD}, establishes the existence of nontrivial Meyer risk measures, provides a full characterisation, and shows that they arise from adjusted risk measures studied in \cite{AlexanderLaudageSass}.
These Meyer risk measures encompass the most well-known adjusted risk measures---Loss VaR of~\cite{bignozzi} and the adjusted ES of~\cite{bignozzi}---as special cases, and simultaneously offer additional flexibility through their dependence on a risk aversion parameter of the underlying CARA threshold utility. On the practical side, we compare all three concepts in several risk management settings, where Meyer risk measures better capture heavy tails (see Figure~\ref{fig:step_functions_1}). 
A connection to SSD-consistent return risk measures (RRMs) tests Meyer risk measures on log-returns of real-world data and relates the concept to further existing results.

Second, $v$-SD orders also align with the recent work of \cite{Muetal} on monotone additive statistics. In Section~\ref{sec:Mu}, we show that such statistics are often Meyer risk measures, particularly under CARA threshold utilities, as formalised by our second main result, Theorem~\ref{thm:Mu}. This highlights the relevance of Meyer risk measures for a broad range of decision-making problems, including those in~\cite{Muetal}.

Finally, in Section~\ref{sec:generalThresholdUtility}, we discuss Meyer risk measures for the general case of an arbitrary threshold utility function. This leads to our third main contribution: a collection of three impossibility results for Meyer risk measures. 
The first, Theorem~\ref{thm:unbounded} shows that the unrealistically conservative worst-case risk measure is the only Meyer risk measure if the threshold risk aversion becomes unbounded in the limit. 
This observation underscores the special role of CARA threshold utilities, since beyond them the class of Meyer risk measures is quite restricted.
The other two impossibility results, Theorems~\ref{thm main 1} and~\ref{thm main 2}, concern additional properties risk measures often enjoy in applications: positive homogeneity (as for ES and VaR) and a positive attitude to diversification. With the exception of the worst-case risk measure, positive homogeneity and $v$-SD-consistency are typically incompatible. 
Likewise, properties such as convexity and star-shapedness force risk to grow at the maximal rate under leverage, so that many convex risk measures, including ES, are not $v$-SD-consistent for nontrivial $v$. For completeness, we provide representations of Meyer risk measures as the lower envelope of a family of ``base risk measures".

The manuscript is structured as follows:  Section~\ref{sec:preliminaries} recalls $v$-SD-orders and defines Meyer risk measures.
Section~\ref{sec:representation_vsd} establishes their existence with respect to CARA threshold utility functions and discusses implications in risk management. Section~\ref{sec:Mu} links Meyer risk measures to monotone additive statistics, while Section~\ref{sec:generalThresholdUtility} develops the impossibility results for general threshold utilities. Section~\ref{sec:conclusionOutlook} concludes and outlines future research directions. 
For the reader's convenience, technical analyses and proofs are collected in appendices.
	     
\section{Preliminaries and notation}\label{sec:preliminaries}

Throughout this paper,  $(\Omega, \mathcal{F}, \mathbb{P})$ denotes a fixed underlying atomless probability space. 
The space of all real-valued (bounded) random variables on $\Omega$ is denoted by $\CL$ ($\CX$, respectively). 
Unless specified otherwise, all (in)equalities between random variables hold almost surely (a.s.)~under $\P$. 
Positive and negative part of a random variable $X$ are denoted by $X^+$ and $X^-$, respectively. 
The subspace of $\CL$ of all random variables for which positive and negative part have finite expectation is $\CL^1$. 
Throughout, we distinguish between increasing and nondecreasing scalar functions and, symmetrically, between decreasing and nonincreasing functions. 

\subsection{Risk measures}

A {\em (monetary) risk measure} is a functional $\rho\colon \mathcal X\to\R$ satisfying
\begin{description}
\itemsep 0em
\item[antitonicity:] $\rho(X)\ge \rho(Y)$ holds for all $X,Y\in\CX$ with $X\le Y$; and 
\item[cash-additivity:]$\rho(X+c)=\rho(X)-c$ holds for all $X\in\CX$ and $c\in\R$.
\end{description}
For a thorough introduction to monetary risk measures, see \cite[Chapter~4]{FoeSch}.
While risk measures have been studied on spaces different from $\CX$ containing unbounded random variables, the present setting is convenient and sufficient for our purposes.

Unless stated otherwise, the argument $X$ of a risk measure $\rho$ represents a random future payoff. The quantity $\rho(X)$ is interpreted as the capital that needs to be held in combination with the financial position $X$ to limit risk to an acceptable level. This can be seen from the representation 
\begin{center}$\rho(X)=\inf\{m\in\R\mid X+m\in\CA_\rho\}$,\end{center}
where the {\em acceptance set}
$\CA_\rho:=\{X\in\CX\mid \rho(X)\le 0\}$
collects all payoffs that are already adequately capitalised and do not require additional capital injections. 
Financial positions $X\in\CX$ model payoffs or net gains. 
Antitonicity reflects that an a.s.~larger payoff should not require more risk capital to reach acceptability.
Cash-additivity means that an additional sure payoff reduces the required capital by precisely that amount.

The interpretation of risk measures as capital requirements makes also other properties economically meaningful. 
A monetary risk measure $\rho$ is called:
\begin{enumerate}[label=\textnormal{(\alph*)}]
\itemsep-1pt 
    \item {\em law invariant} if payoffs with the same distribution under $\P$ are assigned the same risk; 
    \item {\em positively homogeneous} if $\rho(tX)=t\rho(X)$ for all $X\in \CX$ and $t>0$;
    \item {\em convex} if $\rho(\lambda X+(1-\lambda)Y)\le \lambda\rho(X)+(1-\lambda)\rho(Y)$ for all $X,Y\in \CX$ and $\lambda\in [0,1]$; 
    \item {\em normalised} if $\rho(0)=0$;
    \item 
    {\em star-shaped} if it is normalised and $\rho(\lambda X)\ge \lambda\rho(X)$ holds for all $X\in\CX$ and $\lambda\geq 1$.
\end{enumerate}

Positive homogeneity reflects the absence of liquidity risk: the risk increases or decreases proportionally with the exposure to a given payoff.
Convexity means that diversification does not increase the capital requirement and that merge-and-downsize strategies are favoured. 
It is a stronger property than star-shapedness, which encourages downsizing, but not necessarily merging; see \cite{Castagnoli}.
Consequently, star-shapedness, in contrast to positive homogeneity, does not rule out the occurrence of liquidity risk. 

The following are examples of risk measures, with {\em Expected Shortfall} and the {\em worst-case risk measure} being particularly important for the subsequent discussion. 
For $X\in\mathcal L$ define $M(X)$ as the smallest constant a.s.\ upper bound, i.e., 
\begin{equation}\label{def M}M(X):=\inf\{x\in\R\mid X\leq x\},\end{equation}
where as usual $\inf\varnothing =\infty$. The {\em worst-case risk measure} is defined by 
\begin{center}$\worst(X):=M(-X)$\end{center} 
and provides the most conservative risk assessment:  $\rho\le \worst+\rho(0)$ holds for all risk measures $\rho$. 
It has properties (a)--(e), 

Now, let $q_X(p):=\inf\{x\in\R\mid \P(X\le x)\ge p\}$, $p\in(0,1]$, denote the left quantile function of $X\in\CL$.
The {\em Expected Shortfall} (ES) at level $p\in[0,1]$, which satisfies properties (a)--(e), is 
        \begin{align*}
            \es_{p}(X):=\begin{cases}
                -\frac{1}{1-p}\int_{0}^{1-p} q_{X}(r)\diff r, & p\in[0,1),\\[-0.2ex]
                \worst(X), & p=1,
            \end{cases}\qquad X\in\CX.
        \end{align*}
   
\subsection{Stochastic orders}\label{sec:orders}

In this section and throughout the remainder of the paper, 
let $I\subseteq\R$ be a nonempty open interval and let $\mathcal U(I)$ denote the set of twice-differentiable functions on $I$ with positive first derivative. 
We call its elements utility functions. 
The Arrow-Pratt coefficient of absolute risk aversion of $u\in\mathcal U(I)$ is
\begin{equation*}
    R^A_u(x):=-\frac{u''(x)}{u'(x)},\quad x\in I.
\end{equation*}

Consider $X,Y\in\CL$ whose negative parts have finite expectation: $\E[X^{-}],\E[Y^{-}]<\infty$.  
We say that $Y$ dominates $X$ in {\em second-order stochastic dominance (SSD) relation} or {\em increasing concave order} ($X \ssd Y$) if for all $u\in\mathcal U(\R)$ with $R_u^A\ge 0$ such that the expectations of $u(X)$ and $u(Y)$ are finite, we have $\E[u(X)]\le \E[u(Y)]$.\footnote{~For $X \in \CL$, the condition $\E[X^-] < \infty$ is necessary for the existence of some $u \in \mathcal U(\R)$ such that $u(X)\in\CL^1$. 
Without, SSD is vacuous, as we could have $X \ssd Y$ and $Y \ssd X$ for all $Y \in \CL$.}
If $X,Y\in\mathcal L^1$, the above definition is tantamount to requiring that each 
risk-averse expected utility (EU) decision maker weakly prefers $Y$ to $X$; that is, for all nondecreasing and concave, but not necessarily differentiable utility functions $u$ on $\R$ delivering finite expectations of $u(X)$ and $u(Y)$, inequality $\E[u(X)] \le \E[u(Y)]$ holds.

In this paper, we go beyond the classical SSD relation and focus on the more general concept of {\em $v$-SD orders}.
These stochastic orders refine SSD and enable interpolation toward the adjacent first-order stochastic dominance. 
The most relevant theoretical references are \cite{Huang, Fractional} and \cite{Meyer}.\footnote{~Importantly, 
$v$-SD orders differ from alternative stochastic orders that constrain ratios of marginal test utilities, as in \cite{LeshnoLevy} and \cite{Mueller}.
See \cite{DeVecchi} for the connection of such orders to risk measures.}
To define these orders, fix a {\em threshold utility function} $v \in \mathcal U(I)$ and set   
\begin{equation*}
\mathcal U_v(I) := \{u \in \mathcal U(I) \mid R_u^A \ge R_v^A\},
\end{equation*} 
the class of test utility functions representing EU agents at least as risk averse as $v$. Moreover, define $\mathcal L^1_v(I)$ to be the set of all $X \in \mathcal L$ that are $\P$-a.s.\ $I$-valued and for which $v(X)$ has finite expectation. Whenever $I$ is unambiguous from the definition of $v$, we write $\CL^1_v$ for short. 

\begin{definition}\label{def:vSSD}
Given a threshold utility function $v \in \mathcal{U}(I)$ and $X,Y\in\mathcal L^1_v$, $Y$ dominates $X$ in the {\em $v$-SD order} ($X \vsd Y$) if $\mathbb{E}[u(X)] \le \mathbb{E}[u(Y)]$ holds for all $u \in \mathcal{U}_v(I)$ such that $X,Y\in\CL^1_u$.
\end{definition}

The concept of $v$-SD orders generalises the classical SSD order, which can be recovered from Definition~\ref{def:vSSD} by setting $v(x)=x$ for all $x\in\R$. 
In contrast to SSD, however, the threshold utility $v$ or the test utilities $u \in \mathcal U_v(I)$ are not required to be risk averse.  
Equivalently, no sign constraint is imposed on $R_v^A$.  
In particular, $v$ may be (locally) convex, allowing it to represent (locally) risk-loving behaviour.

By \cite[Theorem 8.1.3]{Stoyan}, we can state a parsimonious equivalent characterisation of $\vsd$. For $X,Y\in\CL^1_v$,
\begin{align}\label{complicated equiv}\begin{split}X\vsd Y\quad&\iff\quad v(X)\ssd v(Y)
\end{split}
\end{align}

Two kinds of threshold utility functions are fundamental for the present study. 
The first class are \textit{constant absolute risk aversion (CARA)} utilities defined on $I=\R$. They are represented by the parametrised family $(\mf e_\lambda)_{\lambda\in\R}$ of \textit{exponential utilities} on $\R$. The function $\mf e_\lambda$ is given by 

\begin{equation}\label{ex:orders}\mf e_\lambda(x):=\begin{cases}
    - \lambda e^{-\lambda x}&\text{if }\lambda\neq 0,\\[-0.8ex]
    x&\text{if }\lambda=0.
\end{cases}\end{equation}
Here, the constraint $R^A_{\mf e_\lambda}\equiv\lambda$ is a constant, and we will refer to the resulting stochastic orders as $\lambda$-SD-orders due to their relevance to our study.  
If $\lambda>0$, all test utilities $u\in\mathcal U_{\mf e_\lambda}(\R)$ are strictly concave (equivalently, strictly risk averse), i.e., the respective order is weaker than the classical SSD-relation on $\CX$.
The $0$-SD order agrees with the classical SSD order.
If $\lambda<0$, the set of test utilities $\mathcal U_{\mf e_\lambda}(\R)$ also contains (locally) risk-loving utilities, and the $\lambda$-SD-order is stronger than the SSD-order. 
These orders have been put forward, studied in detail, and generalised to higher orders in \cite{Huang} as a means to interpolate between first- and second-degree (or more general integer degree) stochastic orders. 

Another case of interest are \textit{constant relative risk aversion (CRRA)} utilities defined on ${I=(0,\infty)}$. For a parameter $a\in\R$, they are defined by 
\begin{equation}\label{ex:CRRA}\mf p_a(x) = \begin{cases} x^{a} / a&\quad\text{if }a\neq 0,\\[-0.8ex]
\log(x)&\quad\text{if }a=0.\end{cases}\end{equation}
Their relative risk aversion is given by $1-a$, while $R^A_{\mf p_a}(x)=\frac{1-a}x$ holds for all $x\in I$. 
Consequently, $\mf p_a$ displays strict, but decreasing absolute risk aversion if $a<1$, resulting in the fact that the $\mf p_a$-SD-order is weaker than the SSD order on $\CL^1_{\mf p_a}\cap\CX$. 
More precisely, it holds that 
\begin{equation}\label{eq:nested}\mathcal U_{\mf p_a}(I)\subseteq\mathcal U_{\mf p_b}(I),\quad 0\le a<b.\end{equation}

\subsection{Meyer risk measures}\label{sec:MeyerRMs}

This paper contributes to the extensive literature on the consistency of risk measures (or utility functionals) with stochastic orders; see the discussion below. 
More specifically, we introduce and analyse the consistency of functionals on random variables with $v$-SD orders, or $v$-SD-consistency for short.

\begin{definition}\label{def:consistent}
Let $v\in\mathcal U(I)$ be a threshold utility function and suppose $\mathcal D\subseteq \CX$ is nonempty. 
A functional $\ph\colon \mathcal D\to\R$ is {\em $v$-SD-consistent} if, for all $X,Y\in\mathcal D\cap \mathcal L^1_v$, 
$X\vsd Y$ implies $\ph(X)\ge\ph(Y).$

A {\em Meyer risk measure} is a risk measure for which an open interval $I\subseteq\R$ and a threshold utility function $v \in \mathcal{U}(I)$ exists such that the risk measure is $v$-SD-consistent.
When the threshold utility function $v$ is specified explicitly, we speak of a \textit{$v$-Meyer risk measure}.
\end{definition}

Meyer risk measures exist, although the most immediate example is a trivial one: the worst-case risk measure $\worst$ is $v$-Meyer for all threshold utilities $v$, as shown in Lemma~\ref{lem:worst}.
Also, various threshold utilities lead to consistency concepts  stronger or weaker than SSD, which is why Meyer risk measures encompass and generalise SSD-consistent ones. For example, the discussion below~\eqref{ex:orders} shows the following sequence of implications for functionals $\ph\colon \mathcal D\to\R$ and constant parameters $\gamma\le 0<\lambda$:
\begin{equation}\label{eq:cons1}\lambda\text{-SD-consistency}\quad\implies\quad\text{SSD-consistency}\quad\implies\quad \gamma\text{-SD-consistency}.\footnote{~In line with the convention adopted above, consistency with the $\lambda$-SD-order is abbreviated as $\lambda$-SD-consistency.}
\end{equation}

Likewise, if $\mathcal D$ is a suitable domain of positive random variables and $a\ge 0$, \eqref{eq:nested} yields the implications 
\begin{equation}\label{eq:cons_log}\mf p_a\text{-SD-consistency}\quad\implies\quad\mf p_0\text{-SD-consistency}\quad\implies\quad\text{SSD-consistency}.\end{equation}
 
Crucially, the literature often does not presume SSD-consistency itself, but law invariance of the risk measure in question. If the latter is additionally convex and lower semicontinuous, then the conjunction of these properties implies SSD-consistency. 
This implication is formulated in~\cite[Theorem 5.1]{CerreiaVioglio} for risk measures with domain $\CX$ just as in the present paper, and for domain $\mathcal L^1$ in~\cite[Theorem 4.1]{Dana} and~\cite[Theorem 4.4]{Baeuerle}. 
Not least,~\cite[Theorem 3.6]{Bellini} extend the result to risk measures defined on general spaces of random variables.\footnote{~This result is even more general, because it only requires quasiconvex functionals and does not impose monotonicity. The latter necessitates to use the {\em concave order} rather than SSD.}

Like the present manuscript, some studies link risk measures to stochastic orders different from SSD. 
\cite{YamaiYoshiba} consider risk measures consistent with higher-order stochastic dominance relations. 
The case of the so-called 3-convex order (3CO) is treated in \cite{3convex}. 
Their Theorem 14---extending \cite[Theorem 6.3]{Hurlimann}---demonstrates that, under a mild condition on the underlying probability space, the only 3CO-consistent distortion risk measure is the expectation functional. This triviality result shares the spirit of some of our main results here.

Given the prominence and applicability of the aforementioned results, several natural questions arise in the context of $v$-SD orders. 
Which risk measures are Meyer? How can we construct new classes of Meyer risk measures? 
Which further properties, such as convexity and positive homogeneity, are compatible with $v$-SD-consistency? 
In this study, we address all of these questions. Moreover, the generality of $v$-SD orders allows us to uncover several unifying principles.

\section{Meyer risk measures with constant threshold risk aversion}\label{sec:representation_vsd}

In this section, we focus on the case of CARA (exponential) threshold utility functions and address monetary risk measures which respect the resulting $\lambda$-SD-orders. Our contribution is twofold. First, we establish a representation for $\lambda$-SD-consistent risk measures. Second, we study their performance as a function of the risk aversion parameter $\lambda$ across different financial market setups, showing that they can lead to significant deviations in risk assessment compared to well-established risk measures.

\subsection{Minmax representations of $\lambda$-SD-consistent risk measures}

Theorem~\ref{thm:repEXPSD} characterises $\lambda$-SD-consistent risk measures exhaustively by means of minmax representations.
Result and proof build on and extend the approach of \cite{Consistent} 
to SSD-consistent risk measures.
Representation~\eqref{eq:rhog1} therefore coincides with the result of \cite{Consistent}. 

\begin{theorem}\label{thm:repEXPSD}
Suppose the threshold utility $v$ is CARA, i.e., $v=\mf e_\lambda$ for some $\lambda\in\R$. 
A risk measure $\rho$ is $\lambda$-SD-consistent if and only if there is a family $\mathcal{G}$ of nondecreasing functions $g\colon[0,1]\to(-\infty,\infty]$ with $g\not\equiv \infty$ such that $\rho$ can be represented as
\begin{equation}\label{eq:rhog_new}\rho(X) = \inf_{g\in\mathcal G}\sup_{p\in[0,1]}\big\{ \rho_{\lambda,p}(X)-g(p)\big\},\quad X\in\CX,\end{equation}
where
\[\rho_{\lambda,p}=-\mf e_\lambda^{-1}\circ(-\es_p)\circ \mf e_\lambda.\]
In particular, $\rho$ is SSD-consistent ($0$-SD-consistent) if and only if~\eqref{eq:rhog_new} specialises to 
\begin{equation}\label{eq:rhog1}\rho(X) = \inf_{g\in\mathcal G}\sup_{p\in[0,1]}\big\{\es_p(X)-g(p)\big\},\quad X\in\CX.\end{equation}
Moreover, the family $\CG$ can be chosen as
\[\mathcal G(\rho):=\{p\mapsto \rho_{\lambda,p}(Y)\mid Y\in\CA_\rho\}.\]
\end{theorem}

First, to situate the theorem within the existing literature, note that $\rho_{\lambda,0} = -\mf e_\lambda^{-1}\circ(-\E[\cdot])\circ \mf e_\lambda$ is the well-known \textit{entropic risk measure}. 
In case of $\rho_{\lambda,p}$ with $p>0$, the linear risk measure $-\E[\cdot]$ in the definition of the entropic risk measure is replaced by an ES.
Recall that $-\es_p(X) = \E[X\,|\,X\leq q_{X}(1-p)]$ for continuously distributed $X$. Hence, $-\rho_{\lambda,p}(X)$ is a tail certainty equivalent of the worst $(1-p)$-percent of possible outcomes for~$X$.  \cite{BaeuerleShushi} suggest the term \textit{tail conditional entropic risk measures} and apply these risk measures to various risk management problems, such as capital allocation and optimal reinsurance.\footnote{~The connection between $\rho_{\lambda,p}$ and entropic risk measures can be unified as follows: 
for any $p>0$ and $X\in\CX$, there is some probability measure $\Q_{X,p}$ absolutely continuous with respect to $\P$ such that $\es_p(\mf e_\lambda(X))=-\E_{\Q_{X,p}}[\mf e_\lambda(X)]$. 
Hence, $\rho_{\lambda,p}(X)$ is an entropic risk measure at $X$ under an $X$-specific change of measure.}

Second, keeping $g$ fixed, the inner supremum in~\eqref{eq:rhog1} is called {\em adjusted ES}  by \cite{burzoni}:
\begin{equation}\label{eq:adjustedES}
    \es^g(X):=\sup_{p\in[0,1]}\{\es_p(X)-g(p)\},\quad X\in\CX.
\end{equation}
These maps are the building blocks of SSD-consistent risk measures. 
Adjusted risk measures---obtained by replacing the ES family with an alternative parametric family of risk measures---has been studied in full generality in~\cite{AlexanderLaudageSass}.
The inner suprema in \eqref{eq:rhog_new} define such an adjusted risk measure, with 
$(\rho_{\lambda,p})_{p\in[0,1]}$ as the defining family of risk measures, a case that, to the best of our knowledge, has not yet been studied. 

In the remainder of this section, we examine the risk measures in~\eqref{eq:rhog_new} in more detail and discuss their performance in different (real-world) applications. We begin by discussing several possible choices for $\mathcal{G}$. In line with the literature on adjusted risk measures, we also refer to the functions in $\mathcal{G}$ as \textit{target risk profiles}. 
For each chosen $\mathcal{G}$, we obtain situations in which a $\lambda$-SD-consistent risk measure offers a valid alternative to a Loss VaR or an adjusted ES, highlighting the potential of $\lambda$-SD-consistent risk measures as meaningful alternatives to these established risk measures.

Motivated by the discussion of the adjusted ES in~\cite{burzoni} and~\cite[Example 3.1]{Consistent}, a first natural choice for target risk profiles are step functions. Under this choice, the inner suprema above reduce to maxima of a finite number of risk measures and are therefore easy to implement in practice, for instance to model regulatory capital requirements. 
In the simplest case $g(p) = \infty \cdot \ind_{(q,1]}(p)$, the adjusted ES in~\eqref{eq:adjustedES} simplifies to $\es_q$. Choosing $q = 97.5\%$, one recovers the ES used in the Basel Accords. If the step function has more than one jump, the adjusted ES is given by the maximum of ES values minus the corresponding jump sizes. This case is also of regulatory relevance: a regulator may impose several relevant ES levels, or a company may jointly consider the ES level mandated by regulation and the ES level of an internal model. This choice is analysed for $\lambda$-SD-consistent risk measures of shape~\eqref{for later 2} in Section~\ref{sec:performanceStepFunctions} below.

Second, there is a close connection to ``benchmark risk measures'' of Lemma~\ref{lem:benchmark}, defined for a fixed benchmark payoff $Y\in\CX$ and a suitable preference relation $\peq$ on $\CX$ as
\begin{equation}\label{eq:benchmarkrisk}\rho_\peq(X|Y):=\inf\{m\in\R\mid Y\peq X+m\}.\end{equation} 
If one chooses the function $g$ by $g(p)=\es_p(Y)$, the adjusted ES obtains as benchmark risk measure under choosing SSD for $\peq$:
\begin{equation}\label{for_later_new}\es^g(X)=\rho_{\mathrm{SSD}}(X|Y)=\inf\{m\in\mathbb{R}\mid Y\ssd X+m\}.\end{equation}
This {\em benchmark-adjusted ES} allows to compare the ES profiles of payoffs $X$ and $Y$.
If $Y$ refers to a benchmark index or a comparable financial instrument, such benchmark-adjusted ESs are well suited for portfolio risk management.
If we instead identify $\peq$ with $\le_{\lambda}$ and set $g(p)=\rho_{\lambda,p}(Y)$ for some $Y\in\CX$,
\begin{align}
\label{for later 2}\sup_{p\in[0,1]}\big\{ \rho_{\lambda,p}(X)-g(p)\big\}&=\rho_\lambda(X|Y)=\inf\left\{ m \in \R \mid Y \le_{\lambda} X + m \right\}. 
\end{align}

By Lemma~\ref{lem:base} and Proposition~\ref{prop:Pratt}, the risk measures in~\eqref{for_later_new} and~\eqref{for later 2} are convex, SSD- or $\lambda$-SD-consistent, respectively, and compare the two payoffs in the corresponding stochastic order.  Risk-aversion parameter $\lambda$ in~\eqref{for later 2} controls whether the comparison uses a narrower ($\lambda>0$) or broader ($\lambda<0$) class of test utilities. These risk measures thus form the simplest and most fundamental class of consistent convex risk measures, and we study them in Sections~\ref{sec:benchmarkProfiles}--\ref{sec:motivation}. 

A third option is discussed in Section~\ref{sec:systemicRisk} and defines set $\mathcal G$ as benchmark ES profiles of {\em aggregated} positions. 
This allows to identify the contribution of a single position to the systemic risk of an entire network of firms. 

In sum, we compare the performance of $\lambda$-SD-consistent risk measures with that of well-established risk measures like the adjusted ES and the {\em Loss VaR} of~\cite{bignozzi} and also illustrate their value for various risk-management purposes.

\subsection{Performance under a fixed family of target risk profiles}\label{sec:performanceStepFunctions}

We start by fixing the family $\mathcal{G}$ of target risk profiles and isolate the effect of varying the risk-aversion parameter $\lambda$ on the associated Meyer risk measure.
This yields the following quantitative comparison of $\lambda$-SD-consistent Meyer risk measures with the adjusted ES.

\begin{corollary}\label{cor:orderMeyerRiskMeasures}
    Let $\mathcal{G}$ be a family of nondecreasing functions $g\colon[0,1]\rightarrow(-\infty,\infty]$ satisfying $g\not\equiv\infty$, and let ${\lambda <0<\gamma}$. 
    Then, for all $X\in\CX$,
    \begin{equation}
    \begin{alignedat}{2}
    &\inf_{g\in\mathcal G}\LVaR^g(X)
    &&\leq \inf_{g\in\mathcal G}\sup_{p\in[0,1]}
       \{\rho_{\lambda,p}(X)-g(p)\}\\
    \leq~&\inf_{g\in\mathcal G}\es^g(X)
    &&
    \le \inf_{g\in\mathcal G}\sup_{p\in[0,1]}
       \{\rho_{\gamma,p}(X)-g(p)\}.
    \end{alignedat}
    \label{eq:bounds_e_c}
    \end{equation}
    Here, $\LVaR^g(X) = \sup_{p\in[0,1]}\big\{-q_X(1-p)-g(p)\big\}$, using $q_X(0):=-M(-X)$.
\end{corollary}

Loss VaR is consistent with first-order stochastic dominance (FSD), and the adjusted ES with SSD. For $\lambda<0$---just like the $\lambda$-SD-order---$\lambda$-Meyer risk measures mediate between FSD- and SSD-consistency. This natural extension allows flexible control of how conservative capital requirements should be. For $\lambda>0$, one departs from SSD by excluding EU agents whose risk aversion is too low, leading to a capital requirement that exceeds the adjusted ES.

As mentioned earlier, the case of \textit{step functions} can be used for regulation.
If we use the same step function for all risk measures (independent of the risk aversion $\lambda$), then we are in the situation of Corollary~\ref{cor:orderMeyerRiskMeasures}. 
We test them for simple illustrative balance sheet models, imposing a lognormal distribution on assets, modelling liabilities via gamma (light tail), lognormal (medium tail), and Pareto distributions (heavy tail), and computing equity capital as the value of assets minus the value of liabilities. Table~\ref{tab:step_functions_model_parameters} collects the chosen distribution parameters.\footnote{~The absolute value of equity capital is bounded by $10,000$. This bound ensures that the underlying distribution is essentially bounded, and it is large enough not to affect the simulated equity capitals in our example.}
For a fair comparison, the first two  moments of the gamma and the lognormal liability distributions are calibrated to coincide.  

\begin{table}[ht]
\centering
\footnotesize
\begin{tabular}{lccccc}
  \hline
  & Log-mean & Log-standard deviation & Shape & Scale & Location \\ 
  \hline
Lognormal (assets) & $2.323$ & $0.200$ & - & - & - \\ 
Lognormal (liabilities) & $0.837$ & $0.693$ & - & - & - \\
Gamma & - & - & $2.000$ & $1.333$ & - \\
Pareto & - & - & $3.750$ & - & $1.000$ \\
   \hline
\end{tabular}
\vspace{0.25cm} 
\caption{\footnotesize Distribution parameters to illustrate $\lambda$-SD-consistent risk measures for step functions.}
\label{tab:step_functions_model_parameters}
\end{table}

\begin{figure}[ht]
    \centering
    \subfigure{
        \includegraphics[width=0.45\textwidth]{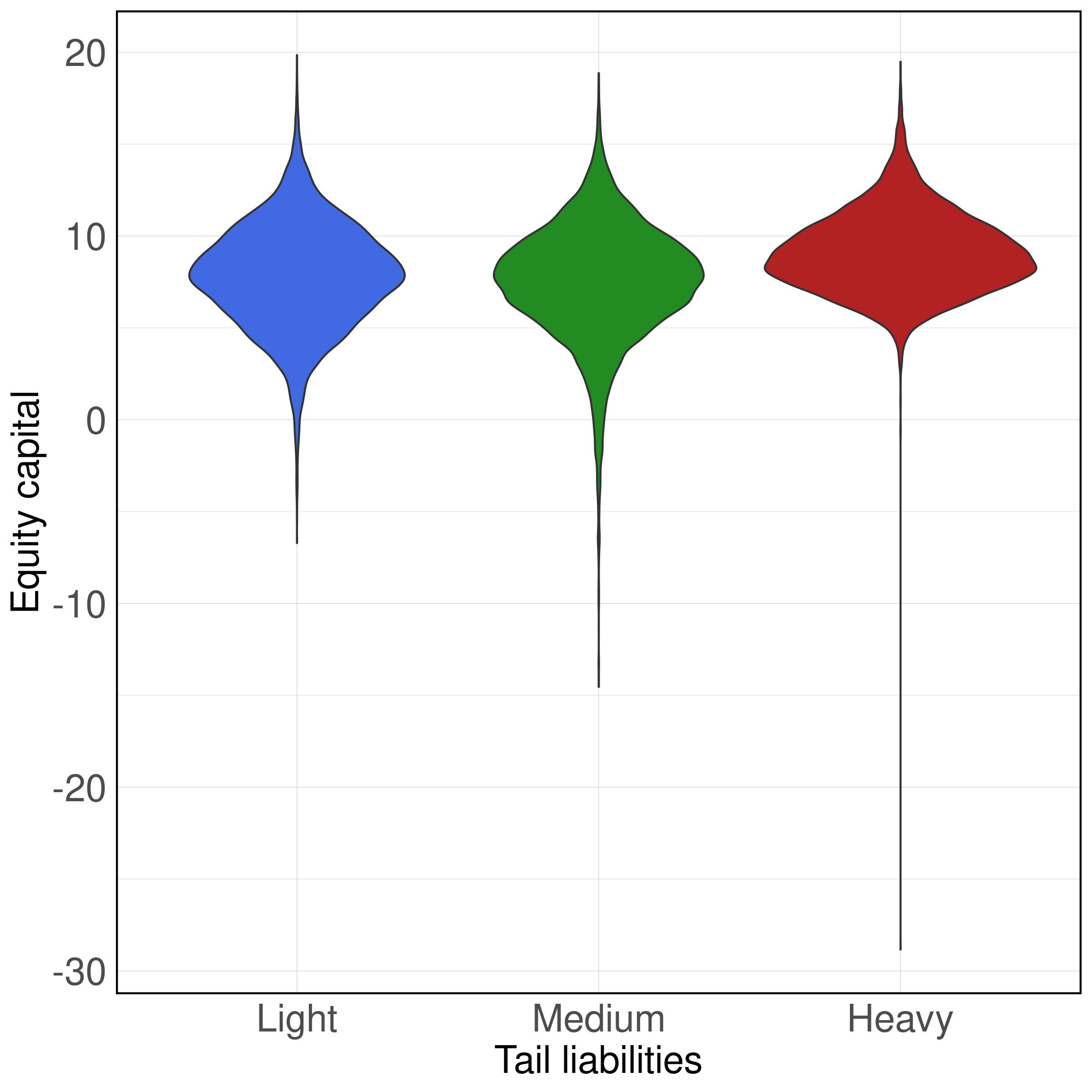}
    }
    \subfigure{
        \includegraphics[width=0.45\textwidth]{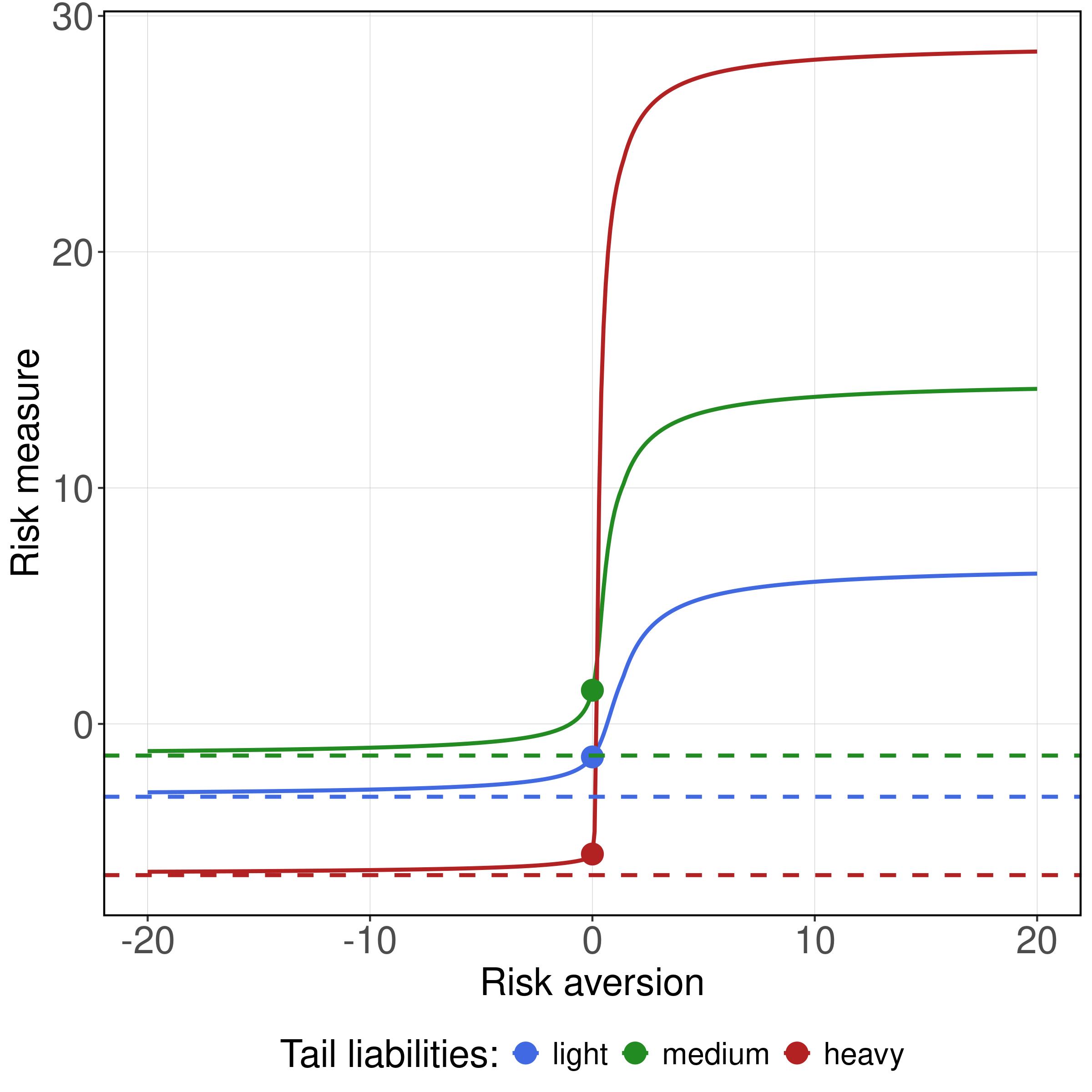}
    }
    \caption{\footnotesize Violin plots of simulated equity capitals (left), and corresponding $\lambda$-Meyer risk measures (right). Dashed lines and dots represent Loss VaR and adjusted ES, respectively.}
    \label{fig:step_functions_1}
\end{figure}

The simulated equity capitals are illustrated on the left-hand side of Figure~\ref{fig:step_functions_1}. All three distributions lead to significantly different minimum values. This difference in the tails will play a crucial role in our analysis.

Let $p_1,p_2\in(0,1)$ with $p_2<p_1$, $c>0$, and $\lambda\in\R$. 
Setting $g(p) = c\cdot\mathbf{1}_{[p_2,p_1)} +\infty\cdot \mathbf{1}_{(p_1,1]}$, we consider the $\lambda$-Meyer risk measure based on $\mathcal{G} = \{g\}$ and risk-aversion parameter $\lambda\in\R$, i.e., 
\[
    \rho(X) = \max\{\rho_{\lambda,p_1}(X)-c,\rho_{\lambda,p_2}(X)\},\quad X\in\CX.
\]
First, we test the values
\[
    p_1 = 0.975,\quad p_2 = 0.9,\quad c = 1.
\]
These risk measures are illustrated on the right-hand side of Figure~\ref{fig:step_functions_1} as a function of $\lambda$. 
Dots and dashed lines indicate the values of Loss VaR and adjusted ES.
By the inequalities in~\eqref{eq:bounds_e_c}, we obtain that the Meyer risk measure with {\em risk-seeking} CARA threshold utility lies between the Loss VaR as a lower bound and the adjusted ES as an upper bound, i.e.,~the Meyer risk measures interpolates between those two values for $\lambda <0$.
The convergence of the curves towards $\LVaR^{g}(X)$ when $\lambda\rightarrow-\infty$ and towards the worst case risk measure $\worst(X)$ when $\lambda\rightarrow-\infty$ is shown in Corollary~\ref{cor:limitResultLVaR}. The latter also shows that such curves are always nondecreasing and continuous. More strikingly, 
the adjusted ES fails to detect the most dangerous tail, namely the heavy tail of the Pareto distribution.
In contrast, the Meyer risk measure at risk-aversion level $\lambda>0$ reacts strongly to the heavier tail and leads to the expected order of capital requirements, i.e.,~largest and smallest capital requirements for Pareto and gamma distribution, respectively. 

The previous observation is a main finding of this study, as it illustrates the potential of $\lambda$-SD risk measures to capture heavy-tailed behavior in cases where an adjusted ES fails. Another advantage is that their calculation is straightforward and not more complex than computing a classical ES. 
This makes the approach suitable for application in practice. Concluding, the new approach is a meaningful add-on for existing risk management procedures, like loss VaR and adjusted ES. 

\begin{figure}[ht]
    \centering
    \subfigure{
        \includegraphics[width=0.45\textwidth]{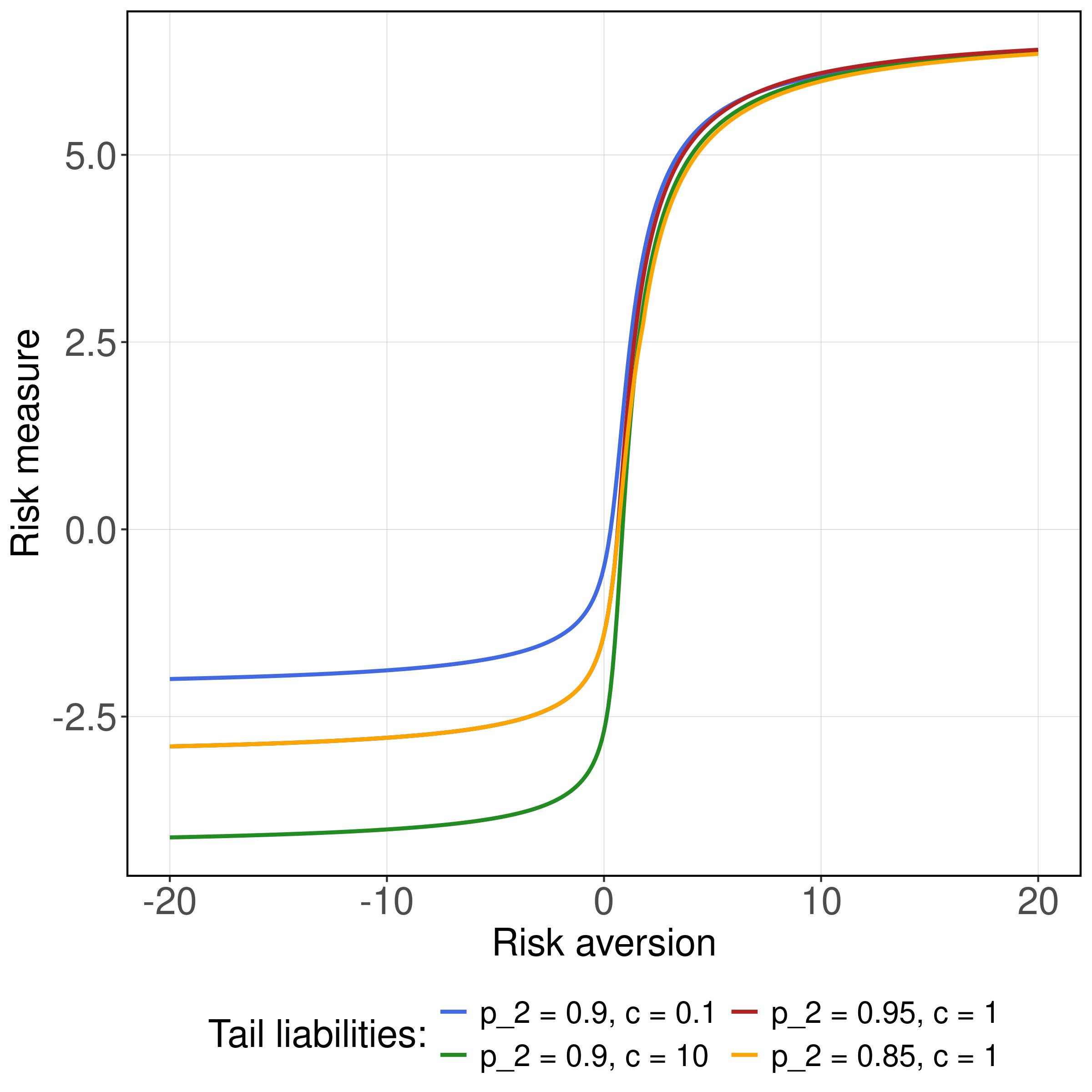}
    }
    \subfigure{
        \includegraphics[width=0.45\textwidth]{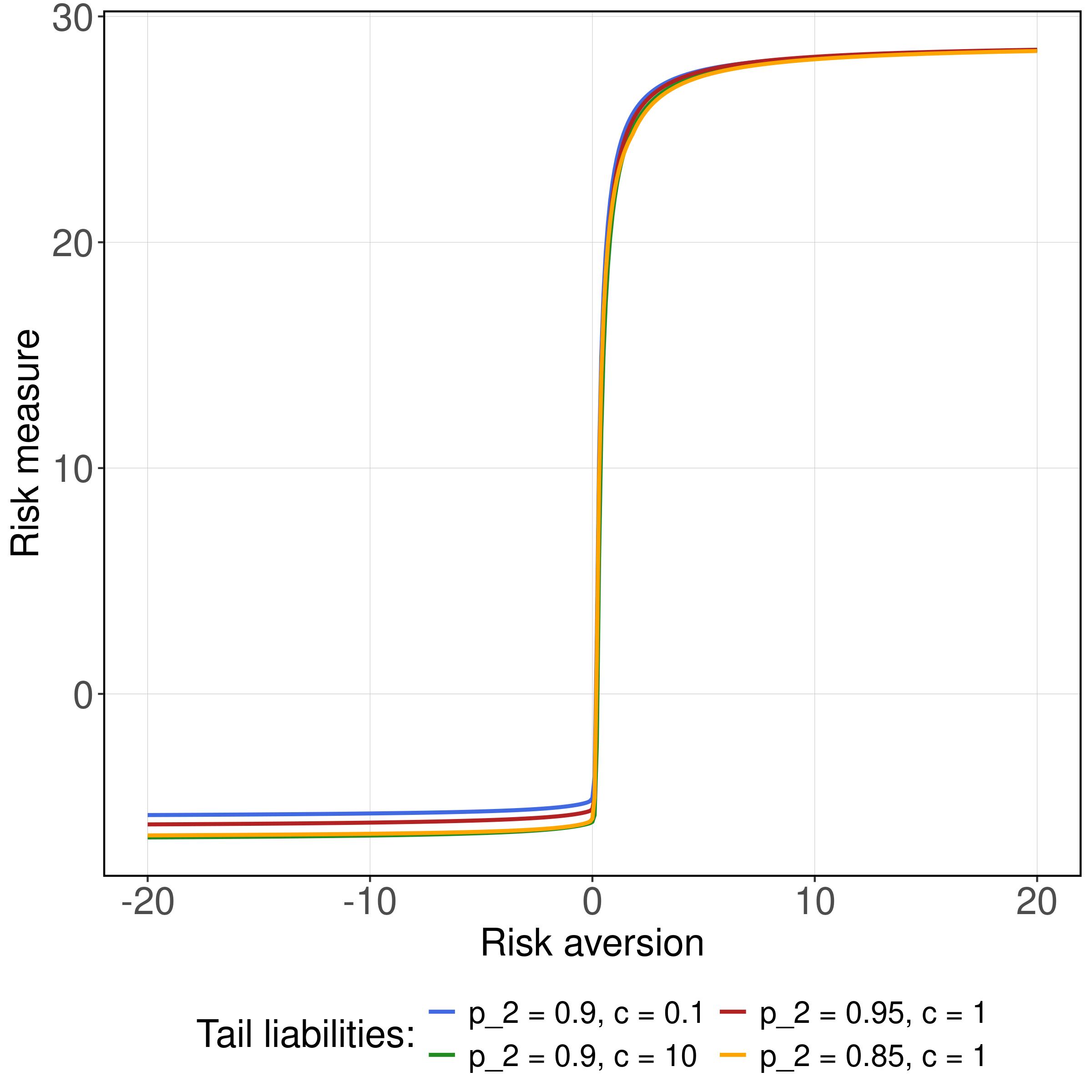}
    }
    \caption{\footnotesize Plots of $\lambda$-SD-consistent risk measures for gamma  (left) and Pareto (right) (liability) distributions.}
    \label{fig:step_functions_2}
\end{figure}

To illustrate robustness of these findings, 
we also alter the parameters $p_2$ and $c$ of the step function $g$; see Figure~\ref{fig:step_functions_2}. In both cases, but particularly for the gamma distribution, changing~$c$ has a strong impact if $\lambda <0$. Nevertheless, the qualitative shape of the curves --- in dependence on $\lambda$ --- does not change. 

\subsection{Performance under benchmark profiles}\label{sec:benchmarkProfiles}

In this section, we show that the behavior described in Section~\ref{sec:performanceStepFunctions} changes when $g$ is also allowed to vary in dependence of the risk aversion. 
The latter is more natural when risk measures are used in an investment context where the investor aims to beat a benchmark index. 
If $Y$ is the future payoff of this index and $\lambda\in\R$ is the risk aversion, one would use $g(p) = \rho_{\lambda,p}(Y)$. 
We illustrate this setting with an artificial portfolio in a classical three-asset Black-Scholes market model and select one of the underlying assets as benchmark. One can think of two stocks included in a specific index and the index itself.

We use stylised market parameters chosen to yield realistic orders of magnitude.
The constant interest rate is assumed to be zero and the market price of risk vector is $\gamma = (0.2\ 0.15\ 0.1)^{\intercal}$. 
Moreover, we assume asset volatilities of $0.40$, $0.15$ and $0.25$, and the correlation matrix of the log-returns being
\begin{equation*}
\begin{pmatrix}
1.00 & 0.00 & 0.90 \\
0.00 & 1.00 & -0.20 \\
0.90 & -0.20 & 1.00
\end{pmatrix},
\end{equation*}
i.e.,~the first two assets are uncorrelated, the first and third asset are strongly correlated and the second and third asset are slightly negatively correlated. 
This can be achieved by the following volatility matrix:
\begin{equation*}
\Sigma =
\begin{pmatrix}
0.40 & 0.00 & 0.00 \\
0.00 & 0.15 & 0.00 \\
0.225 & -0.05 & 0.10
\end{pmatrix}.
\end{equation*}
Moreover, the initial wealth is assumed to be $X_0=10$. 
Then, for the terminal wealth $X_T(\pi)$ at maturity $T>0$ and for a vector $\pi\in\R^3$ (constant portfolio process) it holds that 
\begin{equation*}
X_T(\pi) \overset{d}{\sim} X_0 \exp\left(\left(\pi^{\intercal}\Sigma \gamma - \tfrac{1}{2}\pi^{\intercal}\Sigma\Sigma^{\intercal}\pi \right)T + \sqrt{\pi^{\intercal}\Sigma\Sigma^{\intercal}\pi}\sqrt{T}Z\right),\quad Z\sim N(0,1).
\end{equation*}

The portfolio processes $\pi_1$, $\pi_2$, $\pi_3$ and $\pi^B$ generate three risky payoffs as well as the target risk profile $g(p) = \rho_{\lambda,p}\big(X_T(\pi^{B})\big)$: 
\begin{equation*}
    \pi_1 = \begin{pmatrix}
1.0 \\
0.0 \\
0.0
\end{pmatrix},\quad \pi_2 = \begin{pmatrix}
0.0 \\
1.0 \\
0.0
\end{pmatrix},\quad \pi_3 = \begin{pmatrix}
0.5 \\
0.5 \\
0.0
\end{pmatrix}, \quad \pi^B = \begin{pmatrix}
0.0 \\
0.0 \\
1.0
\end{pmatrix}.
\end{equation*}

\begin{figure}[ht]
    \centering
    \subfigure{
        \includegraphics[width=0.45\textwidth]{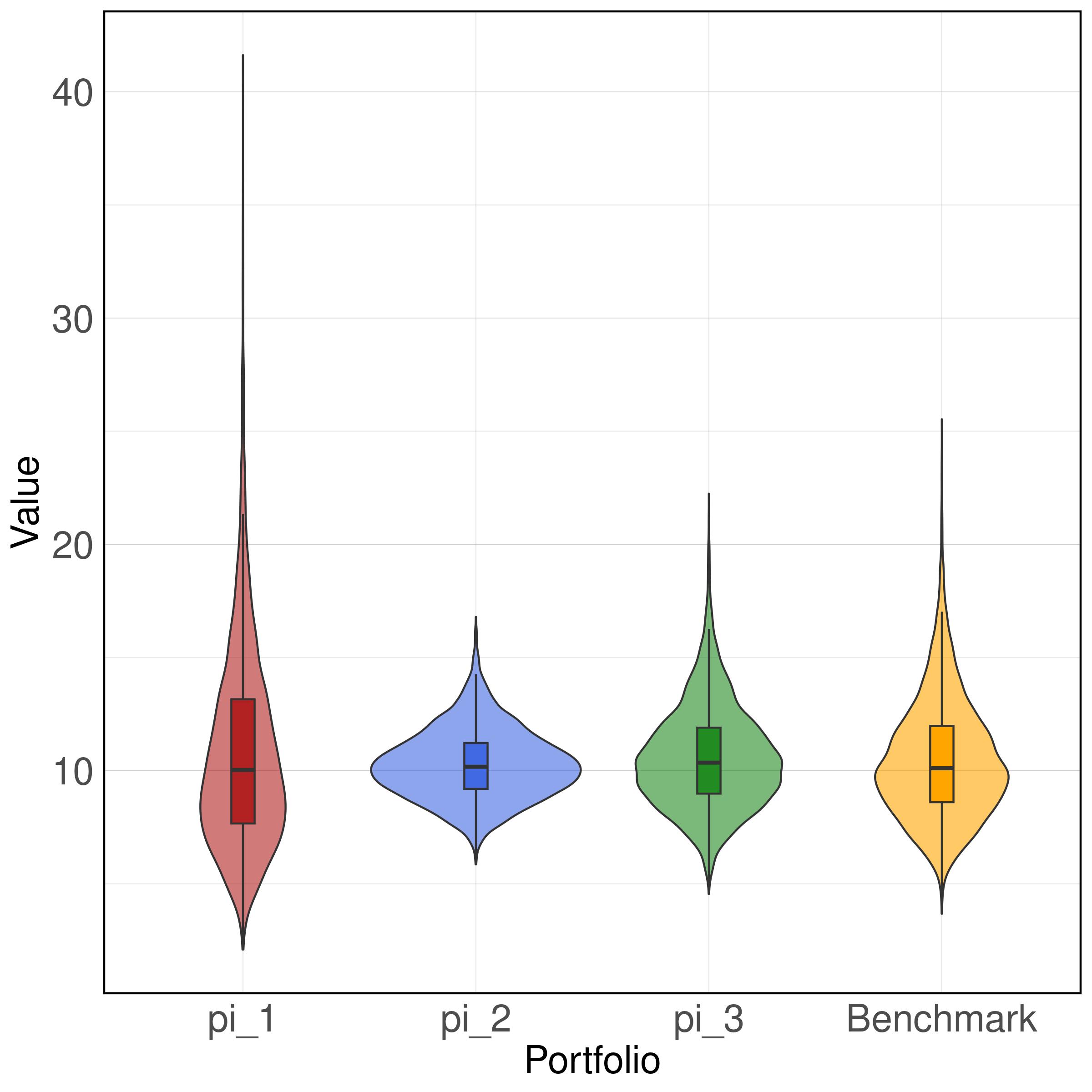}
    }
    \subfigure{
        \includegraphics[width=0.45\textwidth]{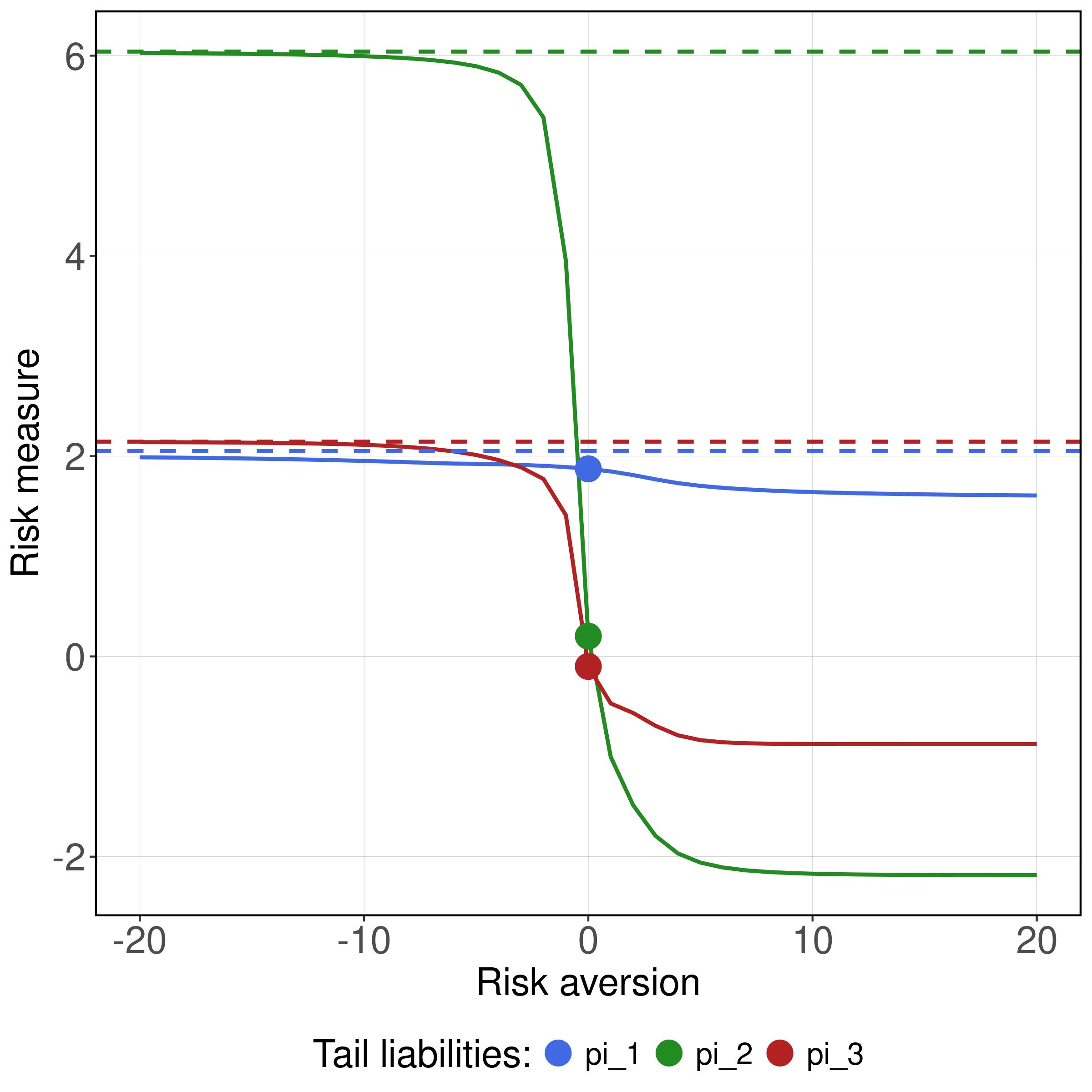}
    }
    \caption{\footnotesize Violin and boxplots of simulated portfolio values (left), and corresponding $\lambda$-Meyer risk measures (right). Dashed lines and dots are the Loss VaRs and adjusted ESs, respectively.}
    \label{fig:benchmark_profiles_1}
\end{figure}

On the left-hand side in Figure~\ref{fig:benchmark_profiles_1}, one can find the violin and boxplots of the corresponding portfolio values. The simulated values corresponding to $\pi_3$ are similar to the ones for the benchmark case. The values related to $\pi_1$ spread the most. 
Interestingly, the new risk measures on the right-hand side behave completely differently from the curves in Figure~\ref{fig:step_functions_1} above. 
In particular, compared to Corollary~\ref{cor:orderMeyerRiskMeasures}, the situation changes drastically.

First, the curve for $\pi_1$ is almost flat in the risk aversion $\lambda$, owing to the high correlation between the first and third asset. Second, for the adjusted ES, portfolio $\pi_3$ requires the smallest reserve. However, when the risk aversion parameter is positive ($\lambda > 0$), portfolio $\pi_2$ results in the smallest capital requirement. Recall that large values of $\lambda$ correspond to very risk-averse agents, who then prefer the portfolio with the smallest volatility. For negative values $\lambda<0$ the pattern reverses: $\pi_2$ now yields the highest capital requirement, as it has the smallest excess return, while $\pi_1$, with the largest excess return, gives the lowest capital requirement. Third, irrespective of the sign of $\lambda$, the smallest capital requirement is not attained by the most diversified portfolio, reflecting that the new risk measures are generally not convex. In time-series applications, where logarithmic returns are used as inputs, this lack of convexity is less relevant; this case is analysed in Section~\ref{sec:motivation} below. 

In conclusion, in the framework where $g$ depends on the risk aversion parameter, using $\lambda$-SD risk measures as portfolio selection  criterion allows an investor to control---through the risk aversion parameter---the extent to which portfolios containing assets with higher volatilities are preferred. Hence, our new $\lambda$-SD-consistent risk measures introduce new flexibility in classical portfolio optimization procedures.

\subsection{$\lambda$-SD-consistent risk measures for logarithmic values}\label{sec:motivation}

In time-series analysis it is common to use logarithmic returns as inputs to classical monetary risk measures; see, e.g.,~\cite{McNeil} and \cite{candia}.  
Notably, the transition from payoffs to logarithmic returns links $\lambda$-SD-consistent risk measures to the study of \emph{return risk measures} (RRMs), first introduced in~\cite{Return}. These RRMs are then consistent with orders arising from power (CRRA) threshold utilities.

The investigation of theoretical properties of RRMs has attracted increasing attention in recent years.  
\cite{Zullino} derive a range of representation results for RRMs, while~\cite{Laudage} propose an extension of RRMs based on achieving a specified acceptability criterion by securitisation involving multiple eligible assets.

Proposition~\ref{prop:ssd1} below demonstrates that an RRM is SSD-consistent if and only if an associated (monetary) risk measure is $-1$-SD-consistent, directly linking RRMs to the new concept of Meyer risk measures.
A key distinction between RRMs and risk measures is that the former are defined on the cone
\[
\CC := \{X \in \CX \mid X > 0~\text{a.s.}\}
\]
of strictly positive elements of $\CX$, rather than the full space.
We interpret the elements of $\CC$ as future payoffs.
For later, we also define the cone of elements of $\CX$ bounded away from 0, 
$\mathcal E=\{e^Y\mid Y\in\CX\},$
which is the norm interior of the positive cone of $\mathcal{X}$. 
Given a risk measure~$\rho$ on $\CX$ and a payoff $X \in \CE$ (i.e., $\log(X) \in \CX$), consider
\begin{equation}\label{rep2}
    \eta(X) := (\exp \circ \rho)(\log(X)).
\end{equation}
This expression captures the core idea behind RRMs: logarithmic values are first passed through a classical risk measure~$\rho$, and the subsequent application of the exponential function restores the monetary scale of~$X$, thereby inducing properties that reflect those of~$\rho$ itself. 
In particular, antitonicity of $\rho$ carries over to $\eta$: all $X, Y \in \CE$ such that $X\leq Y$ satisfy $\eta(X) \ge \eta(Y)$.
Similarly, cash-additivity of $\rho$ becomes $-1$-positive homogeneity of $\eta$, i.e., scaling a payoff reduces the RRM value by the inverse factor---for all $t > 0$ and $X \in \CE$, $\eta(tX) = t^{-1} \eta(X)$. 

These properties are used to define RRMs axiomatically on the set $\CC$, without reference to any specific monetary risk measure. 
A functional $\eta \colon \CC \to (0,\infty)$ is an RRM if it is:
\begin{description}
\itemsep 0em
    \item[antitone:] $\eta(X) \ge \eta(Y)$ when $X, Y \in \CC$ and $X \le Y$; and
    \item[$-1$-positively homogeneous:] For all $t > 0$ and $X \in \CC$, $\eta(tX) = t^{-1} \eta(X)$.
\end{description}
Moreover, every RRM $\eta$ admits a representation as in \eqref{rep2} on $\CE$ via a risk measure $\rho_\eta$ on $\CX$. The latter is given by 
\begin{equation}\label{def rhoeta}\rho_\eta(Y) = (\log \circ \eta)(\exp(Y)).\end{equation} 

The next result establishes the equivalence of $\lambda$-SD-consistency of $\rho_{\eta}$ and $\mf p_{-\lambda}$-SD-consistency of $\eta$. 
This leads to a representation of $\mf p_\lambda$-SD-consistent RRMs analogous to the one of $\lambda$-SD-consistent risk measures in~\eqref{eq:rhog_new}. 
In parallel with $\rho_{\lambda,p}$, we introduce for given $\lambda\in\R$ and $p\in[0,1]$ the functional 
\[
    \eta_{\lambda,p} = \mf p_{\lambda}^{-1}\circ(-1/\es_p)\circ\mf p_{\lambda}\quad\text{on }\CE.
\]
Both $\eta_{\lambda,p}$ as well as the inner map $-1/\es_p$ are easily verified to be RRMs on $\mathcal{E}$.

\begin{proposition}\label{prop:ssd1}
Let $\eta$ be an RRM and suppose that risk measure $\rho_\eta$ is given by \eqref{def rhoeta}. 
Then 
\begin{equation}\label{eq:33a}\rho_\eta\text{ is }{\lambda}\text{-SD-consistent}\quad\iff\quad\eta|_{\CE}\text{ is }\mf p_{-\lambda}\text{-SD-consistent.}\end{equation}
Moreover, if $\lambda\neq 0$ and $\eta$ is $\mf p_{\lambda}$-SD-consistent on $\CE$, it holds that 
\begin{equation}\label{eq:rep_RRM_viaPowerUtility}
    \eta(X) = \inf_{Y\in\mathcal{B}}\sup_{p\in[0,1]}\frac{\eta_{\lambda,p}(X)}{\eta_{\lambda,p}(Y)},\quad X\in\CE,
\end{equation}
where $\mathcal{B}=\{Y\in \CE\,|\,\eta(Y)\leq 1\}$ is the (geometric) acceptance set of $\eta|_{\CE}$. 
\end{proposition}

Together with \eqref{eq:cons_log}, the preceding proposition yields
\begin{equation}\label{eq:moreRA}\text{SSD-consistency of }
\rho_\eta\quad\implies\quad\text{SSD-consistency of }\eta|_{\CE}.\end{equation}
In case $\eta$ is SSD-consistent,~\eqref{eq:rep_RRM_viaPowerUtility} also simplifies to
\[ 
    \eta(X) = \inf_{Y\in\mathcal{B}}\sup_{p\in[0,1]}\frac{\es_p(Y)}{\es_p(X)},\quad X\in\CE,
\]
highlighting the geometric nature of RRMs.
However, the converse implication in \eqref{eq:moreRA} does not hold because SSD-consistency of $\eta$ is equivalent to $-1$-SD-consistency of $\rho_\eta$. 
These observations call for caution when attaching heuristics of risk aversion to RRMs, or risk measurement procedures involving log-returns more generally.

In line with \eqref{eq:rep_RRM_viaPowerUtility}, it is also possible to define consistent RRMs via a set $\mathcal G$ of nondecreasing and positive functions, which allow us to analyse corresponding $\lambda$-SD-consistent risk measures.
To do so, in the subsequent case study, we consider a selection of daily returns $Y_1,...,Y_k\in\CE$ resulting in functions $g_i(p) = \eta_{-\lambda,p}(Y_i)$, $p\in[0,1]$,
and choose $\mathcal{G}\subseteq \{g_1,\dots,g_k\}$ by determining a subset of indices $K\subseteq \{1,\dots, k\}$. 
This leads to a $\mf p_{-\lambda}$-SD-consistent RRM $\eta_{\lambda}$ via the formula
\begin{equation}\label{eq:figure}
    \eta_{\lambda}(X) = \inf_{i\in K}\sup_{p\in[0,1]}\tfrac{g_i(p)}{\eta_{-\lambda,p}(X)},\quad X\in\CE.
\end{equation}
In the following, we use $\eta_\lambda$ to numerically examine the property of $\lambda$-SD-consistency of risk measures applied to log-returns and investigate whether the choice of the risk aversion parameter has also a significant impact on risk assessment for real-world data. To do so, the cumulative distribution functions (CDFs) of the argument $X$ and random variables $Y_i$ following empirical CDFs determined by the corresponding log-returns of the following stock indices:\footnote{~By using the empirical CDFs, we ensure that corresponding random variables lie in $\CE$.}

\begin{table}[ht]
    \centering{\footnotesize
    \begin{tabular}{cl}
        Variable & Stock Index \\
        \midrule
        $Y_1$ & EURO STOXX 50 (STOXX50E) \\
        $Y_2$ & FTSE 100 \\
        $Y_3$ & DAX \\
        $Y_4$ & Dow Jones Industrial Average (DJI) \\
        $Y_5$ & NASDAQ Composite\\
        $Y_6$ & Nikkei 225 \\
        $X$   & S\&P 500 \\
        \bottomrule
    \end{tabular}}
\end{table}

In other words, we are now in a position to compare the performance of the S\&P 500 with that of other common stock indices with the help of $\eta_\lambda$. For reproducibility, let us mention that we use the time-series data based on opening rather than closing prices, i.e.,  prices quoted at the beginning of the trading day. 

\begin{figure}
    \centering
    \resizebox{0.9\textwidth}{!}{ 
    \begin{minipage}{\textwidth}
        \centering
        \subfigure[S\&P 500 vs.\ DAX]{
            \includegraphics[width=0.45\textwidth]{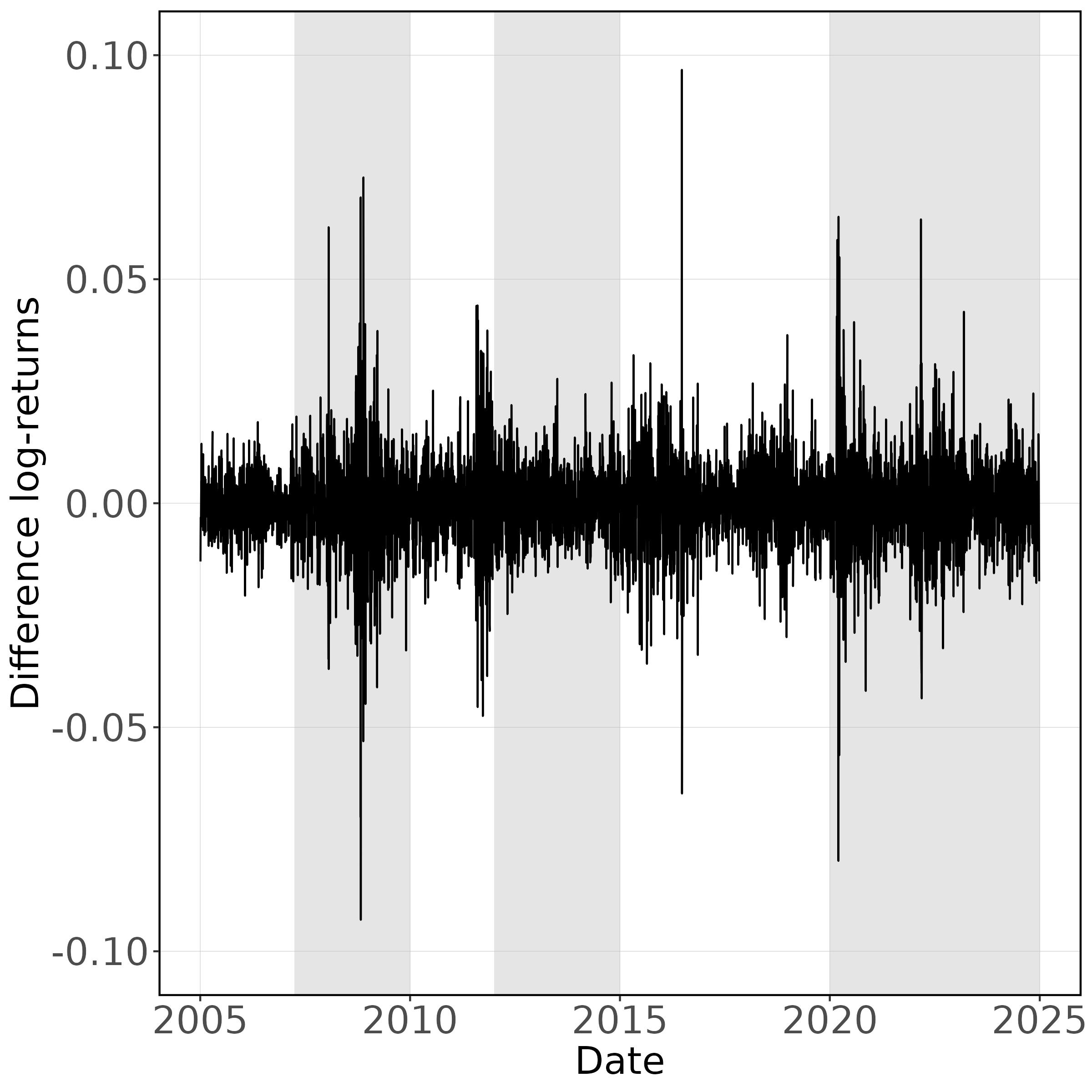}
        }
        \subfigure[S\&P 500 vs.\ DJI]{
            \includegraphics[width=0.45\textwidth]{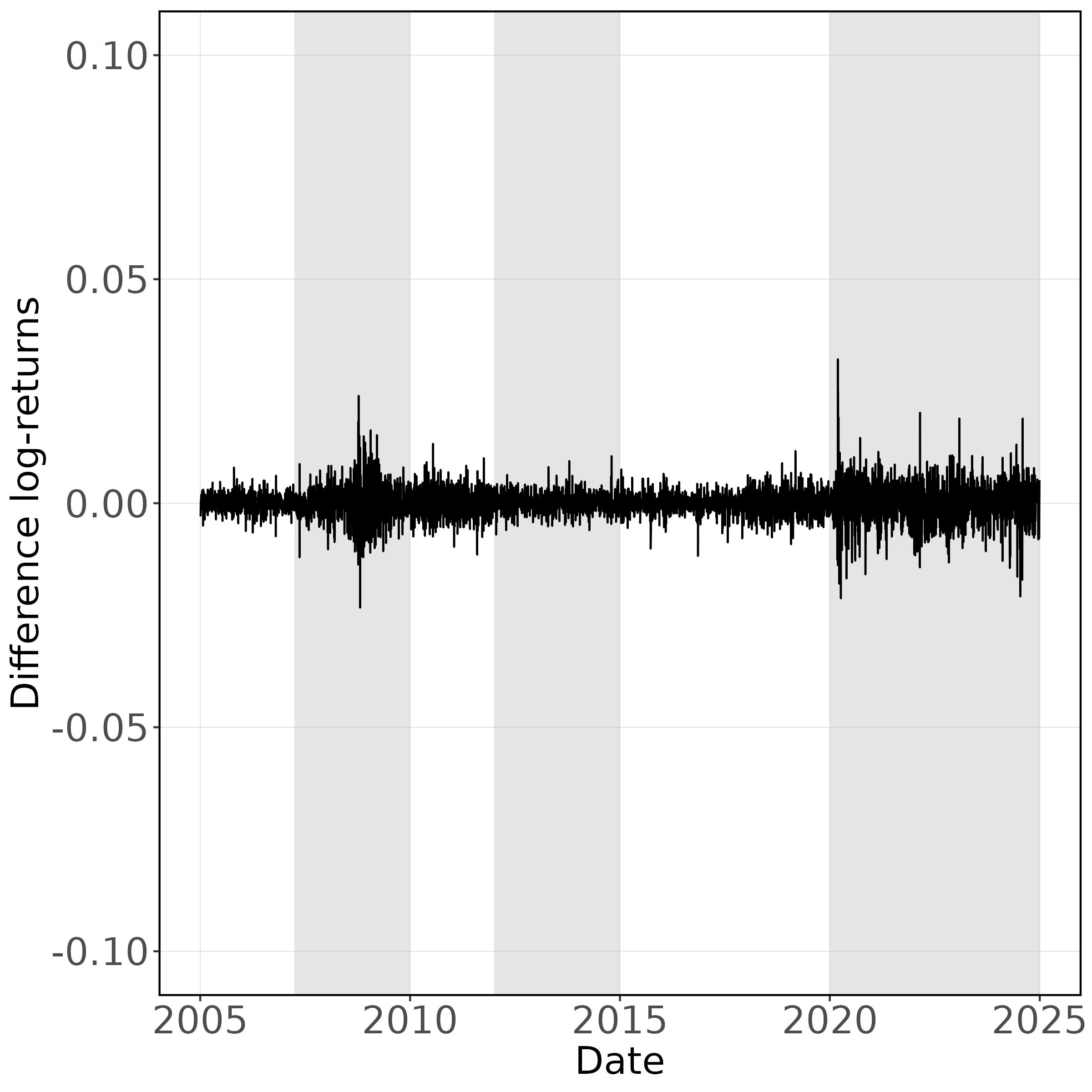}
        }
        \subfigure[S\&P 500 vs.\ FTSE]{
            \includegraphics[width=0.45\textwidth]{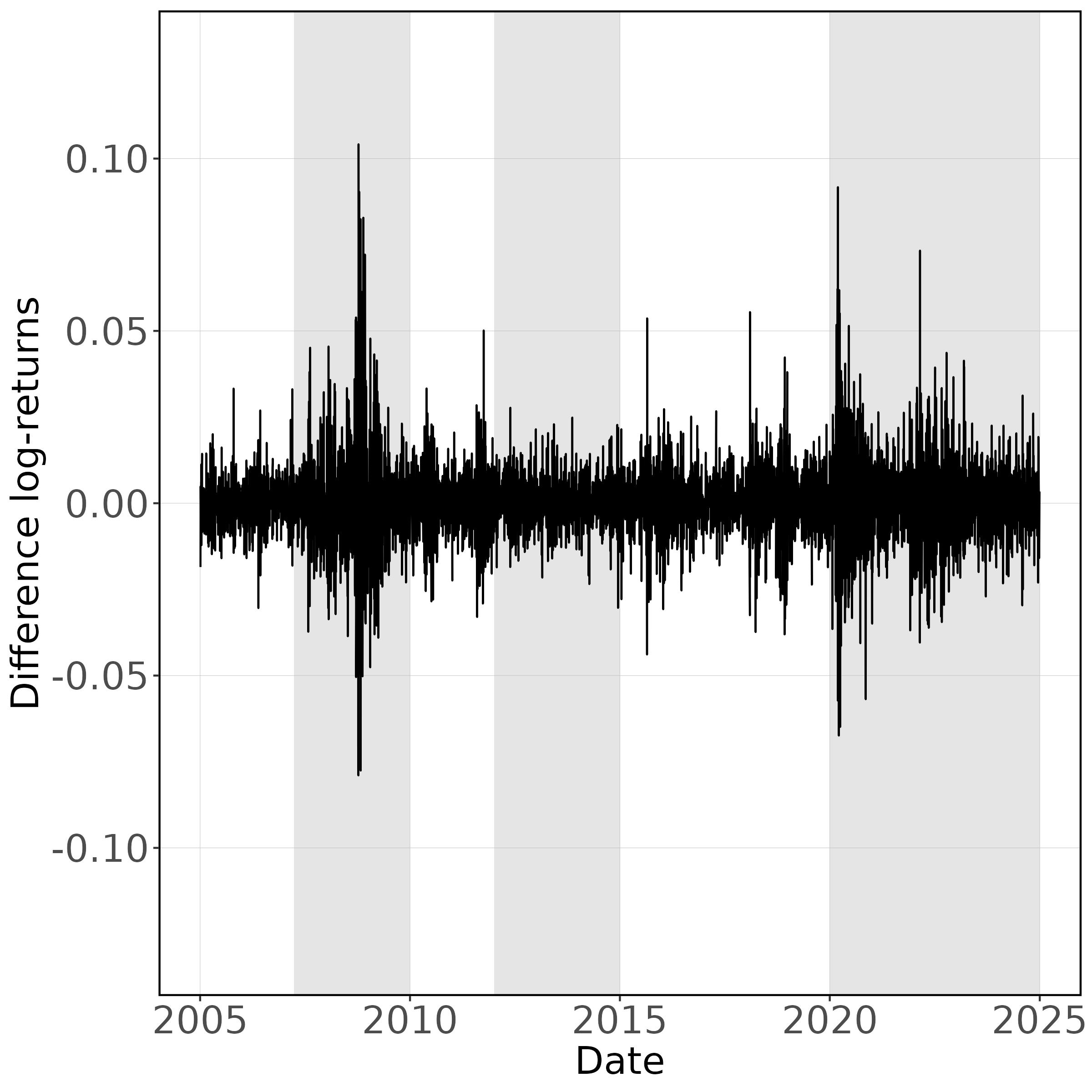}
        }
        \subfigure[S\&P 500 vs.\ N225]{
            \includegraphics[width=0.45\textwidth]{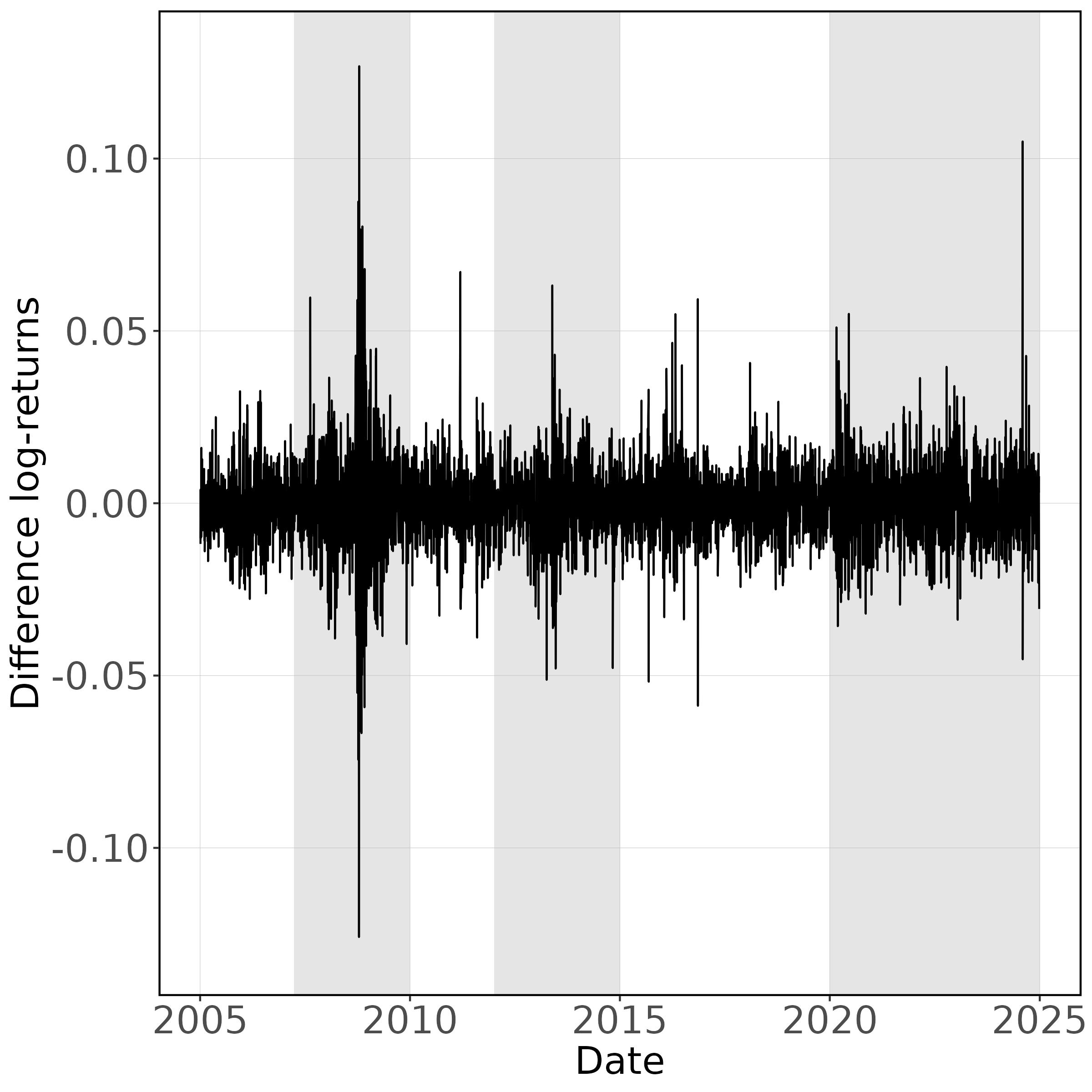}
        }
        \subfigure[S\&P 500 vs.\ NASDAQ]{
            \includegraphics[width=0.45\textwidth]{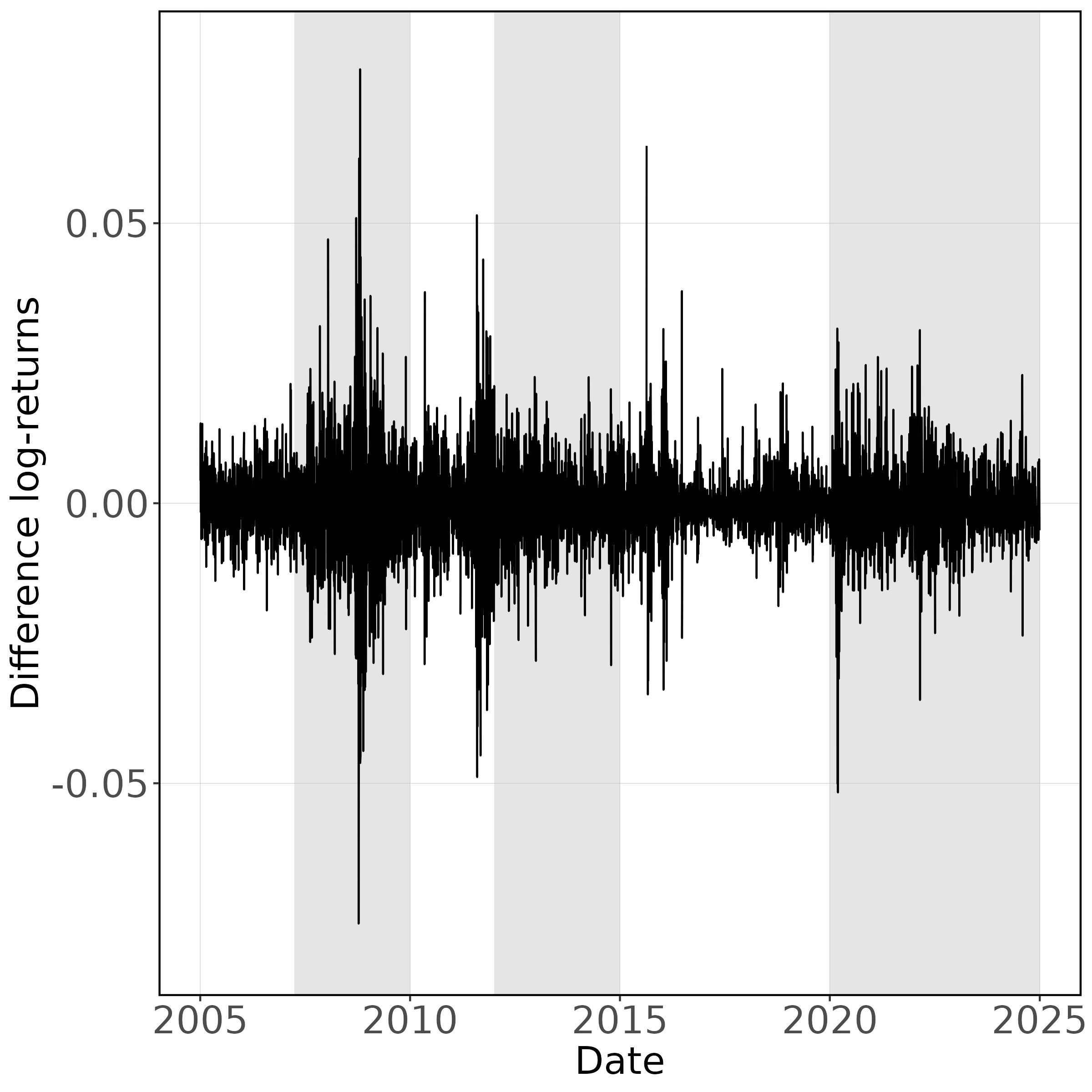}
        }
        \subfigure[S\&P 500 vs.\ STOXX50E]{
            \includegraphics[width=0.45\textwidth]{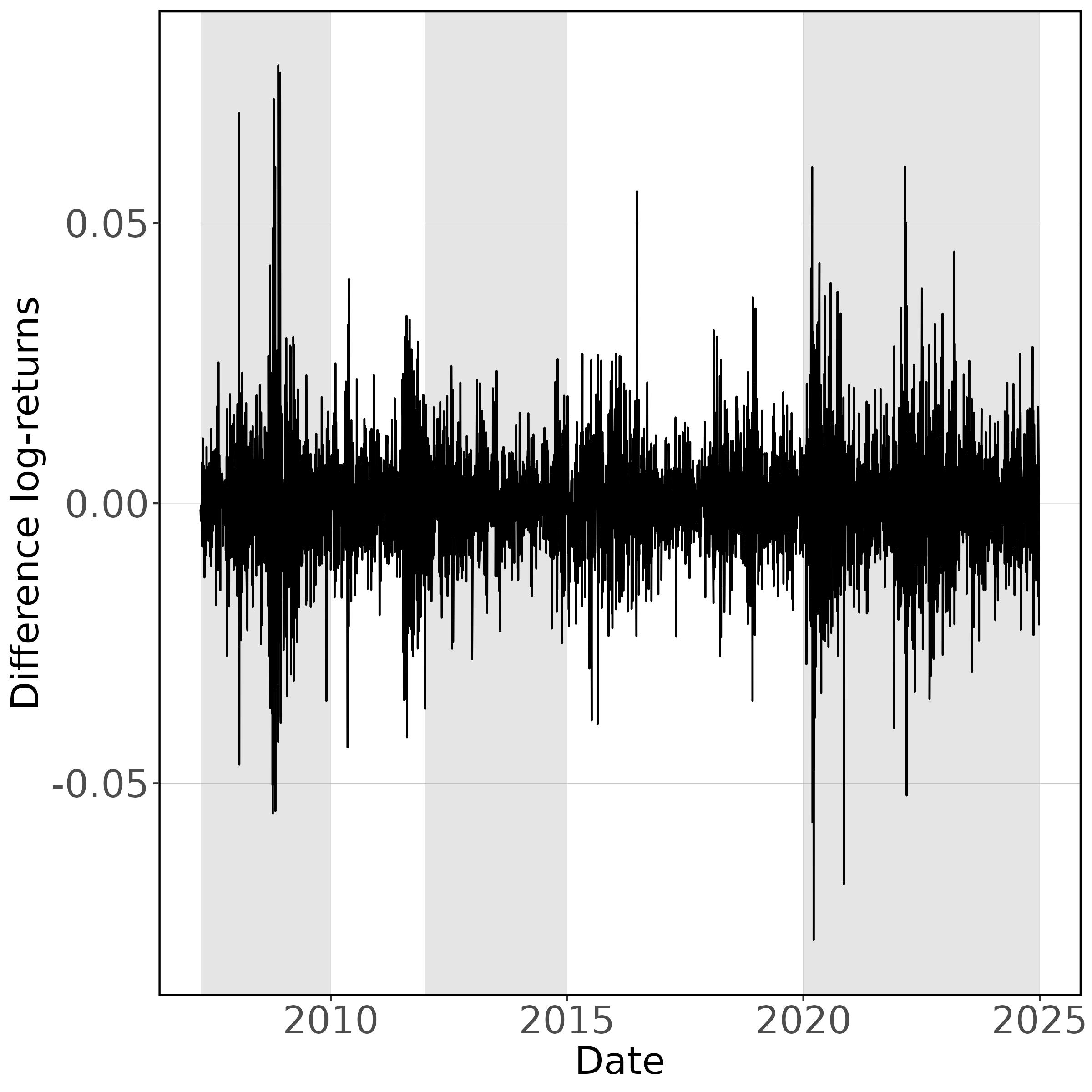}
        }
    \end{minipage}
    }
    \caption{\footnotesize Differences of log-returns based on the S\&P 500 and one other common stock index. The gray boxes indicate the time intervals used to calculate RRMs.}
    \label{fig:diffs_logReturns_indices}
\end{figure}

We examine three different market periods: a crisis period (30 March 2007 -- 31 December 2009), a stable phase (1 January 2012 -- 31 December 2014), and a period covering both crisis and recovery (1 January 2020 – 31 December 2024).
Figure~\ref{fig:diffs_logReturns_indices} illustrates the differences in log-returns between the S\&P 500 and other indices, with key summary statistics shown in Table~\ref{tab:logReturns_keyFigures}.
\begin{center}
\begin{table}[ht]
\centering
\footnotesize
\begin{tabular}{llrrrrrr}
  \hline
  Years & Index & Min. & 1st Qu. & Median & Mean & 3rd Qu. & Max. \\ 
  \hline
2007 -- 2009 & DAX & $-0.0930$ & $-0.0059$ & $-0.0006$ & $-0.0005$ & 0.0057 & 0.0727 \\ 
   & DJI & $-0.0233$ & $-0.0023$ & $-0.0001$ & $-0.0001$ & 0.0021 & 0.0240 \\ 
   & FTSE & $-0.0790$ & $-0.0090$ & $-0.0010$ & $-0.0002$ & 0.0077 & 0.1041 \\ 
   & NASDAQ & $-0.0751$ & $-0.0066$ & $-0.0005$ & $-0.0003$ & 0.0062 & 0.0775 \\ 
   & N225 & $-0.1259$ & $-0.0076$ & 0.0003 & 0.0001 & 0.0080 & 0.1268 \\ 
   & STOXX50E & $-0.0556$ & $-0.0057$ & $-0.0003$ & $-0.0001$ & 0.0055 & 0.0821 \\ 
   \hline
   2012 -- 2014 & DAX & $-0.0247$ & $-0.0043$ & $-0.0001$ & 0.0002 & 0.0049 & 0.0278 \\ 
   & DJI & $-0.0070$ & $-0.0013$ & 0.0002 & 0.0002 & 0.0015 & 0.0105 \\ 
   & FTSE & $-0.0304$ & $-0.0034$ & 0.0003 & 0.0006 & 0.0047 & 0.0277 \\ 
   & NASDAQ & $-0.0289$ & $-0.0040$ & $-0.0003$ & $-0.0001$ & 0.0037 & 0.0226 \\ 
   & N225 & $-0.0512$ & $-0.0065$ & $-0.0002$ & $-0.0003$ & 0.0054 & 0.0632 \\ 
   & STOXX50E & $-0.0278$ & $-0.0041$ & 0.0002 & 0.0004 & 0.0051 & 0.0257 \\ 
   \hline
   2020 -- 2024 & DAX & $-0.0798$ & $-0.0054$ & 0.0000 & 0.0001 & 0.0056 & 0.0639 \\ 
   & DJI & $-0.0212$ & $-0.0025$ & 0.0004 & 0.0002 & 0.0031 & 0.0321 \\ 
   & FTSE & $-0.0674$ & $-0.0070$ & 0.0005 & 0.0004 & 0.0077 & 0.0917 \\ 
   & NASDAQ & $-0.0516$ & $-0.0040$ & $-0.0005$ & $-0.0001$ & 0.0033 & 0.0312 \\ 
   & N225 & $-0.0453$ & $-0.0071$ & $-0.0001$ & 0.0002 & 0.0068 & 0.1050 \\ 
   & STOXX50E & $-0.0780$ & $-0.0057$ & $-0.0000$ & 0.0003 & 0.0061 & 0.0860 \\
   \hline
\end{tabular}
\vspace{0.25cm} 
\caption{\footnotesize Key summary statistics for differences in log-returns of the S\&P 500 and a corresponding second index.}
\label{tab:logReturns_keyFigures}
\end{table}
\end{center}

First, it is worth noting that the S\&P 500 and DJI exhibit very similar behavior, with much smaller differences in their log-returns compared to the other indices. 
In contrast, large deviations during times of crisis (e.g., in 2008) are evident across the remaining indices, especially for the FTSE and N225. 

\begin{figure}
    \centering
    \subfigure[March 30, 2007 -- December 31, 2009]{
        \includegraphics[width=0.45\textwidth]{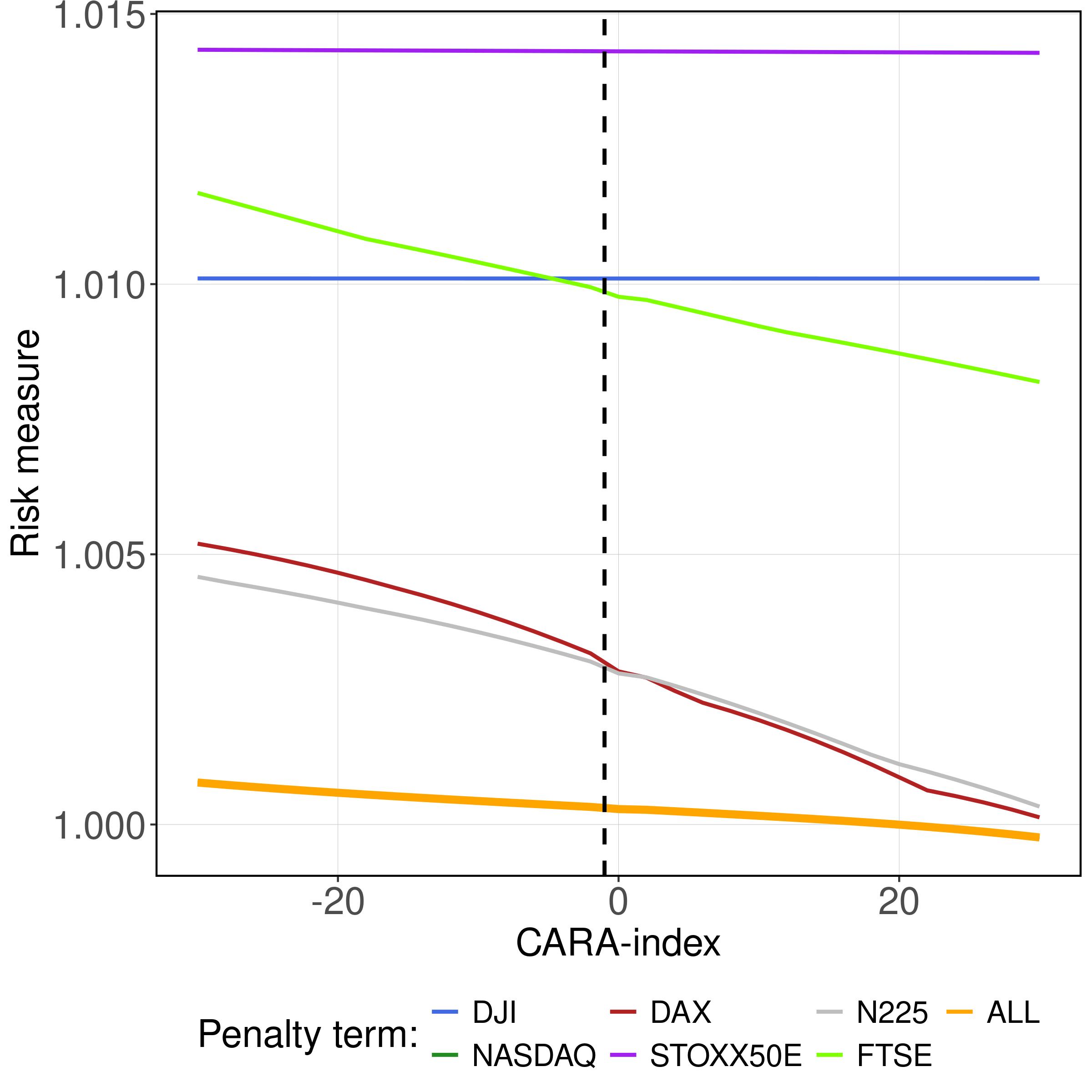}
    }
    \subfigure[January 1, 2012 -- December 31, 2014]{
        \includegraphics[width=0.45\textwidth]{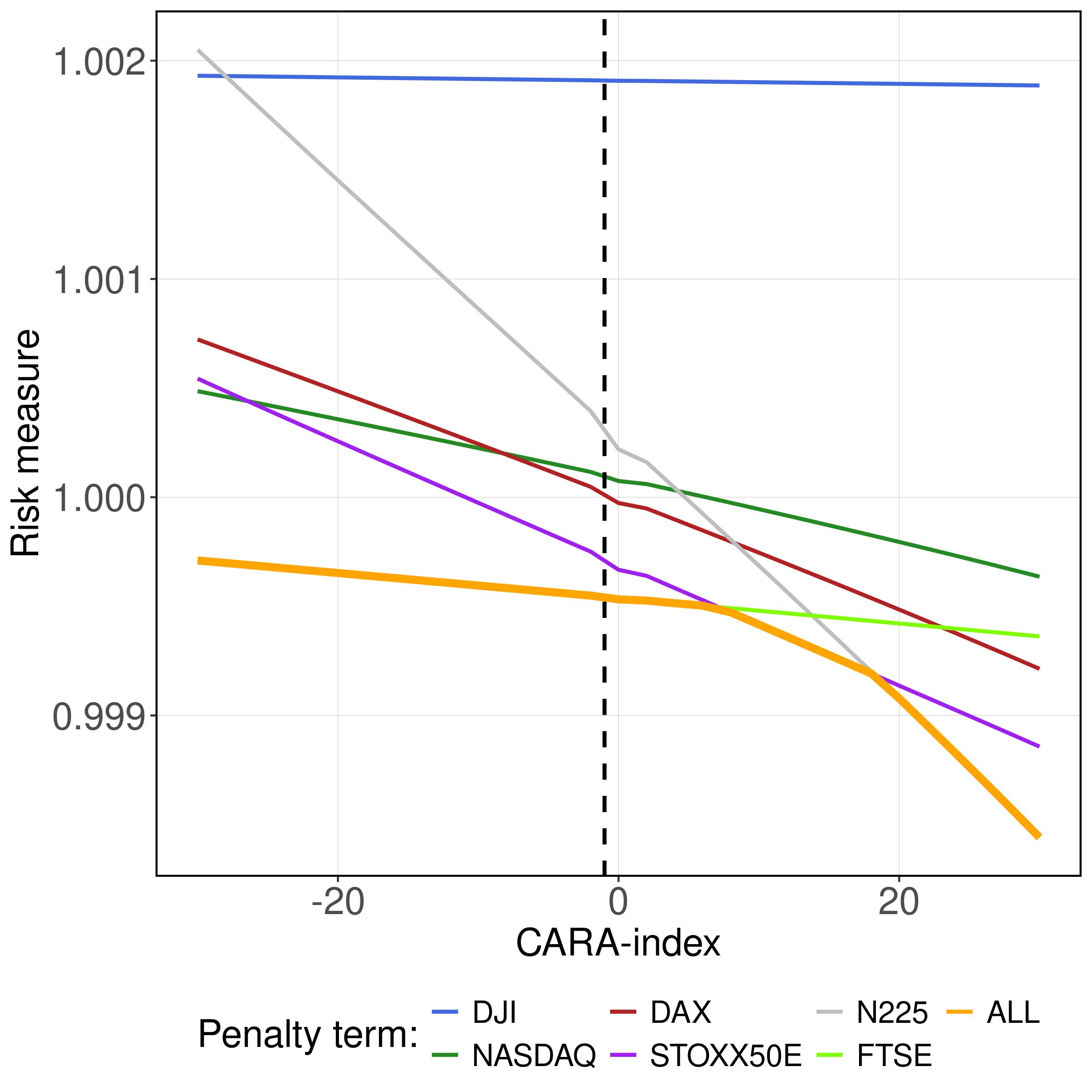}
    }
    \subfigure[January 1, 2020 -- December 31, 2024]{
        \includegraphics[width=0.45\textwidth]{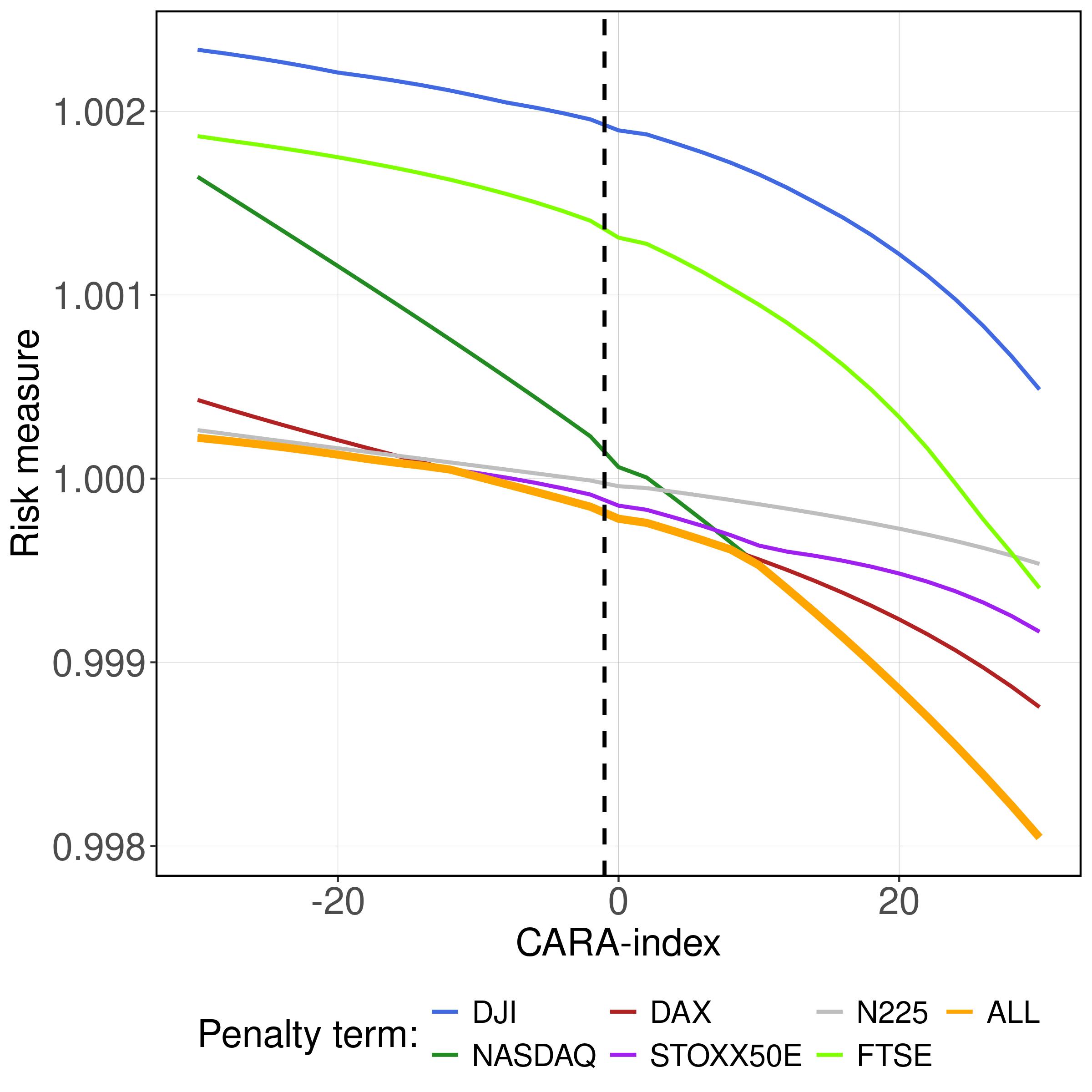}
    }
    \caption{\footnotesize RRMs in \eqref{eq:figure} based on $\lambda$-SD-consistent risk measures, where $\lambda$ is the CARA-index. The dashed line at $\lambda=-1$ marks the SSD-consistent RRMs.}
    \label{fig:e_c_risk_measures_indices}
\end{figure}

Figure~\ref{fig:e_c_risk_measures_indices} displays the function mapping the risk aversion parameter $\lambda$ to $\eta_\lambda(X)$. In addition to the aggregate case $K = \{1, \dots, k\}$, we also show the individual cases $K = \{i\}$ for each $i$, highlighting the influence of individual stock indices. 

\begin{figure}
    \centering
    \subfigure[March 30, 2007 -- December 31, 2009]{
        \includegraphics[width=0.45\textwidth]{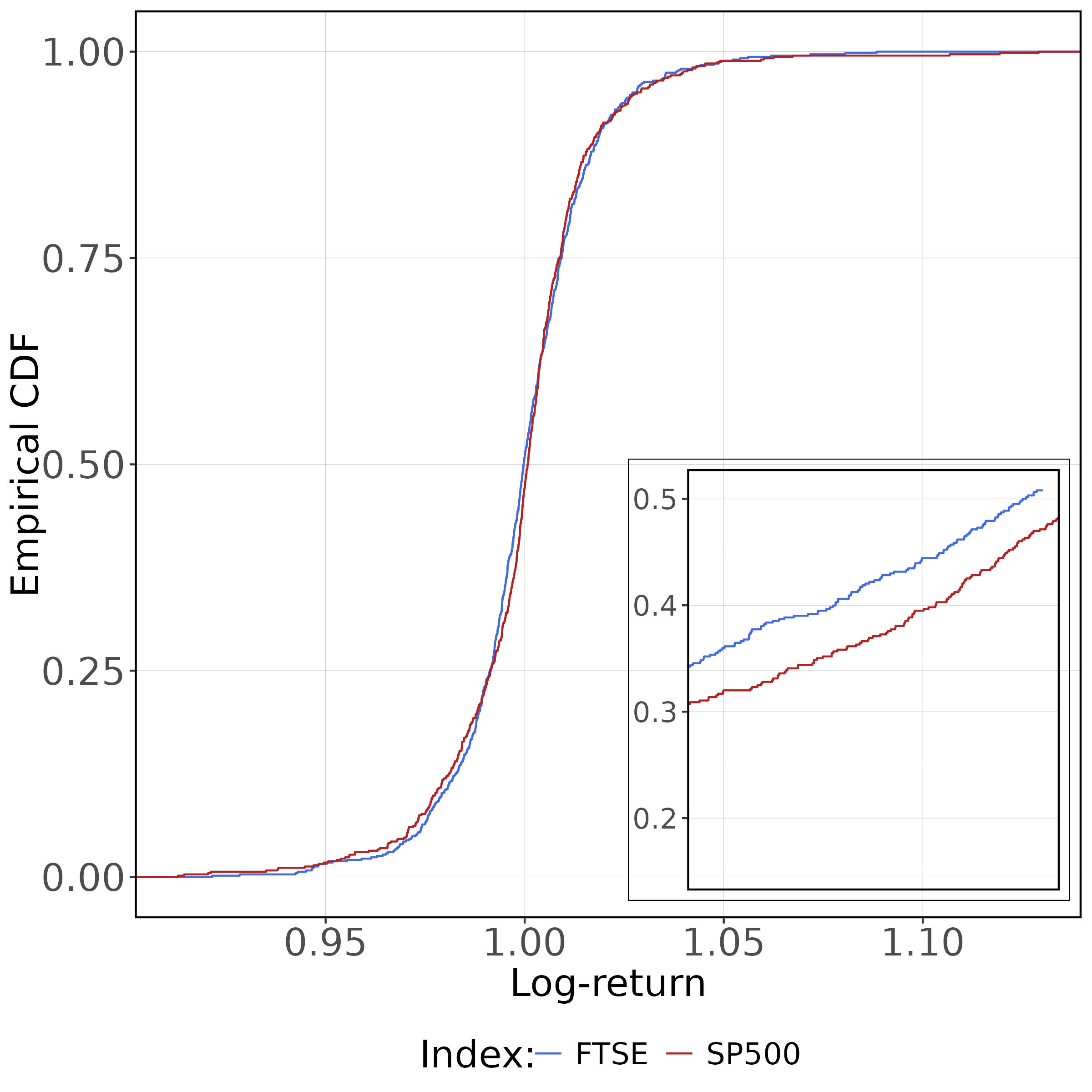}
    }
    \subfigure[January 1, 2012 -- December 31, 2014]{
        \includegraphics[width=0.45\textwidth]{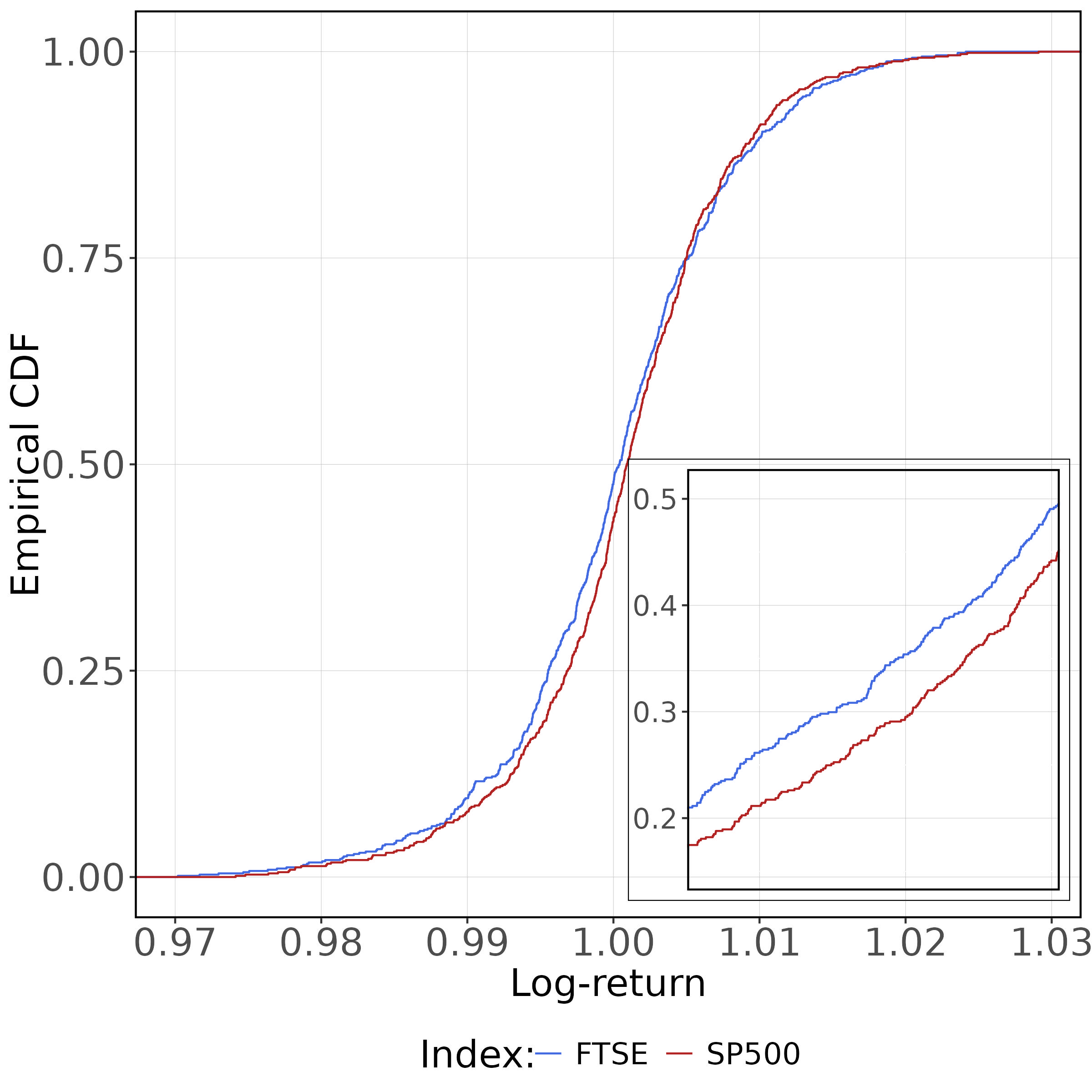}
    }
    \caption{\footnotesize Empirical CDFs of S\&P 500 and FTSE for different time intervals.}
    \label{fig:empiricalCDFs_SP500__FTSE}
\end{figure}

Figure~\ref{fig:e_c_risk_measures_indices}(a) focuses on the financial crisis in 2008, during which the lowest curve corresponding to the $\eta_{\bullet}(X)$ profile results from using the NASDAQ index.
Figure~\ref{fig:e_c_risk_measures_indices}(b) displays times without crisis, when the values of $\eta_\lambda(X)$ are significantly smaller in comparison to the other two time intervals. 
For $\lambda<0$, the FTSE index determines the RRM. This is noteworthy because in Figure~\ref{fig:e_c_risk_measures_indices}(a), the FTSE leads to one of the highest curves. 
The reason for this behavior is that the empirical CDF of the FTSE has more mass in the tail than the S\&P 500 in the years 2012--2014; this is illustrated in Figure~\ref{fig:empiricalCDFs_SP500__FTSE}. Hence, the ES-values for S\&P 500 and FTSE, respectively, differ more markedly during this time period.

If we turn back to Figure~\ref{fig:e_c_risk_measures_indices}(b) and (c), we see that also values below $1$ occur. In this regard, it is worth to point out that the roots of the curves (minus one) are estimators for $\lambda^{\min}_i$, the smallest risk aversion $\lambda$ such that $\log(X)$ stemming from the S\&P 500 $\lambda$-SD dominates the $\log(Y_i)$ estimated from the corresponding index in use. 
Hence, these curves reproduce the estimator for $\lambda^{\min}_i$ developed in~\cite[Proposition 7]{Fractional}.

Finally, let us mention that all curves in Figure~\ref{fig:e_c_risk_measures_indices} are nonincreasing in $\lambda$. This is due to~\eqref{eq:cons1} and the fact that, for $\gamma<\lambda$, the $\gamma$-SD order is stronger than the $\lambda$-SD order.
In case of $\gamma$-SD consistency, two payoffs must be ordered with respect to a greater number of utility functions, which is harder to achieve---larger capital reserves are the consequence. 
This isotonicity can also be used as guideline to choose the risk aversion parameter $\lambda$ by an investor: a risk-averse investor prefers smaller values of $\lambda$, leading to a larger capital reserve and hence to better protection against future losses.

\subsection{Measuring systemic risk}\label{sec:systemicRisk}

Another choice of target risk profiles $\mathcal{G}$ allows to use $\lambda$-SD-consistent risk measures to measure systemic risk. In the situation of considering the log-returns as in Section~\ref{sec:motivation}, we shall test aggregating the index values first using a so-called {\em aggregation function} $\Lambda\colon\R^d\rightarrow\R$
and set
\[
    g(p) = (\rho_{\lambda,p}\circ\Lambda)(Z),\quad Z\in \CX.
\]
This is the conventional way to construct systemic risk measures, see~\cite[Theorem~1]{ChenSystemicRisk} and~\cite[Theorem~1]{Kromer}. 
Given a vector $x\in\R^d$ (representing a realisation of a log-return), the following aggregation functions will be used:  
\[
   \Lambda^{A}(x) = \frac{1}{d}\sum_{i=1}^d (-x_i^{-}),\quad  \Lambda^{G}(x) = \Big(\prod_{i=1}^d (-x_i^{-})\Big)^{\frac{1}{d}}, \quad \Lambda^{E}(x) = 1-\exp\left(-\Lambda^{A}(x)\right),
\]
where for $y\in\R$ we write $y^{-} = -\min\{y,0\}$.
Focusing on the negative parts of $x_i^{-}$ avoids cross-subsidization between firms. The usage of $\Lambda^{A}$ is standard. The additional exponential transform in $\Lambda^{E}$ punishes losses more strongly than $\Lambda^{A}$. Such exponential transforms are used in~\cite[Section 5.1]{Feinstein} and~\cite[Example 2]{Kromer}. Aggregation via the geometric mean $\Lambda^G$ is used as reference model and operates on the same logarithmic return data used in Section~\ref{sec:motivation}. Moreover, we test the minimal and the maximum shortfall by applying the following aggregation functions:
\[
    \Lambda^{\min}(x) = \min_{i}(-x_i^{-})\quad \text{and} \quad \Lambda^{\max}(x) =\max_{i}(-x_i^{-}). 
\]

Our new approach allows to measure the risk of the S\&P 500 compared to the systemic risk of the other indices. Such a use case is important for an investor, because it allows to compare the risk of an investment opportunity to the combined risk of other investments (systemic risk in the corresponding market). 

\begin{figure}
    \centering
    \subfigure[March 30, 2007 -- December 31, 2009]{
        \includegraphics[width=0.45\textwidth]{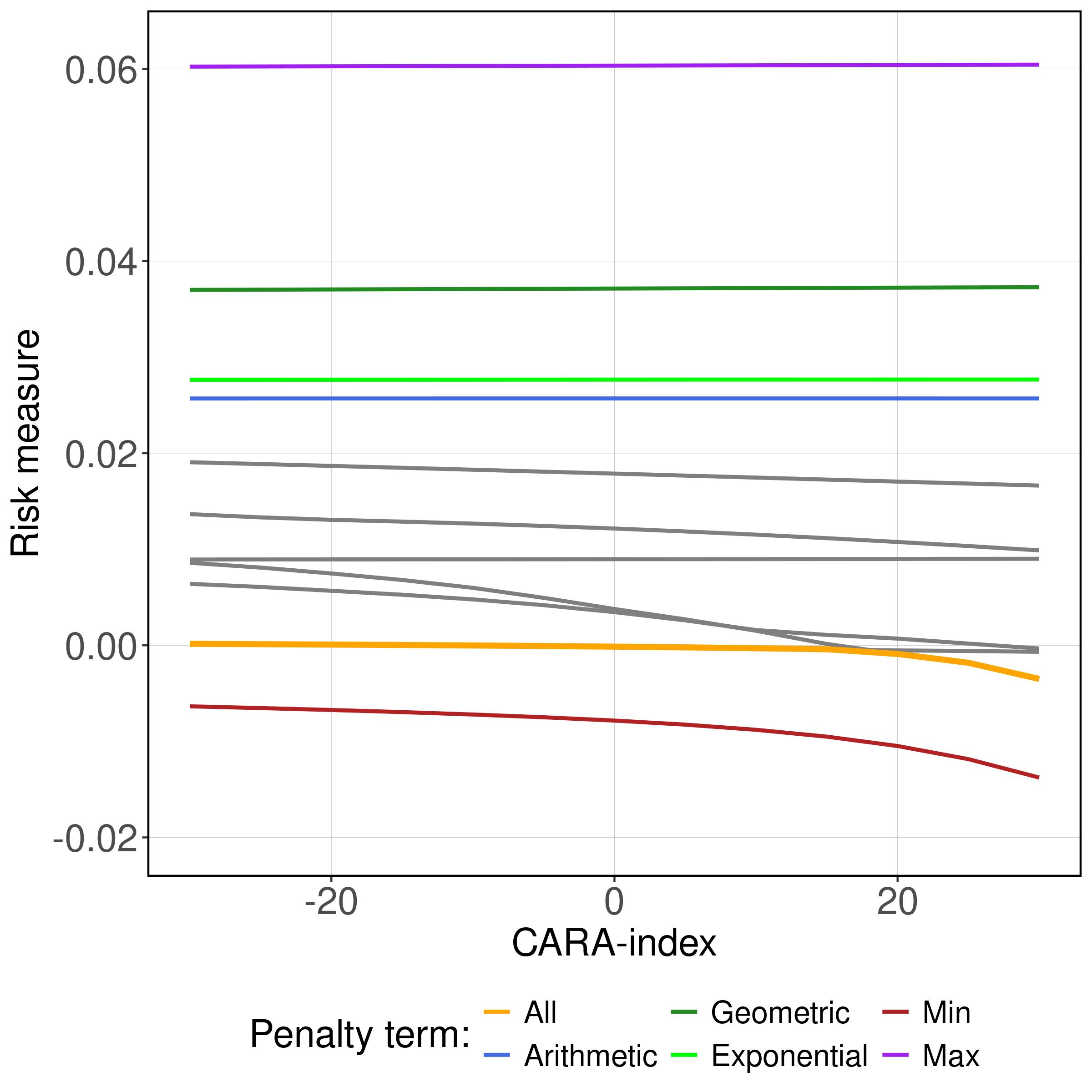}
    }
    \subfigure[January 1, 2020 -- December 31, 2024]{
        \includegraphics[width=0.45\textwidth]{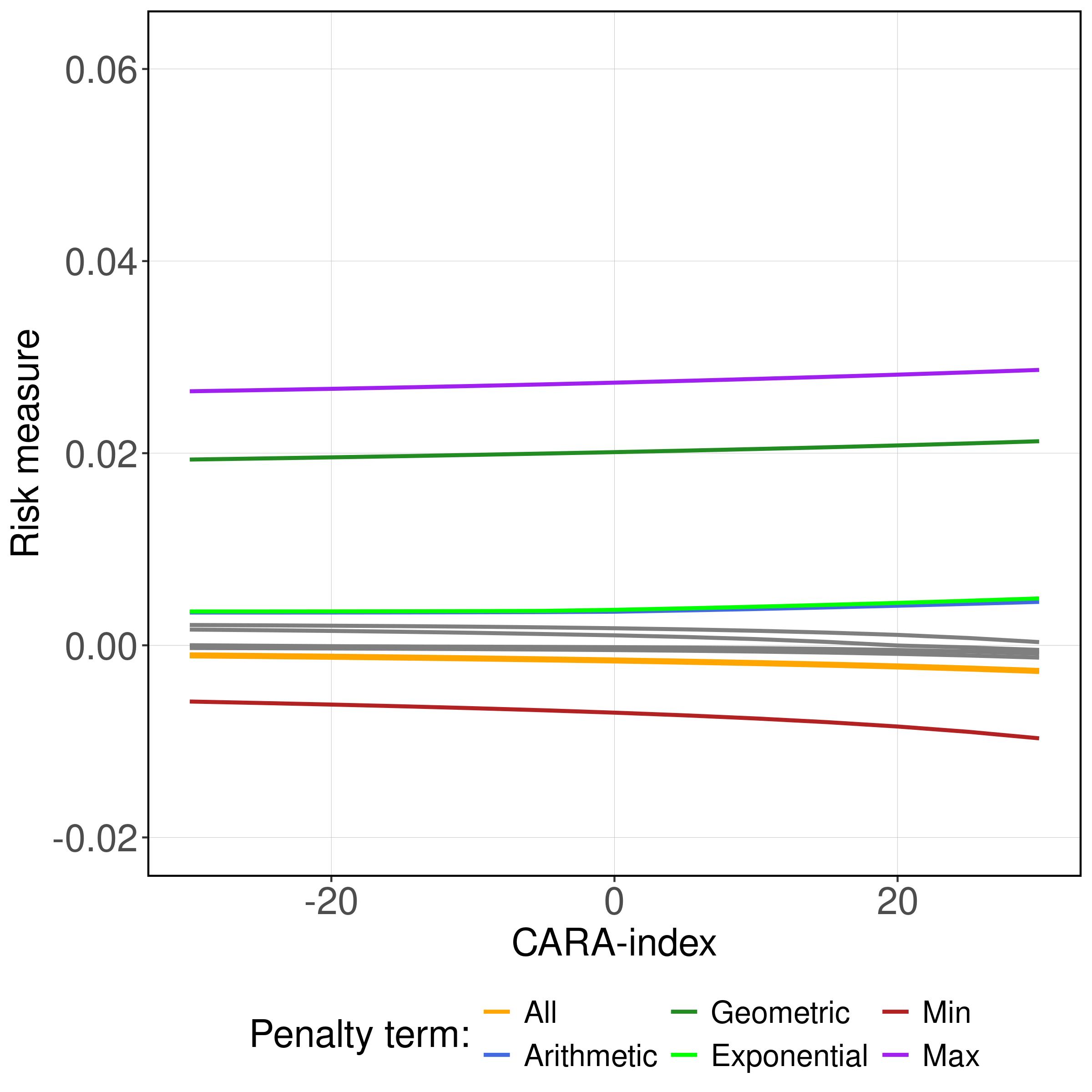}
    }
    \caption{\footnotesize $\lambda$-SD-consistent risk measures, with (CARA) risk aversion parameter $\lambda$, based on different aggregation functions. The gray curves are the corresponding curves of the indizes already presented in Figure~\ref{fig:e_c_risk_measures_indices}.}
    \label{fig:meyer_systemic_variants}
\end{figure}

The results are given in Figure~\ref{fig:meyer_systemic_variants}. For completeness, we also plot the curves from the previous section (orange and gray curves; but without applying the exponential function). First, we observe that the difference between using arithmetic aggregation with or without an exponential transform has only a small impact in this situation. This indicates that logarithmic returns are close to zero such that the approximation $\exp(x)-1\approx x$ is valid. Second, during the COVID crisis, arithmetic, geometric and maximum aggregation functions yield significantly smaller values than for the financial crisis in 2008, while the earlier values (orange curves) are nearly at the same level for both time intervals.   

\begin{figure}
    \centering
    \subfigure[March 30, 2007 -- December 31, 2009]{
        \includegraphics[width=0.45\textwidth]{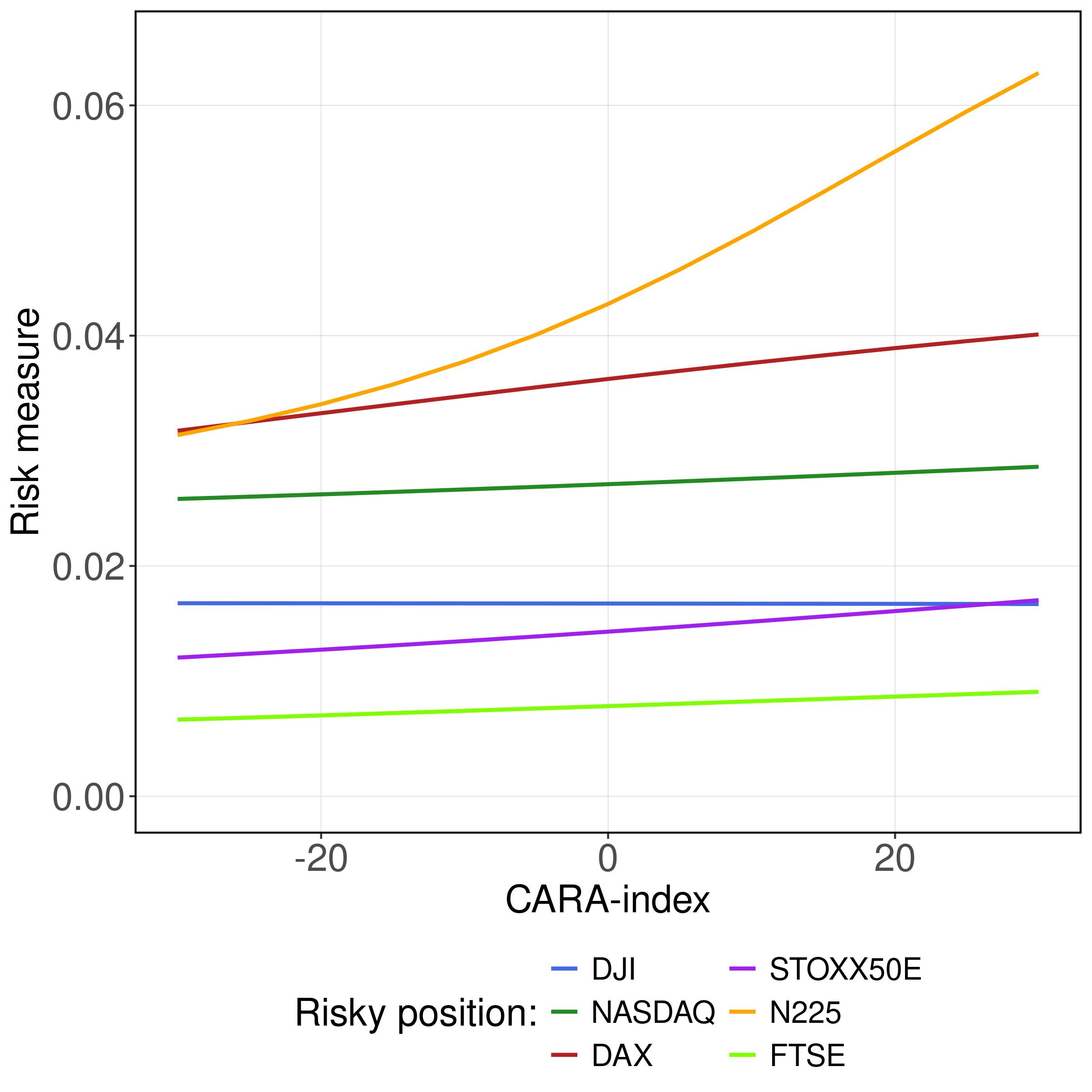}
    }
    \subfigure[January 1, 2020 -- December 31, 2024]{
        \includegraphics[width=0.45\textwidth]{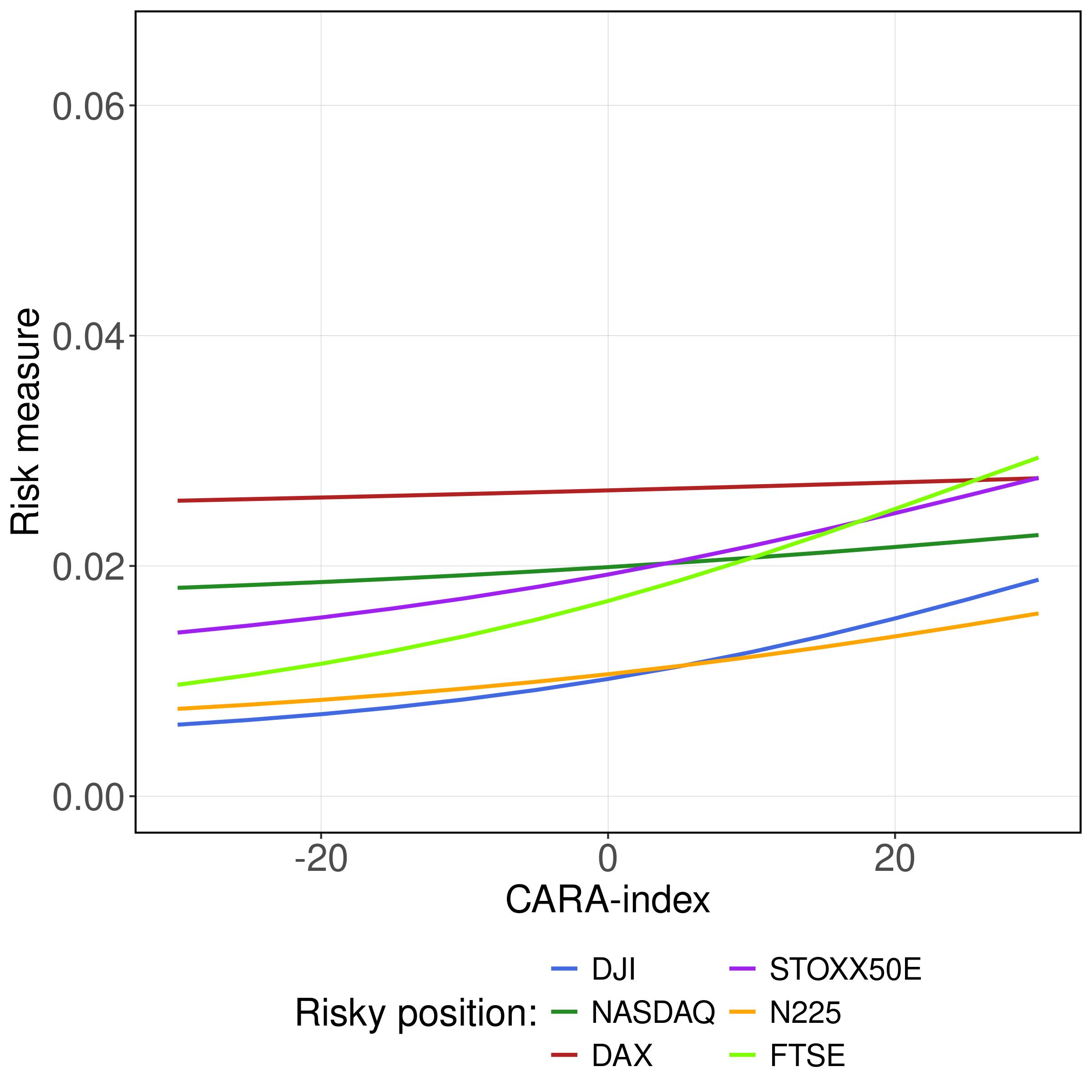}
    }
    \caption{\footnotesize $\lambda$-SD-consistent risk measures, with (CARA) risk aversion parameter $\lambda$, based on different risky positions and the arithmetic aggregation function $\Lambda^{A}$.}
    \label{fig:meyer_systemic_variants_2}
\end{figure}

Finally, using aggregation functions is also relevant for a regulator who aims to compare the risk of a single firm to the systemic risk of a whole network of firms. We artificially analyse this by using the different indices from the benchmark profile also as input for our new risk measures (instead of the different index S\&P 500 as before). The results are demonstrated in~Figure~\ref{fig:meyer_systemic_variants_2}. Compared to the financial crisis in 2008, the systemic impact of N225 is drastically diminished. The only indices that slightly enlarge their impact are FTSE and STOXX50E. These results demonstrate that the new risk measures are naturally suited for identifying main drivers of systemic risk.

\section{Monotone additive statistics}\label{sec:Mu}

In Theorem~\ref{thm:repEXPSD} we have shown that every $\lambda$-SD-consistent risk measure admits a representation as a minmax over penalised risk measures with a particularly simple structure.
These penalised risk measures can locally be interpreted as entropic ones.
While this representation is universal, we now return to the existing literature and investigate a different class of risk measures in which the minmax operation is replaced by a mixture: monotone additive statistics.

In \cite{Muetal}, (descriptive) statistics are law-invariant maps $\Phi\colon \CX\to\R$ satisfying the normalisation property $\Phi(c)=c$ for all $c\in\R$.  
The paper studies {\em monotone additive statistics} (MAS) characterised by the additional properties of isotonicity, i.e., 
\begin{center}$X\le Y\quad\implies\quad\Phi(X)\le\Phi(Y),$\end{center}
and additivity on pairs of independent random variables, a strong and highly restrictive assumption.
MAS are shown to have numerous decision-theoretic applications, e.g., in the study of time preferences. Towards their full characterisation, let $X\in\CX$ and
\[E_c(X):=\begin{cases}-M(-X)&\text{if }c=-\infty,\\[-0.7ex]
\frac 1 c\log\big(\E[e^{cX}]\big)&\text{if }c\in\mathbb{R}\setminus\{0\},\\[-0.7ex]
\E[X]&\text{if }c=0,\\[-0.7ex]
M(X)&\text{if }c=\infty.\end{cases}\]
Recall that functional $M$ is defined by \eqref{def M}, and $E_c(X)$ with $c\in\mathbb{R}$ is the certainty equivalent for exponential utility $\mf e_{-c}$. For $c\in \R$ it also holds that $E_c(X) = \rho_{c,0}(-X)$ for $X\in \CX$.
\cite[Theorem 1]{Muetal} demonstrates that a map $\Phi$ is a MAS if and only if there exists a unique Borel probability measure $\mu$ on the extended real line $\overline\R=[-\infty,\infty]$ such that $\Phi$ can be represented as a mixture: 
\begin{equation}\label{eq:Mu}\Phi(X)=\int_{\overline\R}E_c(X)\mu(\diff c),\quad X\in\CX.\end{equation}
Functionals of shape \eqref{eq:Mu} have been studied before as ``mixed exponential premium principles'' by \cite{Goovaerts}. 

MAS $\Phi$ fit the topic of this manuscript because the sign-changed versions $\rho_\Phi:=-\Phi$ are risk measures that often are Meyer, provided threshold utility $v$ is suitably chosen.

\begin{theorem}\label{thm:Mu}
    Set $c_-:=\inf\supp(\mu)$ and $c_+:=\sup\supp(\mu)$,
    and suppose $c_+<\infty$. Then 
    \[\{\lambda\in\R\mid \rho_\Phi~\lambda\text{-Meyer}\}=(-\infty,\lambda^\star],\]
    for some $\lambda^\star\in\R$ satisfying $-c_+\le \lambda^\star\le -c_-$.
\end{theorem}

Theorem~\ref{thm:Mu} applies directly in two settings considered in \cite{Muetal}. 
Call an MAS {\em extendable} if it admits a finite extension to all random variables $X\in\mathcal L$ with finite moment-generating function. 
By
\cite[Theorem~2]{Muetal}, this is equivalent to the representing probability $\mu$ in \eqref{eq:Mu} having compact support in $\R$, resulting in $c_+<\infty$. 
Moreover, \cite[Proposition~5]{Muetal} states that $\Phi\le \E[\cdot]$ if and only if $c_+\le 0$. 
Equivalently, $\rho_\Phi$ then satisfies the property of {\em loadedness} (or \textit{expectation boundedness} in the risk-measure literature), i.e., $\rho_\Phi\ge -\E[\cdot]$. 
Consequently, if $c_+<0$, Theorem~\ref{thm:Mu} yields $-c_+$-SD-consistency of $\rho_\Phi$, which is a more precise statement than mere SSD-consistency.
Moreover, note that Theorem~\ref{eq:rhog_new} in combination with Theorem~\ref{thm:Mu} gives an alternative representation for a MAS. 

Unfortunately, Theorem~\ref{thm:Mu} is only an existence statement. 
While  Example~\ref{exam:MAS} shows that $\lambda^{\star} = -c^{+}$ can hold for simple Borel probability measure, the question how to generally identify $\lambda^\star$ remains open to us. Nevertheless, the result shows that MAS induce many Meyer risk measures if the threshold utilities $v$ show upper bounded risk aversion, excluding unbounded aversion to severe and increasing losses.

\begin{corollary}\label{cor:Mu}
Let $v\in\mathcal U(I)$ be a threshold utility and define $c_+,\lambda^{\star}$ as in Theorem~\ref{thm:Mu}.
\begin{enumerate}[label=\tn{(\alph*)}]
\item Risk measure $\rho_\Phi$ is $v$-Meyer if threshold utility $v$ satisfies $R_v^A\le \lambda^\star$.
\item $\rho_\Phi$ is not $v$-Meyer if 
\begin{equation}\label{eq:condMAS1}R_v^A\text{ is continuous, }c_-<\infty,\text{ and }R_v^A(x)>-c_-\text{ for some }x\in I,\end{equation}
or
\begin{equation}\label{eq:condMAS2}\text{there exists $\varepsilon>0$ such that }\big\{R_v^A\ge \lambda^\star+\eps\big\}\text{ includes an unbounded interval}.\end{equation}
\end{enumerate}
\end{corollary}

While Corollary~\ref{cor:Mu} does not answer for all threshold utilities $v$ whether or not a given MAS is $v$-Meyer, it shows that many nontrivial $v$-Meyer risk measures exist. 
This begs for a closer study of this more general class, keeping the $v$ as general as possible. 

\section{Meyer risk measures with general threshold risk aversion}\label{sec:generalThresholdUtility}

While Sections~\ref{sec:representation_vsd} and~\ref{sec:Mu} demonstrate the fundamental importance of the $\lambda$-SD orders of~\cite{Huang}, this section treats $v$-Meyer risk measures for general threshold utilities $v$. 
The following arguments justify this general approach.

\begin{figure}
    \centering
    \subfigure[Logistic utility]{
        \includegraphics[width=0.475\textwidth]{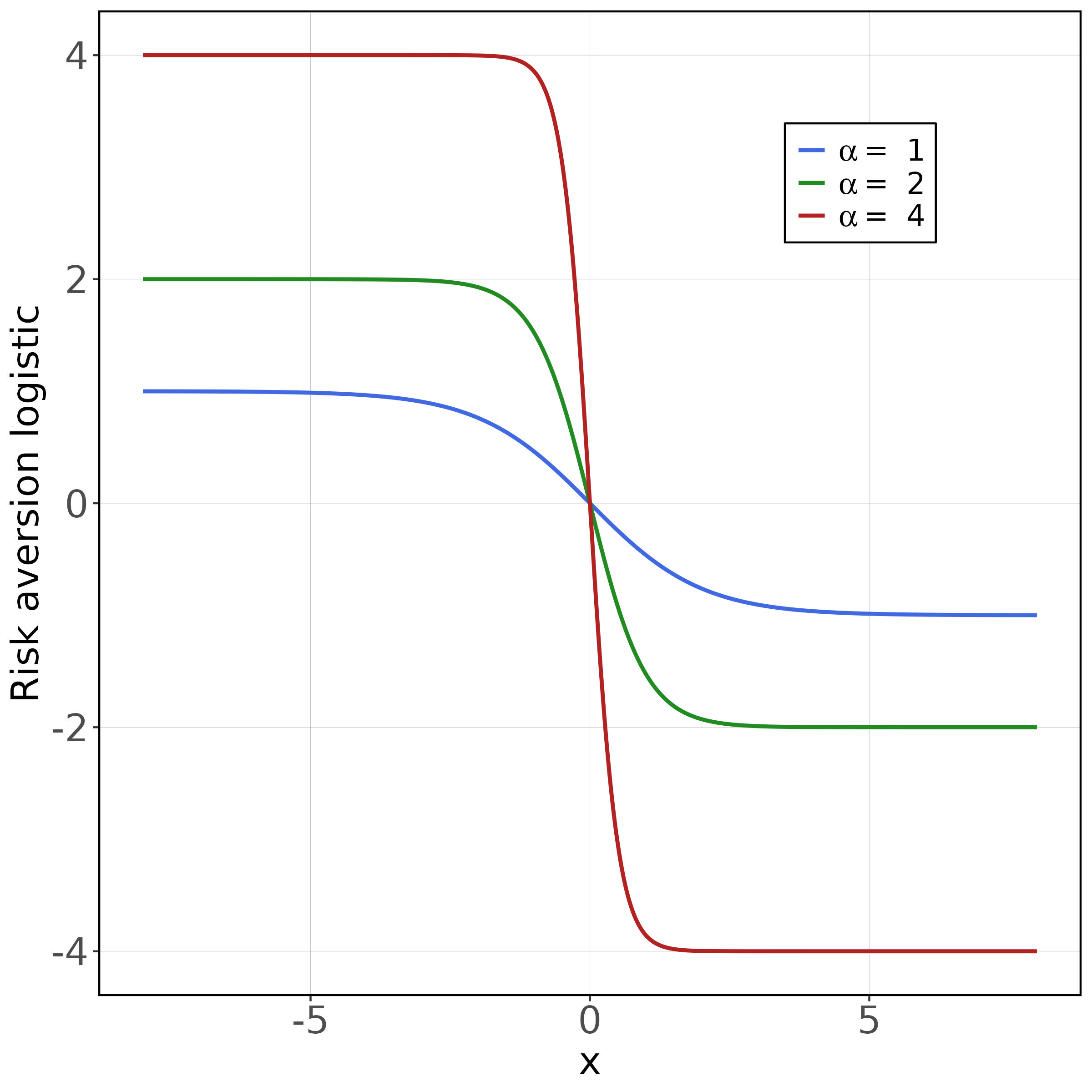}
    }
    \subfigure[SAHARA utility]{
        \includegraphics[width=0.475\textwidth]{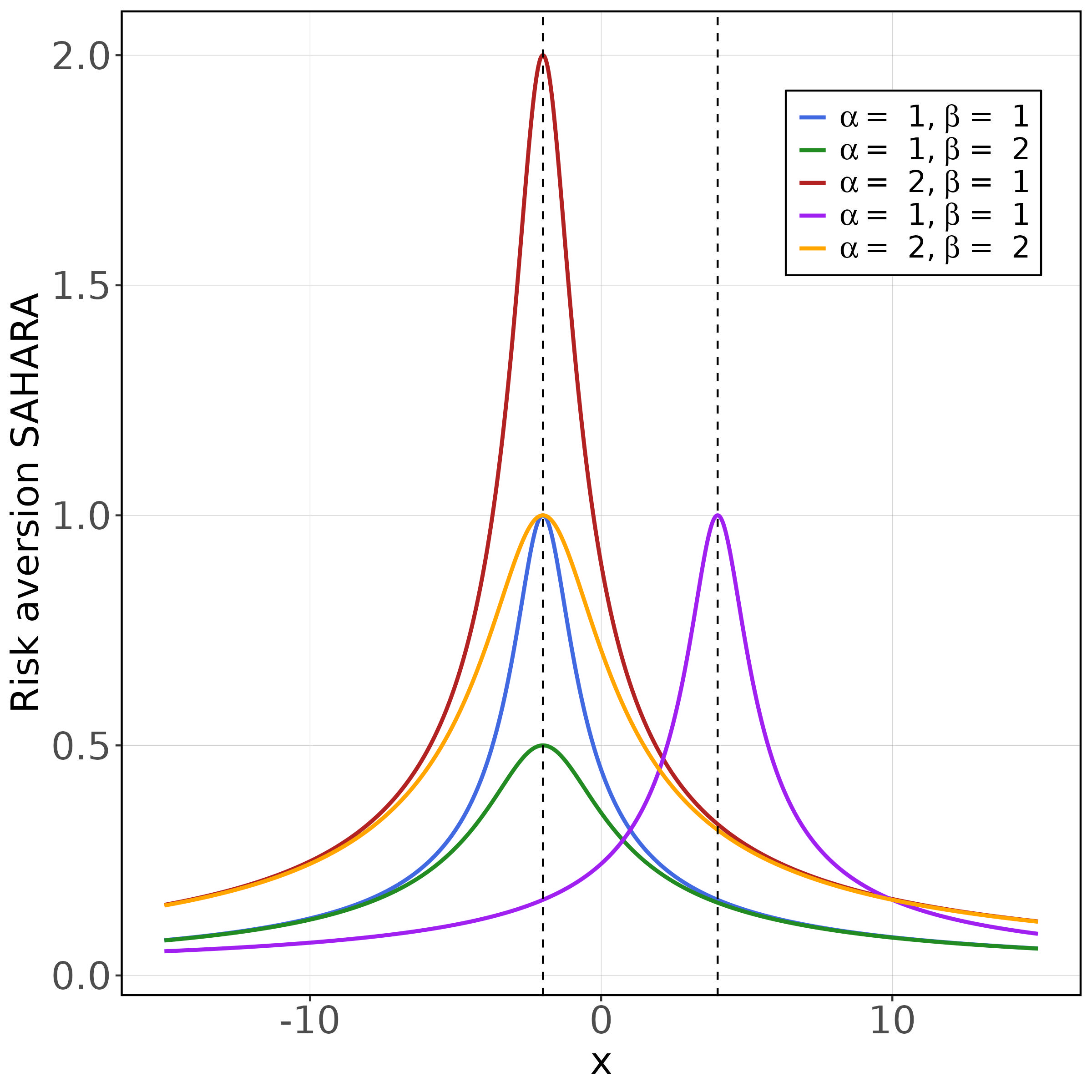}
    }
    \caption{\footnotesize Risk aversion for logistic and SAHARA utility functions, depending on different parameters. The dashed vertical lines on the right-hand side mark the values of the parameter~$d$.}
    \label{fig:logisitic_sahara}
\end{figure}

First, consider a firm with a broad ownership represented by EU agents like in the famous Arrow-Lind Theorem; see \cite{ArrowLind} and \cite[Chapter 19D]{Book}. 
A risk measure used for internal risk management can well be thought to be $v$-Meyer for a threshold utility $v$ representing an aggregate of the individual risk aversions of decision makers. 
The more heterogeneous the ownership, the more restrictive the assumption becomes that $R_v^A$ can only be selected from the set of constants. 
Second, while the natural domain of definition of $\mf e_\lambda$ is $\R$, the definition of $v$-SD orders in general also admits smaller domains of definition $I\subsetneq \R$. This flexibility was relevant, for example, for the  RRMs discussed in Section~\ref{sec:motivation}. 
Third, the risk aversion defined by $\mf e_\lambda$ is always constant. If $\lambda>0$, test utilities $u\in\mathcal U_{\mf e_\lambda}(\R)$ are necessarily strictly risk averse at all levels of wealth. 
If $\lambda<0$, i.e., $\mathcal U_{\mf e_\lambda}(\R)$ contains locally risk-loving test utilities, then $\mathcal U_{\mf e_\lambda}(\R)$ fails to represent a consensual aversion to very large losses. In contrast, a non-CARA $v$ encodes more nuanced risk attitudes. 
Examples are {\em logistic} and {\em SAHARA utilities},  whose risk aversion profiles for various parameter choices are depicted in Figure~\ref{fig:logisitic_sahara}. 

Since they are important examples in this section, we introduce logistic and SAHARA utilities in more detail. For a parameter $\alpha>0$, the logistic utility function is a smooth and S-shaped utility given by
    \begin{equation}\label{def logistic}v(x)=\frac 1{1+e^{-\alpha x}},\quad x\in\R.\end{equation} 
The $v$-SD-order can be approximated by the $\alpha$-SD-order in that $X,Y\in\CX$ satisfy $X\le_{\alpha} Y$ if $X\vsd Y$. 
This approximation is crude though. The risk aversion of every test utility in $\mathcal U_{\mf e_{\alpha}}(\R)$ is lower bounded by $\alpha$ and fails to reflect that $v$ is risk loving at every positive wealth level. Larger $\alpha$ makes this approximation more risk averse, but the risk affinity of $v$ more pronounced. 

SAHARA utility functions have been introduced by \cite{SAHARA}; threshold utility $v$ belongs to this class if there are parameters $\alpha,\beta>0$ and $d\in\R$ such that 
\begin{align}\label{eq:sahara}
    R_v^A(x)=\frac{\alpha}{\sqrt{\beta^2+(x-d)^2}},\quad x\in\R.
\end{align}
Again, the $v$-SD-order could be crudely approximated by the $\lambda$-SD-order, $\lambda:=\frac \alpha\beta$, indicating strict risk aversion. However, the test utilities in $\mathcal U_{\mf e_\lambda}(\R)$ fail to reflect that $v$, and many elements of $\mathcal U_v(\R)$, are near risk neutrality for wealth levels fairly close to anchoring point $d$, the only level at which absolute risk aversion equals $\lambda$. For instance, $R_v^A(d\pm 10\beta)<-0.1 \lambda$.

We start the discussion for general threshold utilities $v$ asking if nontrivial $v$-Meyer risk measures can exist; see Lemma~\ref{lem:worst}. 
A necessary condition is that the risk aversion of a $v\in\mathcal U(\R)$ does not blow up at either tail. Otherwise, the only $v$-Meyer risk measure is the worst-case risk measure---an unrealistically conservative risk assessment which necessarily respects all $v$-SD orders. This is our first impossibility result.

\begin{theorem}\label{thm:unbounded} 
The only normalised $v$-Meyer risk measure is $\worst$ if threshold utility $v\in\mathcal U(\R)$ satisfies 
\begin{equation}\label{eq:liminf}\liminf_{x\to\infty}R_v^A(x)=\infty\quad\text{or}\quad\liminf_{x\to -\infty}R_v^A(x)=\infty.\end{equation}
\end{theorem}

The proof of Theorem~\ref{thm:unbounded} confirms a central insight: cash-additivity and $v$-SD-consistency are in fundamental tension. 
Moreover, finding nontrivial $v$-Meyer risk measures requires to understand when a threshold utility $v$ fails \eqref{eq:liminf}. 
Typically, this is the case when $R_v^A$ is bounded above by some constant $\lambda\in\R$---exceptions being pathological and economically unsound threshold risk aversion profiles such as  $R_v^A(x)=x^2(\sin(x)+1)$, $x\in\R$.
In this situation, the $v$-SD order is weaker than the $\lambda$-SD order, and every $\lambda$-Meyer risk measure constructed in Theorem~\ref{thm:repEXPSD} is also $v$-Meyer. 

The situation of general threshold utilities (bounded from above) is further discussed in Section~\ref{sec:baseRMs}. 
In Sections~\ref{sec:impossibilityPosHom} and~\ref{sec:convex}, we present two further impossibility results under the economically sound assumptions of positive homogeneity and convexity of a risk measure.  

\subsection{Base risk measures}\label{sec:baseRMs}

Theorem~\ref{thm:repEXPSD} shows that the particularly simple risk measures in~\eqref{for_later_new} and \eqref{for later 2} are the fundamental building blocks for $\lambda$-SD-consistent risk measures. 
They arise from setting binary relation $\peq$ in \eqref{eq:benchmarkrisk} to be the SSD/$\lambda$-SD order, but alternative choices among general $v$-SD-orders are conceivable. 
This embeds the functionals in~\eqref{for_later_new} and \eqref{for later 2} in an equally simple, but larger class of risk measures: 

\begin{definition}
    Let $v\in\mathcal U(\R)$ be a threshold utility. For $Z\in\CX$, the {\em base risk measure} under~$v$, is the functional \begin{equation}\label{eq:Z}
    \rho_v(X|Z):=\inf\{m\in\R\mid Z\vsd X+m\},\quad X\in\CX.
    \end{equation}
\end{definition}

\begin{lemma}\label{lem:base}
    Given a threshold utility $v\in\mathcal U(\R)$, each base risk measure under $v$ is a risk measure. 
    Moreover, $v$ is concave if and only if all base risk measures under $v$ are convex.  
\end{lemma} 

Theorem~\ref{thm:repEXPSD} extends seamlessly to the case of general $v\in\mathcal U(\R)$; base risk measures are the fundamental building blocks of all Meyer risk measures.

\begin{proposition}\label{prop:rep general}
 Given a threshold utility $v\in\mathcal U(\R)$, any $v$-Meyer risk measure $\rho$ on $\CX$ can be represented as the lower envelope of base risk measures:
\begin{align}\label{eq:base rep}\rho(X)=\inf_{Y\in\CA_\rho}\rho_v(X|Y),\quad X\in\CX.\end{align}
\end{proposition}

Together, Theorem~\ref{thm:repEXPSD}, Theorem~\ref{thm:unbounded} and Proposition~\ref{prop:rep general} offer an almost complete picture of general Meyer risk measures. First, for nontrivial $v$-Meyer risk measures to exist, the threshold risk aversion $R_v^A$ typically has to be bounded from above by a constant $\lambda$ by Theorem~\ref{thm:unbounded}. Then, Theorem~\ref{thm:repEXPSD} allows us to construct nontrivial $\lambda$-Meyer---and thus $v$-Meyer---risk measures. 
Conversely, if $R_v^A$ is bounded below by constant $\gamma$ and $\rho$ is a $v$-Meyer risk measure, $\rho$ is also $\gamma$-Meyer. 
In conclusion, we can either represent $\rho$ in terms of exponentials (Theorem~\ref{thm:repEXPSD}), avoiding explicit recurrence to $v$, or in terms of $v$-base risk measures (Proposition~\ref{prop:rep general}). 

The preceding construction takes the detour via CARA-SD-orders because, crucially, Proposition~\ref{prop:rep general} is not constructive. It is not an equivalence and {\em does not} claim that each risk measure of shape \eqref{eq:base rep} is $v$-Meyer. 
While this would follow if each base risk measure were $v$-Meyer, Proposition~\ref{prop:Pratt} and Theorem~\ref{thm:MARA} will show that $v$-SD-consistency can only be expected if $v$ is CARA. We start with Proposition~\ref{prop:Pratt}, which demonstrates that base risk measures are not generally $v$-SD-consistent, with the exception of $v$ being CARA. 

\begin{proposition}\label{prop:Pratt}
    The following statements are equivalent for a threshold utility  $v\in\mathcal U(\R)$. 
    \begin{enumerate}[label=\textnormal{(\alph*)}]
        \item $X\vsd Y$ implies for arbitrary $m\in\R$ that  $X+m\vsd Y+m$.
        \item For each $Z\in\CX$, the base risk measure $\rho_v(\cdot|Z)$ is $v$-Meyer.
        \item $R^A_v$ is constant. 
    \end{enumerate}
\end{proposition}

One might argue that the lack of $v$-SD consistency in some base risk measures is negligible if most satisfy it. However, this is not the case for threshold utility functions satisfying:

\begin{assumption}\label{ass:MARA}
There is $a\in\R$ and an interval $J\in\{(a,\infty),(-\infty,a)\}$ such that $v|_{J}$ is three times  continuously differentiable and the absolute risk aversion $R_v^A$ is increasing or decreasing throughout $J$.
\end{assumption}

Apart from the continuous differentiability assumption, Assumption~\ref{ass:MARA} means $v$ exhibits strictly increasing or decreasing absolute risk aversion (IARA or DARA) for all sufficiently large holdings or losses. The DARA assumption is economically sound, as argued by \cite{Arrow71}. Moreover, this class of utilities is rich: Figure~\ref{fig:logisitic_sahara} shows that SAHARA and logistic utilities introduced above satisfy Assumption~\ref{ass:MARA}. Another S-shaped example follows:

\begin{example}\label{ex:utility}
For fixed parameters $\alpha,\beta\in(0,1)$, define  \begin{equation}\label{eq:KT}v(x)=\begin{cases}x^\alpha,&x\ge 0,\\[-0.8ex]
    -(-x)^\beta,&x<0,\end{cases}\qquad x\in\R.\end{equation}
    This $v$
    \`a la \cite{Tversky} is not twice differentiable and thus not in $\mathcal U(\R)$. However, a slight adjustment at wealth levels near 0 makes it three times continuously differentiable and fit Assumption~\ref{ass:MARA}. 
    In that case,~\eqref{eq:KT} can be assumed to hold for all $x\in\R$ with $|x|\ge \eps$. Generally, any utility function being is CRRA for large holdings satisfies Assumption~\ref{ass:MARA}.
\end{example}

Now, given a threshold utility $v$, each $Z\in\CX$ defines a ``location family'' of base risk measures $\rho_v(\cdot|Z+a)$ parametrised by $a\in\R$. 
In case of constant $R_v^A$, this operation is a mere additive shift of the base risk measure $\rho_v(\cdot|Z)$, i.e., $\rho_v(\cdot|Z+a)=\rho_v(\cdot|Z)+a$ for each $a\in\R$, and each element of the family is $v$-SD-consistent. 
If $v$ complies with Assumption~\ref{ass:MARA}, these families contain base risk measures lacking $v$-SD-consistency. 

\begin{theorem}\label{thm:MARA}
    If $v\in\mathcal U(\R)$ complies with Assumption~\ref{ass:MARA} and $Z \in \CX$ is not $\P$-a.s.\ constant, then there exists $a \in \R$ such that 
    $\rho_v(\cdot|Z+a)$ is not $v$-Meyer.
\end{theorem}

By Theorem~\ref{thm:MARA}, we cannot infer easily that risk measures of the form~\eqref{eq:base rep} are $v$-Meyer. The only guarantee for the existence of nontrivial $v$-Meyer risk measures, discussed after Proposition~\ref{prop:rep general}, is an upper bound on threshold risk aversion, in line with Theorem~\ref{thm:unbounded}.
Whether there exist $v$-Meyer risk measures that are not also $\lambda$-SD-consistent for some $\lambda\in\mathbb{R}$ remains open, although our results tentatively suggest that none do.

\subsection{Further properties of threshold utilities}

Before stating our remaining impossibility results, it is necessary to introduce two additional assumptions for $v$. The first is a mild Inada-type growth condition on marginal threshold utility $v'$, which will turn out to exclude positive homogeneity of $v$-Meyer risk measures.

\begin{assumption}\label{ass:Inada}
    Threshold utility $v\in\mathcal U(I)$ satisfies 
\[\liminf_{x\uparrow \sup I}v'(x)=0\quad\text{or}\quad\limsup_{x\downarrow \inf I}v'(x)=\infty.\]
\end{assumption}

Table~\ref{table:utilities} shows that the preceding assumption is satisfied in many common examples. An alternative sufficient criterion based on risk aversion is recorded in the following lemma.

\begin{lemma}\label{lem:utility2}
    If $I$ is unbounded to the right and there is $c>0$ such that $R_v^A(x)\ge c$ holds for all $x$ large enough, then $v$ satisfies Assumption~\ref{ass:Inada}.
\end{lemma}

The second assumption concerns the inverse function $v^{-1}$ rather than the derivative. 

\begin{assumption}\label{ass:star}
    For all $\gamma,\delta>0$ there exists $C\in \R$ and an increasing sequence $(t_n)\subseteq v(I)$ such that $t_1>\frac C{1+\delta}$, $C-\delta t_n\in v(I)$ for all $n\in\N$, and 
    \[\lim_{n\to\infty}v^{-1}(t_n)+\gamma v^{-1}(C-
    \delta t_n)=\infty.\]
\end{assumption}

As $v^{-1}$ is nondecreasing, Assumption~\ref{ass:star} can only hold for thresholds $v\in\mathcal U(I)$ for which the sequence $v^{-1}(t_n)$ diverges to $\infty$. More precisely, the following conditions are necessary.  
\begin{enumerate}[label=(\alph*)]
\itemsep0em
    \item $I$ is unbounded to the right.
    \item If $v(I)$ is unbounded above, then $v(I)=\R$.
\end{enumerate}
Assumption~\ref{ass:star}, though seemingly unwieldy, is satisfied by many examples; see Table~\ref{table:utilities}. 

\begin{table}[ht]
\centering
{
\renewcommand{\arraystretch}{1.5} 
\footnotesize 
\begin{tabular}{lcc}
& Assumption~\ref{ass:Inada} & Assumption~\ref{ass:star} \\
\midrule
$v$ concave 
& $\iff~\begin{cases}
\lim\limits_{x\to\sup I}v'(x)=0\text{ or}\\
\lim\limits_{x\to\inf I}v'(x)=\infty
\end{cases}$ 
&  $\iff~\begin{cases}\text{Ass.~\ref{ass:Inada} satisfied and}\\
v(I)\text{ unbounded below}
\end{cases}$ \\

$v$ convex 
& No 
& No \\

$\mf e_\lambda,~\lambda\in \R$ (Eq.~\eqref{ex:orders})
& $\iff\,\lambda>0$ 
& $\iff\,\lambda>0$ \\

$\mf p_a,\,a\ge 0$ (Eq.~\eqref{ex:CRRA})
& $\iff~a<1$ 
& $\iff~a=0$ \\

Kahneman-Tversky, $0<\alpha,\beta<1$ (Ex.~\ref{ex:utility}) 
& Yes 
& $\iff\,\alpha<\beta$ \\

Logistic utility (Eq.~\eqref{def logistic})
& Yes 
& Yes \\

SAHARA, $\alpha,\beta>0$, $d\in\R$ (Eq.~\eqref{eq:sahara})
& Yes 
& Yes \\
\bottomrule
\end{tabular}
\vspace{0.9em} 
\caption{\footnotesize Examples of threshold utilities that do or do not satisfy Assumptions~\ref{ass:Inada} and~\ref{ass:star}. Notably, Assumption~\ref{ass:Inada} for concave $v$ is equivalent to one of the Inada conditions, which are common in EU maximisation problems; see~\cite{kramkov, kramkov2}.}
\label{table:utilities}
} 
\end{table}

\subsection{Impossibility of positive homogeneity}\label{sec:impossibilityPosHom}

The second impossibility theorem considers the additional property of positive homogeneity. 
Under Assumption~\ref{ass:Inada}, it turns out that the only positively homogeneous 
$v$-Meyer risk measure is the worst-case risk measure.

\begin{theorem}\label{thm main 1}
If threshold utility $v\in\mathcal U(I)$ satisfies Assumption~\ref{ass:Inada}, the only positively homogeneous $v$-Meyer risk measure is $\worst$. 
\end{theorem}

Theorem~\ref{thm main 1} is of theoretical significance and complements various results from the existing literature.
\cite[p.\ 4632]{Huang} observes that the $\lambda$-SD order ($\lambda<0$) is invariant under additive translations, but not changes of scale. While reminiscent of Theorem~\ref{thm main 1}, our result addresses risk measures consistent with the stochastic order rather than the stochastic order itself. Additionally, threshold utilities $\mf e_\lambda$ with parameter $\lambda<0$ do not comply with Assumption~\ref{ass:Inada} and are not covered by Theorem~\ref{thm main 1}.

Another instructive comparison can be drawn between Theorem~\ref{thm main 1} and one of the main results from \cite{Fractional}, Theorem 3. 
By the first statement in said theorem, for $\lambda\le 0$, a positively homogeneous risk measure is $\lambda$-SD-consistent if and only if it is SSD-consistent.
This implies that many $\lambda$-SD-consistent, positively homogeneous risk measures exist. 
Conversely, Theorem~\ref{thm main 1} shows that for many threshold utilities, $v$-SD-consistency, cash-additivity and positive homogeneity do not go together at all unless one deals with the worst-case risk measure.
This asymmetry is fully in line with Table~\ref{table:utilities}, which records that $\mf e_\lambda$ complies with Assumption~\ref{ass:Inada} if and only if $\lambda>0$. 
Hence, we also see that Assumption~\ref{ass:Inada} cannot be dropped without risking Theorem~\ref{thm main 1} to fail. 
Another clarifying observation is the second statement from \cite[Theorem 3]{Fractional}, asserting that every SSD-consistent risk measure is $\mf p_a$-Meyer for suitable parameters $a$. These threshold utilities are defined on~$(0,\infty)$ rather than the whole real line. 
In this way, Theorem~\ref{thm main 1} complements the results of \cite{Fractional} and deepens the understanding of the relationship between risk measures and stochastic orders.

\subsection{Impossibility of star-shapedness}\label{sec:convex}

We now turn to the class of law-invariant star-shaped risk measures studied by \cite{Castagnoli}, which encompasses all 
positively homogeneous and normalised convex risk measures. 
A key concept is the associated {\em asymptotic functional}~$\rho^\infty$, a standard object from convex analysis:
\[\rho^\infty(X):=\lim_{t\to\infty}\tfrac{\rho(tX)}{t},\quad X\in\CX,\]
where the preceding limit is necessarily isotone. 
By construction, $\rho^\infty$ is itself a positively homogeneous risk measure with $\rho\le \rho^\infty$. 

Theorem~\ref{thm main 2} asserts that, if $v$ complies with Assumption~\ref{ass:star} and $\rho$ is star-shaped, each $v$-Meyer risk measure $\rho$ satisfies $\rho^\infty=\worst$. 
The value of this necessary---but not sufficient---condition is that it excludes many common star-shaped risk measures from being $v$-Meyer.

\begin{theorem}\label{thm main 2}
If threshold utility $v\in\mathcal U(I)$ satisfies Assumption~\ref{ass:star}, a law-invariant star-shaped risk measure $\rho$ is $v$-Meyer only if $\rho^\infty=\worst$.  
\end{theorem}

Note that Theorem~\ref{thm main 2} applies to every normalised, convex, law-invariant risk measure. 
However, also in that case the condition $\rho^\infty=\worst$ is not sufficient. 
Indeed, consider for $\lambda>0$ the risk measure $\rho_{\lambda}=\rho_{\lambda,0}$---which satisfies $\rho_\lambda^\infty=\worst$---and set $v=\mf e_{1}$. Then $\rho_\lambda$ is $v$-Meyer if and only if $\lambda\ge 1$. 

In a nutshell, the results in the preceding sections show that assuming a risk measure is $v$-SD-consistent shrinks the class of admissible functionals immensely, and often excludes other typically assumed properties.  
Theorem~\ref{thm main 1} and Theorem~\ref{thm main 2}
illustrate the latter tension for positive homogeneity and star-shapedness. 

\section{Conclusion and outlook}\label{sec:conclusionOutlook}

We study {\em Meyer risk measures}, defined by their consistency with a stochastic order from the $v$-SD class, $v$ being a threshold utility function. This generalises SSD-consistent risk measures.
Among the possible choices for $v$, the family $(\mf e_\lambda)_{\lambda\in\R}$ of exponential utilities, displaying constant absolute risk aversion, is crucial. With this choice, formulae for general risk measures are readily available, demonstrating the abundance of nontrivial examples. Several realistic risk management frameworks show that these Meyer risk measures can outperform existing risk measures, like Loss VaR or adjusted ES. We also empirically evaluate the performance of $\lambda$-SD-consistent risk measures using logarithmic returns from real-world financial data, leading to a direct connection to return risk measures. Then, by proving that monotone additive statistics are typically Meyer risk measures, we obtain that the latter is also relevant for several decision-making applications. While $v$-SD-consistent risk measures, for a general threshold utility $v$, can be represented as the lower envelope of a class of so-called base risk measures, the latter are typically not $v$-SD consistent if risk aversion of $v$ is not constant. 
This reveals a deeper link between the axiomatic structure of monetary risk measures and SSD than previously acknowledged, which becomes more pronounced when additional properties are imposed. 
Notably, it is often only the worst-case risk measure which achieves both positive homogeneity and $v$-SD consistency. Similarly, the classes of convex or star-shaped Meyer risk measures are quite limited.

Open questions persist and present opportunities for future research. Firstly, in our study we always fix a threshold utility function $v$ and inspect if $\rho$ is $v$-Meyer. The inverse route, identifying for a given risk measure $\rho$ all threshold utilities $v$ for which it is consistent, remains open.
Secondly, our results rely on the cash-additivity of monetary risk measures , and relaxing this assumption would give rise to promising new technical challenges.
Finally, further applications of Meyer risk measures, such as their use in risk-sharing problems or portfolio optimisation, are beyond the scope of this paper.

\bigskip

\appendix

\section{The case of alternative Meyer orders}\label{sec:otherMeyer}

The $v$-SD-orders introduced in Definition~\ref{def:vSSD} rank random prospects via a consensus among EU agents {\em at least} as risk averse as a threshold case. However, \cite{MeyerJET} and \cite{LiuMeyer2025} also explore stochastic orders resulting from excluding {\em excessively} risk-averse EU agents. 
Specifically, they impose upper bounds on relative or absolute risk aversion instead of lower bounds, or combine the two approaches by sandwiching risk aversion between two fixed thresholds. 
Here, we show that risk measures consistent with these alternative stochastic orders are linked directly to Meyer risk measures and thus covered by the results of our paper.

Given an open interval $I$, fix two threshold utilities $v,w\in\mathcal U(I)$ and introduce
\begin{equation*}\mathcal U^v(I):=\{u\in\mathcal U(I)\mid R_u^A\le R_v^A\}\and \mathcal U_v^{w}(I):=\{u\in\mathcal U(I)\mid R_v^A\le R_u^A\le R_{w}^A\}.\end{equation*} 
The test utilities in 
$\mathcal U^v(I)$ are at least as risk-seeking as $v$, and the risk aversion of the ones in $\mathcal U_v^{w}(I)$  is bounded between those of $v$ below and $w$ above. 
Adapting Definition~\ref{def:vSSD}, both sets seamlessly lead to stochastic orders on $\CX$ we shall denote by $\le^v$ and $\le_v^{w}$. 

\begin{proposition}
    If a risk measure $\rho$ is consistent with $\le^v$, $\bar\rho:=-\rho(-\cdot)$ is Meyer. If $\rho$ is consistent with $\le^w_v$, then $\rho$ is $v$-Meyer. 
\end{proposition}
\begin{proof}
Set $J:=-I$ and note that 
$T\colon \mathcal U(I)\to\mathcal U(J)$ defined by $T(v):=-v(-\cdot)$
is a bijection. 
Moreover, for all $u\in\mathcal U(I)$ and $y\in J$, $R_{T(u)}^A(y)=-R_{u}^A(-y)$.
Hence, utility functions $u,h\in\mathcal U(I)$ and a payoff $X\in\mathcal X$ satisfy (a) $X\in\mathcal L^1_{T(u)}$ if and only if $-X\in\mathcal L^1_u$, (b) $R_{T(u)}^A\ge R_{T(h)}^A$ if and only if $R_{u}^A\le R_{h}^A$, and (c) $X\le^v Y$ if and only if $-Y\le_{T(v)} -X$.

Now suppose that $\rho$ is a risk measure on $\CX$ that is consistent with $\le^v$, i.e., $X\le^v Y$ implies $\rho(X)\ge \rho(Y)$. 
The functional $\bar\rho(X):=-\rho(-X)$, $X\in\CX$, is also a risk measure.
Whenever $X,Y\in\CX$ satisfy $X\le_{T(v)} Y$, property (c) implies $-Y\le^v -X$, which entails that 
$\rho(-Y)\ge \rho(-X)$ and  $\bar\rho(Y)\le\bar\rho(X)$.
Hence, $\bar\rho$ is $T(v)$-Meyer. 

Regarding the ``sandwiched'' order $\le^{w}_v$, the definition immediately implies that  
\[X\vsd Y~\text{or}~X\le^{w}Y\quad \implies\quad X\le^{w}_v Y.\]
Hence, risk measures consistent with $\le^{w}_v$ are $v$-Meyer. 
\end{proof}

As the original risk measure $\rho$ can be recovered from $\bar\rho$ without effort, meaning that all our results transfer easily to risk measures consistent with $\le^v$. 
In case of $\le^w_v$, either threshold utility $v$ or $w$ can serve as the starting point of the analysis. 

\section{Some auxiliary results}

For an open interval $I\subseteq\R$, $u\in\mathcal U(I)$, and $X\in \CL^1_u(I)$,  $K_u(X):=u^{-1}\big(\E[u(X)]\big)\in I$ is the certainty equivalent of $X$ under $u$.
The next lemma follows from \cite[Theorem 1]{Pratt}. 

\begin{lemma}\label{lem:certainty}
    \begin{enumerate}[label=\tn{(\alph*)}]
        \item For all $v\in\mathcal U(I)$ and $X\in\CL^1_v$, $X\vsd K_v(X)$.
        \item For all $v\in\mathcal U(I)$, $u\in\mathcal U_v(I)$, and $X,Y\in \CL^1_u\cap\CL^1_v$, $X\vsd Y$ implies $K_u(X)\le K_u(Y)$. 
    \end{enumerate}
\end{lemma}

\begin{lemma}\label{lem:benchmark}
    Let $\peq$ be a binary relation on $\CX$ with the following properties: 
    \begin{enumerate}[label=\tn{(\roman*)}]
    \itemsep-1pt
        \item $X\peq Y$ and $(X,Y)$ having the same marginals as $(X',Y')$ implies $X'\peq Y'$.
        \item $X\le Y$ a.s.\ implies $X\peq Y$.
        \item $x,y\in\R$ with $x<y$ satisfy $x\not\peq y$.
    \end{enumerate}
    If $Y\in\CX$ is arbitrary but fixed, $\rho_\peq(\cdot|Y)$ defined by \eqref{eq:benchmarkrisk}
    is a law-invariant risk measure. 
\end{lemma}
\begin{proof}
    Law invariance, antitonicity and cash-additivity follow directly from properties (i) and (ii). 
    Let $X,Y\in\CX$.
    As $X+\|X\|_\infty+\|Y\|_\infty\ge Y$, the set $\{m\in\R\mid X+m\succeq Y\}$ is nonempty by (ii) and $\rho_\peq(X|Y)>-\infty$. Moreover, $\rho_\peq(X|Y)<\|X\|_\infty+\|Y\|_\infty$. Indeed, for $m<-\|X\|_\infty-\|Y\|_\infty$, 
    \[X+m\le \|X\|_\infty+m<-\|Y\|_\infty\le Y.\]
    Assuming $X+m\succeq Y$ would imply $-\|Y\|_\infty\peq \|X\|_\infty+m$ and contradict (iii). 
\end{proof}

\begin{lemma}\label{lem:worst}
For $I\subseteq\R$ and $v\in \mathcal U(I)$, the worst-case risk measure  $\worst$ is $v$-Meyer. 
\end{lemma}
\begin{proof}
    We have to prove for $X,Y\in\mathcal \CX\cap L^1_v(I)$ with $X\vsd Y$ that $\worst(X)\ge \worst(Y)$.
    {\em A priori}, $\worst(X),\worst(Y)\le -a$ for  $a:=\inf I$.
    Thus, if $\worst(X)=-a$, we can already conclude that $\worst(X)\ge \worst(Y)$. 
    Otherwise, consider the nonpositive, concave and nondecreasing function 
    \begin{center}$u(x)=\min\{x-v(-\worst(X)),0\},\quad x\in\R.$\end{center}
    By~\eqref{complicated equiv}, $v(X)\ssd v(Y)$, which means that $0=\E[(u\circ v)(X)]\le 
    \E[(u\circ v)(Y)]$. 
    Hence, $v(Y)\ge v(-\worst(X))$, from which the desired inequality $\worst(X)\ge\worst(Y)$ follows immediately. 
\end{proof}

\section{Proofs accompanying Section~ \ref{sec:representation_vsd}}\label{appendix:4}

For the proof of Theorem~\ref{thm:repEXPSD}, we recall from \cite[Definition 3.1]{BaeuerleShushi} the \textit{tail quasi-linear mean} under a utility function $v\in\mathcal U(\R)$ and a confidence level $p\in[0,1]$, which is the functional
\[
    \rho_{v,p}(X):= -(v^{-1}\circ(-\es_p)\circ v)(X),\quad X\in\CX.
\]
This functional is well-defined because $v(a)\le -\es_p(v(X))\le v(b)$ whenever $a\le X\le b$, i.e., $-\es_p(v(X))\in v(\R)$ for all $X\in\CX$. 

\begin{lemma}
    For a threshold utility $v\in\mathcal U(\R)$ and a $v$-Meyer risk measure $\rho$ it holds that
    \begin{align}\label{eq:tailQuasiLinearMap_rep}
        \rho(X)=\inf_{Y\in\CA_\rho} \inf\big\{m\in\R\mid\forall\,p\in[0,1]:~\rho_{v,p}(Y)\geq \rho_{v,p}(X+m)\big\}, \quad X\in\CX.
    \end{align}
\end{lemma}

\begin{proof}
    For all $Y\in\CA_\rho$, $X\in\CX$, and $m\in\R$, 
$Y\vsd X+m$ is equivalent to $(\es_p\circ v)(Y)\ge (\es_p\circ v)(X+m)$ for  all $p\in[0,1]$. Using that $-v^{-1}$ is decreasing, we obtain that this statement is equivalent to $\rho_{v,p}(Y)\geq \rho_{v,p}(X+m)$ for all $p\in[0,1]$. Hence,~\eqref{eq:tailQuasiLinearMap_rep} is a consequence of~\eqref{eq:base rep}.
\end{proof}

\begin{proof}[Proof of Theorem~\ref{thm:repEXPSD}] The case $\lambda = 0$ is \cite[Theorem 3.1]{Consistent}. Now, let $\lambda \neq 0$. If $v = \mf e_{\lambda}$, following the argumentation of~\cite[Theorem 2.2]{Certainty} delivers cash-additivity of  the tail quasi-linear mean $\rho_{v,p} = \rho_{\lambda,p}$, irrespective of $p\in[0,1]$. 
From representation~\eqref{eq:tailQuasiLinearMap_rep}, we obtain 
\[
    \rho(X) = \inf_{Y\in\CA_{\rho}}\sup_{p\in[0,1]} \{\rho_{\lambda,p}(X)-\rho_{\lambda,p}(Y)\},\quad X\in\CX.
\]

Now, note that for each $Y\in\mathcal{A}_{\rho}$ the map $p\mapsto -\es_p\big(\mf e_{\lambda}(Y)\big)$ is nonincreasing. Hence, the set $\mathcal G:=\{g_Y\mid Y\in\CA_\rho\}$ with $g_Y(p):=\rho_{\lambda,p}(Y)$ only contains nondecreasing functions, attaining values in $\R\subseteq (\infty,\infty]$. 

Conversely, by antitonicity of $\es_p$ and $\mf e_\lambda$ being nondecreasing, we obtain that $X\mapsto\rho_{\lambda,p}(X)$ is a risk measure, i.e.,~antitone and cash-additive. Furthermore, every risk measure of shape~\eqref{eq:rhog_new} is $\lambda$-SD-consistent because of \eqref{complicated equiv} and SSD-consistency of ES. 
\end{proof}

\begin{proof}[Proof of Corollary~\ref{cor:orderMeyerRiskMeasures}]
    Combine \cite[Theorem 3.5]{BaeuerleShushi} with Jensen's inequality applied to $\es_p(\mf e_\lambda(X))=-\E_{\Q_X}[\mf e_\lambda(X)]$.
\end{proof}

\begin{lemma}\label{lem:riskMeasureDependenceRiskAversion}
    For fixed $X\in\CX$ and $p\in[0,1]$, the map $h(\lambda):=\rho_{\lambda,p}(X)$ is nondecreasing, continuous, and satisfies  
    \begin{equation*}\lim_{\lambda\rightarrow -\infty}h(\lambda) = -q_X(1-p),\quad h(0) = \es_p(X),\quad\text{and}\quad\lim_{\lambda\rightarrow \infty}h(\lambda) = \worst(X).\end{equation*}
\end{lemma}

\begin{proof}
    If $p=1$, it holds that $\rho_{\lambda,p}(X) = \es_1(X)=\worst(X)$ for arbitrary $\lambda\in\R$. As the assertions of the lemma are easy to check in this case, we assume $p\in[0,1)$ throughout the rest of the proof. 
    The identity $h(0)=\es_p(X)$ is clear. 
    To prove the remaining statements, invoke \cite[Theorem 4.47 and Remark 4.48]{FoeSch} to find a probability measure $\Q$ absolutely continuous with respect to $\P$, independent of $\lambda$, and satisfying $\{\frac{\diff \Q}{\diff \P}>0\}= \{X\leq q_X(1-p)\}$ such that   
    \[
        h(\lambda) = -\mf e_\lambda^{-1}\big(\E_{\Q}[\mf e_{\lambda} (X)]\big) = \frac{1}{\lambda}\log\left(\E_{\Q}[e^{-\lambda X}]\right).
    \]
    Hence, $h(\lambda)$ is the negative certainty equivalent under $\Q$ and $\mf e_\lambda$. 
    As $\mf e_{\gamma}\in\mathcal U_{\mf e_\lambda}(\R)$ whenever $\gamma\ge\lambda$, $h$ being nondecreasing follows with \cite[Theorem 1]{Pratt}. 
    
    Continuity of $h$ on $\R\setminus\{0\}$ follows with the norm continuity of $\E_{\Q}[\cdot]$ on $\CX$. 
    For continuity in $0$, let $\widetilde h$ be the function obtained by replacing $X$ with $-X$ in the definition of $h$; then $h(\lambda)=-\widetilde h(-\lambda)$. By \cite[Example 1.3.1]{Kaas}, 
    \[\lim\limits_{\lambda\downarrow 0}h(\lambda) = h(0)\and \lim\limits_{\lambda\uparrow 0}h(\lambda)=-\lim\limits_{\lambda\downarrow 0}\widetilde h(\lambda)=0.\]
        
    For the limit behaviour, note that  $\lim_{\lambda\to\infty}h(\lambda)=M^\Q(-X)=\worst(X)$ by \cite[Equation (4)]{follmerKnispel}. 
    Here, $M^{\Q}(Y)=\inf\{x\in\R\,|\,Y\leq x~\Q\text{-a.s.}\}$, $Y\in\CX$. 
    Moreover, using $\widetilde h$ once again, 
    \[\lim_{\lambda\to-\infty}h(\lambda)=-\lim_{\lambda\to\infty}\widetilde h(\lambda)=-M^\Q(X)=-q_X(1-p).\]
\end{proof}

\begin{corollary}\label{cor:limitResultLVaR}
    Let $g\colon[0,1]\rightarrow (-\infty,\infty]$ be a nondecreasing function with $g\not\equiv\infty$ and $X\in\CX$ with continuous 
    quantile function $q_X$. 
    Then the function  
    $H(\lambda):= \sup_{p\in[0,1]}\{\rho_{\lambda,p}(X)-g(p)\}$
    is nondecreasing, continuous, and satisfies 
    \[
        \lim_{\lambda\rightarrow -\infty} H(\lambda) = \LVaR^g(X)\quad\text{and}\quad 
        \lim_{\lambda\rightarrow \infty}H(\lambda) = \worst(X)-g(0).
    \]
\end{corollary}

\begin{proof}
    $H$ being nondecreasing is a direct consequence of Lemma~\ref{lem:riskMeasureDependenceRiskAversion}. 
    Set
    \[
        f_{\lambda}(p):=\rho_{\lambda,p}(X)-g(p)\quad\text{and}\quad f(p):=-q_{X}(1-p)-g(p),\quad \lambda\in\R,~p\in[0,1].
    \]
    Because of the continuity of $p\mapsto \rho_{\lambda,p}(X)$ for fixed $\lambda$, we can tacitly assume that $g$ is left-continuous on $(0,1]$, which in turn leads to upper semicontinuity of  
     each $f_\lambda$. \cite[Theorem 2.43]{Ali} admits to select for all $\lambda\in\R$ a $p_\lambda\in[0,1]$ such that $H(\lambda)=\rho_{\lambda,p_\lambda}(X)-g(p_\lambda)$. 
    Standard arguments now show that $H$ is continuous.
    
   Regarding the limit behaviour, note that $q_X$ is continuous.
    By Lemma~\ref{lem:riskMeasureDependenceRiskAversion}, $\rho_{\lambda,p}(X)\downarrow -q_X(1-p)$ holds for all $p\in[0,1]$ as $\lambda\downarrow-\infty$.  
    By Dini's Theorem~\cite[Theorem 2.66]{Ali}, $\lim_{\lambda\to-\infty}\sup_{p\in[0,1]}|\rho_{\lambda,p}(X)+q_{X}(1-p)|=0$. Finally, setting $I:=\{p\in[0,1]\mid g(p)<\infty\}$,
    \[\limsup_{\lambda\to-\infty}\sup_{p\in I}|f_{\lambda}(p)-f(p)|\le  \lim_{\lambda\to-\infty}\sup_{p\in[0,1]}|\rho_{\lambda,p}(X)+q_{X}(1-p)|=0.
    \]
    The other limit is obtained analogously by employing   Lemma~\ref{lem:riskMeasureDependenceRiskAversion}.
\end{proof}

For the proof of Proposition~\ref{prop:ssd1}, we first generalise the point of view of the Section~\ref{sec:motivation}. 
The objects of interest are a risk measure $\rho$ on $\CX$, open intervals $I,J\subseteq \R$, and utility functions $v\in\mathcal U(I)$ and $u\in\mathcal U(J)$ with $v(I)=u(J)$.  
The latter condition permits the definition of utility functions $h:=v^{-1}\circ u\in\mathcal U(J)$ and $h^{-1}:=u^{-1}\circ v\in\mathcal U(I)$.
Moreover, set $\mathcal D:=\{X\in\CX\mid X\in J,~h(X)\in\CX\}$.
We are interested in comparing the two functionals $\rho$ on $\CX$ and 
$\eta:=\rho\circ h$ on $
\mathcal D$.
A relevant example is $v=\id_\R$ and $u=\log\in\mathcal U\big((0,\infty)\big)$, in which case $h=\log$ and  $\exp\circ\eta$ is an RRM.

\begin{theorem}\label{relationMeyerRRMs}
    $\rho$ is $v$-Meyer if and only if $\eta$ is $u$-SD-consistent. 
\end{theorem}
\begin{proof}
    Assume $\rho$ is $v$-Meyer and let $X,Y\in \mathcal L^1_u$ such that $X\le_{u}Y$. The latter means that $u(X)\ssd u(Y)$. 
    In particular, $X':=(v^{-1}\circ u)(X)=h(X)\in\mathcal L^1_v$, $Y':=(v^{-1}\circ u)(Y)=h(Y)\in\mathcal L^1_v$, and 
    \[v(X')=u(X)\ssd u(Y)=v(Y'),\]
    meaning that $X'\vsd Y'$.
    Hence, $\eta(X)=\rho(X')\ge \rho(Y')=\eta(Y)$. The converse implication is shown analogously, replacing $h$ by $h^{-1}$. 
\end{proof}

\begin{proof}[Proof of Proposition~\ref{prop:ssd1}]
We aim to apply Theorem~\ref{relationMeyerRRMs} with $v=\mf e_\lambda$, $\lambda\in\R$.
To this end, note that 
\[\mf e_\lambda(\R)=\begin{cases}(0,\infty)&~\text{if }\lambda<0,\\[-0.8ex]
\R&~\text{if }\lambda=0,\\[-0.8ex]
(-\infty,0)&~\text{if }\lambda>0.
\end{cases}\]
Set $J=(0,\infty)$ and 
\[u_\lambda(x):=\begin{cases}\mf p_0(x)=\log(x)&~\text{if }\lambda=0,\\[-0.8ex]
\lambda^2\mf p_{-\lambda}(x)=-\lambda x^{-\lambda}&~\text{if }\lambda\neq 0,\end{cases}\quad x\in J.\]
It is easy to see that $\mf e_\lambda(\R)=u_\lambda(J)$ and that $h=\mf e_\lambda^{-1}\circ u_\lambda=\log$. 
Moreover, $R^A_{u_\lambda}=R^A_{\mf p_{-\lambda}}$, meaning that the $u_\lambda$-SD order and the $\mf p_{-\lambda}$-SD order coincide. 

For the second part, note that by $-1$-positive homogeneity of $\eta$, we obtain that 
\[
    \eta|_{\CE}(X) = \inf\{m\in(0,\infty)\,|\, mX\in\CB\},\quad X\in\CE.
\]
In parallel with with the proof of Theorem~\ref{thm:repEXPSD}, one shows
\[
    \eta|_{\CE}(X) = \inf_{Y\in\CB}\inf\Big\{m\in(0,\infty)\,\Big|\,\forall\,p\in[0,1]:~\eta_{\lambda,p}(Y) \geq \eta_{\lambda,p}(mX)   \Big\}.
\]
From positive homogeneity of $\eta_{\lambda,p}$ we then obtain~\eqref{eq:rep_RRM_viaPowerUtility}.
\end{proof}

\section{Proofs accompanying Section~\ref{sec:Mu}}

\begin{proof}[Proof of Theorem~\ref{thm:Mu}]
Define $\Lambda:=\{\lambda\in\R\mid \rho_\Phi~\lambda\text{-Meyer}\}$.
First, $\rho_\Phi$ being $\lambda$-Meyer implies for all $c\le \lambda$ that $\rho_\Phi$ is also $c$-SD-consistent. Hence, $\Lambda$ is an interval.

Next, we claim that $-c_+\in\Lambda$. Indeed, for all $c\in\supp(\mu)$, we have $-c\ge -c_+$. 
As $E_\lambda(X)$ is the certainty equivalent under $\mf e_{-\lambda}$, $\lambda\in\R$, the domination
$X\le_{-c_+}Y$ implies $E_c(X)\le E_c(Y)$ for all $c\in\supp(\mu)$. 
Consequently, $\rho_\Phi(X)\ge \rho_\Phi(Y)$ in that case. 

Define $\lambda^\star:=\sup\Lambda$ and assume towards a contradiction that $\lambda^\star=\infty$. This would mean that $\N\subseteq\Lambda$. 
Consequently, because $X\le_n E_{-n}(X)$ holds for arbitrary $X\in\CX$ and $n\in\N$, we have 
\[\rho_\Phi(X)\ge -E_{-n}(X)=\frac 1 n\log\big(\E[e^{-nX}]\big).\]
Taking the limit $n\to\infty$, $\rho_\Phi(X)\ge M(-X)$, which means that $\Phi(X)\le -M(-X)$. 
The latter would imply $\supp(\mu)=\{-\infty\}$, which is a contradiction. 

Now let $X,Y\in\CX$ be such that $X\le_{\lambda^\star}Y$, i.e., $\mf e_{\lambda^\star}(X)\ssd\mf e_{\lambda^\star}(Y)$. For $\lambda^\star$-SD-consistency, we need to show $\rho_\Phi(X)\ge \rho_\Phi(Y)$. To this end, let $c<\lambda^\star$ and rewrite the above statement as
\[\mf e_c\big((\mf e_c^{-1}\circ \mf e_{\lambda^\star})(X)\big)\ssd \mf e_c\big((\mf e_c^{-1}\circ \mf e_{\lambda^\star})(Y)\big),\]
i.e., $X_c:=(\mf e_c^{-1}\circ \mf e_{\lambda^\star})(X)$ and $Y_c:=(\mf e_c^{-1}\circ \mf e_{\lambda^\star})(X)$ satisfy $X_c\le_cY_c$. 
Consequently, 
\[\rho_\Phi(X_c)\ge \rho_\Phi(Y_c),\quad c<\lambda^\star.\]
As $c\uparrow\lambda^\star$, $X_c$ and $Y_c$ converge to $X$ and $Y$ in norm, respectively. 
Hence, we have 
$\rho_\Phi(X)\ge \rho_\Phi(Y)$. 

It remains to show that $\lambda^\star\le-c_-$, which is clear if $c_-=-\infty$. 
Otherwise, pick an arbitrary $\lambda>-c_-$, i.e., the certainty equivalent $K_\lambda(X)$ for an arbitary nonconstant $X\in\CX$ satisfies 
\[K_\lambda(X)<E_c(X),\quad c\in\supp(\mu).\]
Consequently, we obtain $\rho_\Phi(K_\lambda(X))>\rho_\Phi(X)$, contradicting $\lambda$-SD-consistency. 
\end{proof}

\begin{example}\label{exam:MAS}
    Let $a_1,\dots,a_n\in \R$ with $0<a_1<\dots<a_n$ and let $\pi_1,\dots,\pi_n\in(0,1)$ with ${\sum_{i=1}^n\pi_i = 1}$.  
    Let $\mu$ be the Borel probability measure assigning to each singleton $\{a_i\}$ probability $\pi_i$. 
    The resulting MAS-induced risk measure is
    \[
        \rho_{\Phi}(X) = -\Phi(X) = -\int_{\overline\R} \rho_{c,0}(-X) \mu(\diff c) = -\sum_{i=1}^n\pi_i\,\rho_{a_i,0}(-X),\quad X\in\CX.
    \]
    We claim that $\lambda^\star=-a_n$. 
    By Theorem~\ref{thm:Mu}, $\lambda^\star\ge-a_n$, and for the converse inequality, it suffices to show for $0<r<a_n$ that $\rho_{\Phi}$ is not $-r$-SD-consistent, i.e., exhibiting random variables $X,Y\in \CX$ satisfying $X\leq_{-r} Y$ and $\rho_{\Phi}(X)<\rho_{\Phi}(Y)$. 
    Using the transformations $U=e^{X}$ and $V=e^{Y}$, one can equivalently construct random variables $U,V\in\CE$ with $U\ssd V$ and 
    \begin{align}\label{eq:examMAS}
        \prod_{i=1}^{n}\E[ V^{a_i}]^{\pi_i/a_i}<\prod_{i=1}^{n}\E[U^{a_i}]^{\pi_i/a_i}. 
    \end{align}
    To this end, choose $V = 1$ a.s.~and for $M>1$ determined below, suppose $U$ satisfies that 
    \[
        \P\big(U=M^{1/r}\big) = 1 - \P\big(U= 2^{-1/r}\big) =\frac 1{2M}.
    \]
    Observe that $U^r\ssd\E\big[U^r\big] = 1-\frac 1{4M}<1 = V^{r}$. Moreover, for all $1\le i\le n-1$,
    \begin{align*}
        \E[U^{a_i}]&= \tfrac 12 M^{\frac{a_i}{r}-1} + 2^{-\frac{a_i}{r}}\big(1-\tfrac{1}{2M}\big) > 2^{-\frac{a_i}{r}-1}\quad\text{for all }1\le i\le n-1,\\
        \text{and}\qquad \E[U^{a_n}]&=\tfrac 1 2M^{\frac{a_n}{r}-1} + 2^{-\frac{a_n}{r}}\big(1-\tfrac{1}{2M}\big) > \tfrac 1 2M^{\frac{a_n}{r}-1}.
    \end{align*}
    Together, we obtain the lower bound
    \[
        \prod_{i=1}^{n}\E[U^{a_i}]^{\pi_i/a_i} > \Big(\prod_{i=1}^{n-1}2^{-\frac{\pi_i}{r}-\frac{\pi_i}{a_i}}\Big)2^{-\frac{\pi_n}{a_n}}M^{\frac{\pi_n}{r}-\frac{\pi_n}{a_n}}\rightarrow\infty
    \]
    which becomes arbitrarily large as $M\rightarrow \infty$.
    Hence, there exists $M>1$ such that~\eqref{eq:examMAS} is satisfied and therefore, $\rho_{\Phi}$ is not $-r$-SD-consistent.
\end{example}

\begin{proof}[Proof of Corollary~\ref{cor:Mu}]
For statement (a), note that each $u\in\mathcal U_{\mf e_{-\lambda^\star}}(\R)$ satisfies $R_v^A(x)\le -\lambda^\star\le R_u^A(x)$, $x\in I$. Hence, $u|_I\in\mathcal U_v(I)$. Consequently, all $X,Y\in \CL^1_v\cap \CX$ satisfying $X\vsd Y$ also satisfy $X\le_{-\lambda^\star}Y$. Conclude with Theorem~\ref{thm:Mu}. 

For statement (b), assume first that \eqref{eq:condMAS1} holds and select an open interval $J\subseteq I$ on which $\inf_JR_v^A\ge \lambda>-c_-$. Hence, for binary random variable $X$ with values in $J$, $K_v(X)\le K_\lambda(X)$. If $\rho_\Phi$ were $v$-SD-consistent, the proof of Theorem~\ref{thm:Mu} would deliver the contradiction 
\[\rho_\Phi(X)\ge \rho_\Phi(K_v(X))>\rho_\Phi(X).\]

Now suppose \eqref{eq:condMAS2} holds. Theorem~\ref{thm:Mu} allows to select $X,Y\in\CX$ such that $X\le_{\lambda^\star+\eps}Y$, but $\rho_\Phi(X)>\rho_\Phi(Y)$. 
Let $J\subseteq\{R_v^A\ge \lambda^\star+\eps\}$ be an unbounded open interval.
By shifting $X$ and $Y$ along the same additive constant, we can guarantee that they attain values in $J$ and that the supports of their respective distributions have positive distance to the boundary of $J$ (provided the boundary is nonempty). Hence, we can assume $X,Y\in\mathcal L^1_v$, and from $X\le_{\lambda^\star+\eps}Y$ we infer $X\le_vY$. $\rho_\Phi$ cannot be $v$-Meyer. 
\end{proof}

\section{Proofs accompanying Section~\ref{sec:generalThresholdUtility}}

\begin{proof}[Proof of Theorem~\ref{thm:unbounded}]
The inequality $\rho\le \worst$ holds for every normalised risk measure; thus, we only need to prove the converse inequality. 
Without loss of generality, we assume that the first condition in \eqref{eq:liminf} holds: for every $\lambda>0$, we can find $k>0$ such that $R_v^A\ge \lambda=R_{\mf e_{\lambda}}^A$ on $[k,\infty)$.  
Let $\rho$ be $v$-Meyer, $X\in\CX$ be arbitrary, and choose $m$ large enough such that $X+m\ge k+1$.
By Lemma~\ref{lem:certainty}, 
\[X+m\vsd K_v(X+m)\le K_{\mf e_{\lambda}}(X+m)=-\tfrac 1 {\lambda}\log\big(\E[e^{-\lambda X}]\big)+m.\]
By $v$-SD-consistency, 
\[\rho(X)=\rho(X+m)+m\ge \tfrac 1 {\lambda}\log\big(\E[e^{-\lambda X}]\big).\]
As this argument holds for arbitrary $\lambda$, taking the limit $\lambda\to\infty$ delivers 
$\rho(X)\ge \worst(X).$
\end{proof}

\begin{proof}[Proof of Lemma~\ref{lem:base}]
Each $\rho_v(\cdot|Z)$ is a risk measure by Lemma~\ref{lem:benchmark}.     
If each base risk measure under $v$ is convex, fix  $Z\in\CX$.
As $\rho_v(\cdot|Z)$ is law invariant, it must be SSD-consistent as well.
From $Z\ssd \E[Z]$, we obtain 
\[\E[Z]=\rho_v(Z|Z)+\E[Z]\ge \rho_v(\E[Z]|Z)+\E[Z]=\rho_v(0|Z).\]
        From Lemma~\ref{lem:certainty}(a), we deduce  
        $\rho_{v}(0|Z)=\inf\{m\in\R\mid Z\vsd m\}=K_v(Z).$
       Taken together, $\E[Z]\ge K_v(Z)$ for all $Z\in\CX$.
       By \cite[Proposition 1.2]{Eeckhoudt}, the latter is equivalent to $v$  being concave.

        Conversely, assume that $v$ is concave. Suppose that $X,Y\in\CX$, $0<\lambda<1$, and $m,n\in\R$ are such that $Z\vsd X+m$ and $Z\vsd Y+n$.
        By concavity of $v$, 
        \[v\big(\lambda(X+m)+(1-\lambda)(Y+n)\big)\ge \lambda v(X+m)+(1-\lambda)v(Y+n).\]
        For all $p\in[0,1]$, antitonicity, convexity and SSD-consistency of $\es_p$ imply 
        \begin{align*}
            \es_p\big(v\big(\lambda(X+m)+(1-\lambda)(Y+n)\big)\big)&\le \lambda \es_p\big(v(X+m)\big)+(1-\lambda)\es_p\big(v(Y+n)\big)\\
            &\le \es_p\big(v(Z)\big).
        \end{align*}
    As this estimate holds for all $p$, we have $Z\vsd \lambda X+(1-\lambda)Y+\lambda m+(1-\lambda) n$. 
    Letting $m\downarrow \rho_{v}(X|Z)$ and $n\downarrow \rho_{v}(Y|Z)$, the estimate 
    $$\rho_{v}(\lambda X+(1-\lambda)Y|Z)\le \lambda\rho_{v}(X|Z)+(1-\lambda)\rho_{v}(Y|Z)$$
    follows. This is convexity of $\rho_{v}(\cdot|Z)$.
\end{proof}

\begin{proof}[Proof of Proposition~\ref{prop:rep general}]
First, $v$-SD consistency of $\rho$ entails for $X\in\CX$ that $X\in\CA_\rho$ if and only if there exists $Y\in\CA_\rho$ such that $Y\vsd X$. 
Hence, using cash-additivity for the first identity,  
\begin{align*}\rho(X)&=\inf\{m\in\R\mid \exists\,Y\in\CA_\rho:~X+m=Y\}=\inf\{m\in\R\mid \exists\,Y\in\CA_\rho:~Y\vsd X+m\}\\
&=\inf\{m\in\R\mid \exists\,Y\in\CA_\rho:~\rho_v(X|Y)\le m\}=\inf_{Y\in\CA_\rho}\rho_v(X|Y).\end{align*}
\end{proof}

\begin{remark}\label{rem:tech1}
If $I$ is any open interval and $v\in\mathcal U(I)$, base risk measure $\rho_{v}(\cdot|Z)$ is well-defined and cash-additive if $Z\in\mathcal L^1_v$, but not necessarily finite-valued. 
It is a risk measure if $\sup I=\infty$, which covers the case of CRRA threshold utilities in~\eqref{ex:CRRA}. 
In that case, Lemma~\ref{lem:base} holds, but Proposition~\ref{prop:rep general} may fail because $\CA_\rho$ might not be a subset of $\mathcal L^1_v$. 
\end{remark}

\begin{proof}[Proof of Proposition~\ref{prop:Pratt}]
    (a) implies (b): Fix $Z\in\CX$ and let $X\in\CX$ be arbitrary. 
    By definition, $Z\vsd X+m$ holds for all $m>\rho_{v}(X|Z)$. Whenever $u\in\mathcal U_v(\R)$, dominated convergence implies
    \[\E[u(Z)]\le\inf_{m>\rho_{v}(X|Z)}\E[u(X+m)]=\E[u(X+\rho_{v}(X|Z))].\]
    Consequently, $Z\vsd X+\rho_{v}(X|Z)$. Now, if $X\vsd Y$, (a) implies that also 
    \[
        Z\vsd X+\rho_{v}(X|Z)\vsd Y+\rho_{v}(X|Z).
    \]
    By definition of $\rho_{v}(\cdot|Z)$, $\rho_{v}(X|Z)\ge \rho_{v}(Y|Z)$ holds as desired. 

    (b) implies (c): Suppose that $R^A_v$ is not constant, i.e., we find $x,d\in\R$, $d\neq 0$, such that $R^A_v(x+d)>R^A_v(x).$
    The underlying probability space is atomless. Hence, for every $s>0$ there is a random variable $Z_s$ with $\P(Z_s=x+s)=\P(Z_s=x-s)=\tfrac 1 2$. 
    The Arrow-Pratt approximation of the certainty equivalent---see \cite[p.\ 11]{Eeckhoudt}---delivers  
    \[K_v(Z_s)\approx x-\tfrac 1 2s^2R^A_v(x)\quad\text{and}\quad K_v(Z_s+d)\approx x+d-\tfrac 1 2s^2R^A_v(x+d).\]
    By choosing $s>0$ small enough, we can guarantee that \begin{equation}\label{eq:Pratt3}
        K_v(Z_s)>K_v(Z_s+d)-d.
    \end{equation}
    From the proof of Lemma~\ref{lem:base},
    $\rho_{Z_s,v}(0)=K_v(Z_s).$
    Using Lemma~\ref{lem:certainty} for the inequality, 
    $$-d=\rho_{v}(Z_s+d|Z_s)\ge \rho_{v}\big(K_v(Z_s+d)|Z_s\big)=\rho_{v}(0|Z_s)-K_v(Z_s+d)=K_v(Z_s)-K_v(Z_s+d).$$
    The latter contradicts \eqref{eq:Pratt3}. 
    
    (c) implies (a): This follows from \cite[Theorem 8.1.4]{Stoyan}. 
\end{proof}

\begin{remark}\,
    Regarding Proposition~\ref{prop:Pratt}, \cite[Theorem 8.1.4]{Stoyan} establishes the equivalence of (c) with the following variant of (a): \begin{center}If $X \vsd Y$ and $Z \in \CX$ is independent of both $X$ and $Y$, then $X + Z \vsd Y + Z$.\end{center}
    However, we do not see how this equivalence helps to establish the implication (b) $\Rightarrow$ (c).
    The proof presented here is direct and self-contained, though it could have been shortened by invoking \cite[Theorem 2.2]{Certainty} or \cite[Proposition 2.46]{FoeSch}. 
    Another related contribution is \cite{MWZ}, which characterises fractional degree stochastic dominance through invariance properties.
Item (a) in Proposition~\ref{prop:Pratt} corresponds to their notion of translation invariance. A precise connection lies beyond our scope, as \cite{MWZ} does not fall within the Meyer framework.
\end{remark}

\begin{proof}[Proof of Theorem~\ref{thm:MARA}]
We shall focus solely on the case where Assumption~\ref{ass:MARA} is satisfied for $J=(a,\infty)$. 
The other case is treated analogously. 

By the computation on \cite[p.\ 17]{Eeckhoudt}, the function $R_v^A$ satisfies  
\[(R_v^A)'(x)=R_v^A(x)\big(R_v^A(x)+\tfrac{v'''(x)}{v''(x)}\big),\quad x\in\R.\]

\textsc{Case 1:} $R_v^A$ is increasing. 
In this case, $R_v^A$ has at most one root in $J$, and there is a countable set $E\subseteq J$ such that $(R_v^A)'(x)>0$ for all $x\notin E$. 

\textsc{Case 1.1:} $R_v^A(x)<0$ for all $x\notin E$.
As $R_v^A$ is increasing, we must have $R_v^A(x)<0$ for all $x>a$, resulting in the fact that $v'$ is strictly increasing and twice differentiable on $J$ and can be interpreted as a utility function. 
Moreover, for all $x\notin E$,
\[R_v^A(x)<-\tfrac{v'''(x)}{v''(x)}=R_{v'}(x).\]
Consider the function 
$h:=v'\circ v^{-1}$ on $v(J)$. One shows that 
$h''(y)<0$ holds for all $y\notin v(E)$. 
As $v(E)$ is countable, this implies strict concavity of $h$. 
Consequently, Jensen's inequality shows for all $J$-valued $Y\in\CX$ that are not $\P$-a.s.\ constant that $K_v(Y)>K_{v'}(Y)$. Last, set $J_0:=J$.

\textsc{Case 1.2:} There exists $a^*\ge a$ such that $R_v^A(x)>0$ for all $x>a^*$. For $J_0:=(a^*,\infty)$, we have  $-v'$ is a utility function on $J_0$ and 
\[R_v^A(x)>-\tfrac{v'''(x)}{v''(x)}=R_{-v'}(x),\quad x\in J_0\setminus E.\]
Arguing like in Case 1.1, $h:=v\circ (-v')^{-1}$ defined on $(-v')(J_0)$ is strictly concave. 
Similar to Case 1.1, we obtain for all $J_0$-valued $Y\in\CX$ that are not $\P$-a.s.\ constant that $K_v(Y)<K_{-v'}(Y)$.

\textsc{Case 2:} $R_v^A$ is decreasing. Again, $R_v^A$ has at most one root in $J$, and there is a countable set $E\subseteq J$ such that $(R_v^A)'(x)<0$ for all $x\notin E$. 

\textsc{Case 2.1:} $R_v^A(x)>0$ for all $x\notin E$.
As $R_v^A$ is decreasing, we must have $R_v^A>0$ on $(a,\infty)$, resulting in $-v'$ being a utility function on $J$. 
Moreover, 
$R_v^A(x)<-\tfrac{v'''(x)}{v''(x)}=R_{-v'}(x)$ for all $x\notin E$. 
The function 
$h:=-v'\circ v^{-1}$ on $v(J)$ is once again strictly concave. 
Consequently, 
for all $J$-valued $Y\in\CX$ that are not $\P$-a.s.\ constant, we obtain $K_v(Y)>K_{-v'}(Y)$. Set $J_0:=J$.

\textsc{Case 2.2:} There exists $a^*\ge a$ such that $R_v^A(x)<0$ for all $x>a^*$. 
For $J_0:=(a^*,\infty)$,  $v'$ is a utility function on $J_0$ and $R_v^A(x)>-\tfrac{v'''(x)}{v''(x)}=R_{v'}(x)$ for all $x\in J_0\setminus E$.
Arguing like above, all $J_0$-valued $Y\in\CX$ that are not $\P$-a.s.\ constant satisfy $K_v(Y)<K_{v'}(Y)$.

Now fix a $Z\in\CX$ which is not $\P$-a.s.\ constant and assume towards a contradiction 
the function $\Psi(t):=K_v(Z+t)$, $t\in\R$, satisfies $\Psi'\equiv 1$.
Let $t_0>0$ be large enough such that $Z+t_0$ takes values in $J_0$ defined in each of the cases above. 
By the proof of \cite[Theorem 2.2]{Certainty}, 
$$K_v(Z+t)=\begin{cases}K_{v'}(Z+t)&~\text{if }v'|_{J_0}\in\mathcal U(J_0)\\[-0.6ex]
K_{-v'}(Z+t)&~\text{if }-v'|_{J_0}\in\mathcal U(J_0)\end{cases},\qquad t\ge t_0.
$$ 
However, this contradicts the conclusions in Cases 1 and 2 above. 
Consequently, there must be $d>0$ such that $K_v(Z+d)\neq K_v(Z)+d$. 

If $K_v(Z)+d>K_v(Z+d)$, the proof of Proposition~\ref{prop:Pratt} shows that $\rho_{v}(\cdot|Z)$ is not $v$-Meyer. 
Else, let $\widetilde Z:=Z+d$ and note that 
$K_v(\widetilde Z-d)<K_v(\widetilde Z)-d$. 
Hence, $\rho_{v}(\cdot|Z+d)$ is not $v$-Meyer. 
\end{proof}

Throughout the rest of this appendix, we abbreviate the inverse function of the threshold utility by $h:=v^{-1}$. 

\begin{proof}[Proof of Lemma~\ref{lem:utility2}]
Let $J\subseteq I$ be an interval unbounded to the right on which $R_v^A\ge c$. In particular, $v$ is concave on $J$, $v'$ is nonincreasing, and $\lambda:=\lim_{x\to\infty}v'(x)$ exists. 
For all $x\in J\cap(0,\infty)$,
\[\frac{v'(2x)-v'(x)}x=v''(\xi)\le -cv'(\xi)\le -c\lambda,\] 
where $\xi\in (x,2x)$ is suitably chosen. Letting $x\to\infty$ and noting that $c>0$ implies $\lambda=0$.
\end{proof}

\subsubsection*{Mathematical details of Table~\ref{table:utilities}}

{\bf $v$ concave.}
As $v'$ is nonincreasing, the claimed equivalence to Assumption~\ref{ass:Inada} is clear. 
For Assumption~\ref{ass:star}, 
assume first that $v(I)$ is bounded below. This means that also $I$ must be bounded below by concavity. 
Let $C\in\R$ and $\gamma,\delta>0$ be arbitrary, and suppose an increasing sequence $(t_n)\subseteq v(I)$ can be chosen such that $t_1>\frac{C}{1+\delta}$ and $C-\delta t_n\in v(I)$ for all $n\in\N$. 
In particular, $t_n\le (C-\inf v(I))/\delta$ and 
$$h(t)+\gamma h(C-\delta t)\le h\big(\tfrac{C-\inf v(I)}\delta\big)+\gamma h(C-\delta t_1)<\infty.$$
Assumption~\ref{ass:star} fails. 
If $v(I)$ is unbounded below, we must have $v(I)=\R$. 
For $\delta>0$ and $t>C/(1+\delta)$, let 
$f(t):=\vinverse(t)+\gamma \vinverse(C-\delta t)$,  and note that its derivative is 
\begin{equation}\label{eq:f'}f'(t)=\frac {1}{v'(\vinverse(t))}-\frac{\gamma\delta}{v'(\vinverse(C-\delta t))}.\end{equation}
By choosing $\gamma>0$ large enough, this can be made negative if $v'$ is bounded above and away from 0. In this case, Assumption~\ref{ass:star} fails as well. 

{\bf $v$ convex.} As $v'$ is nondecreasing, 
$\limsup_{x\downarrow \inf I}v'(x)\le \liminf_{x\uparrow \sup I}v'(x)$
and Assumption~\ref{ass:Inada} is never satisfiable. 
Assumption~\ref{ass:star} is not satisfied either. For $t>C/(1+\delta)$ with $C-\delta t\in v(I)$, the derivative in \eqref{eq:f'} satisfies
$$f'(t)\le \frac{1-\gamma\delta}{v'(\vinverse(C-\delta t))},$$
and the upper bound is negative for $\gamma>\frac 1\delta$.

{\bf Kahneman-Tversky utility.}
Suppose $\alpha<\beta$ and fix $\gamma,\delta>0$ and set $C=0$. For all $t$ large enough, 
$\vinverse(t)+\gamma \vinverse(-\delta t)=t^{1/\alpha}-\gamma(\delta t)^{1/\beta}$ diverges because $\alpha<\beta$.
Else, if $\alpha\ge \beta$, set $\gamma=2$ and $\delta=1$. No sequence satisfying Assumption~\ref{ass:star} can be bounded. Hence, for any $C\in\R$, any sequence $(t_n)$ with $\lim_{n\to\infty}t_n=\infty$, and all $n$ large enough,
\[\vinverse(t_n)+2\vinverse(C-t_n)=t_n^{1/\alpha}-2(t_n-C)^{1/\beta}.\]
This is seen to diverge to $-\infty$ if $\alpha\ge \beta$.

{\bf Logistic utility function.} The inverse of $v$ is given by
$\vinverse(y)=\tfrac 1 \alpha \log\big(\tfrac{y}{1-y}\big)$ for $y\in(0,1)$.
Choose $\delta<C<\delta+1$ and $C/(\delta+1)<t<1$. 
Then 
\[\vinverse(t)+\gamma \vinverse(C-\delta t)=\tfrac 1 \alpha \log\big(\tfrac{t}{1-t}\big)+\tfrac 1 \alpha \log\big(\tfrac{C-\delta t}{1-C+\delta t}\big)\to \infty,\quad t\uparrow 1.\]
This shows that Assumption~\ref{ass:star} is satisfied. 

{\bf SAHARA utility.} 
By \cite[eq.\ (4)]{SAHARA}, 
threshold utility function $v$ has derivative 
\[v'(x)=\beta^{-\alpha}\exp\big(-\alpha\cdot\arcsinh\big(\tfrac{x-d}\beta\big)\big),\quad x\in\R,\] 
up to a constant. 
As $\lim_{y\to\pm\infty}\arcsinh(y)=\infty$, the derivative vanishes as $x\to\infty$. 
Assumption~\ref{ass:Inada} is satisfied.
Also, SAHARA utilities are strictly concave. In view of the first row of Table~\ref{table:utilities}, Assumption~\ref{ass:star} is satisfied because $v$ is defined on the whole real line and thus $v(\R)$ must be unbounded below.\hfill\qed

\smallskip

To prove Theorems~\ref{thm main 1} and~\ref{thm main 2}, we isolate two shared arguments into auxiliary lemmas.

\begin{lemma}\label{lem:essinf}
    If $\rho$ is positively homogeneous and $\rho(\ind_A)=0$ for all events $A\in\CF$ with $\P(A)<1$, then
    $\rho=\worst.$
\end{lemma}
\begin{proof}
    The identity $\rho(X)=\worst(X)$ obviously holds for all $X\in\CX$ that are a.s.\ constant. 
    Moreover, as $\worst$ is the most conservative normalised risk measures, the inequality
    $\rho(X)\le \worst(X)$
    holds for all $X\in\CX$. 
    For the converse inequality, fix an arbitrary $X\in\CX$ that is not a.s.\ constant and consider the nonnegative random variable $Y:=X+\worst(X)$. Moreover, for all $n\in\N$ large enough, we have that 
    $Y\le \tfrac{n^2-1}n\ind_{\{Y>1/n\}}+\tfrac 1 n$, $\P(Y>\frac 1n)<1$, and 
    \[\rho(X)-\worst(X)=\rho(Y)\ge \rho\big(\tfrac{n^2-1}n\ind_{\{Y>1/n\}}+\tfrac 1 n\big)=\tfrac{n^2-1}n\rho(\ind_{\{Y>1/n\}})-\tfrac 1 n=-\tfrac 1n.\]
    In the preceding estimate, we have used positive homogeneity, antitonicity and cash-additivity of $\rho$.
    Rearranging the inequality and letting $n\to\infty$ delivers $\rho(X)\ge \worst(X)$. 
\end{proof}

In the second lemma and throughout the proofs of Theorems~\ref{thm main 1} and~\ref{thm main 2}, we use for a fixed event $A\in\CF$ the notation 
$$Y_{x,y}:=x\ind_A+y\ind_{A^c},\quad x,y\in\R.$$
Recall also that $h = v^{-1}$.

\begin{lemma}\label{lem:spread}
    Let $v\in\mathcal U(I)$ be a threshold utility. A law-invariant risk measure $\rho$ is {\em not} $v$-Meyer if there exist $s,s',t,t'\in v(I)$ and $A\in\CF$ such that 
     $s'<s<t<t'$, $\E[Y_{s,t}]=\E[Y_{s',t'}]$, and 
     \begin{equation}\label{eq:contr}\rho\big(\vinverse(Y_{s,t})\big)>\rho\big(\vinverse(Y_{s',t'})\big).\end{equation}
\end{lemma}
\begin{proof}
    By construction, $Y_{s',t'}$ is a mean-preserving spread of $Y_{s,t}$. Hence, 
    $Y_{s',t'}\ssd Y_{s,t},$
    which translates by~\eqref{complicated equiv} as 
    $\vinverse\big(Y_{s',t'}\big)\vsd \vinverse\big(Y_{s,t}\big).$
    If $\rho$ were $v$-Meyer, we would observe 
    $\rho(\vinverse(Y_{s,t}))\le \rho(\vinverse(Y_{s',t'}))$ in contradiction to \eqref{eq:contr}. 
\end{proof}

\begin{proof}[Proof of Theorem~\ref{thm main 1}]
    The fact that $\worst$ is $v$-Meyer is Lemma~\ref{lem:worst}. 
    For the converse implication, assume that $\rho\neq\worst$. 
    By Lemma~\ref{lem:essinf}, we may fix an event $A\in\CF$ with $p:=\P(A)\in(0,1)$ and $r:=|\rho(\ind_A)|>0$.
    For $x,y\in\R$ with $x<y$, cash-additivity and positive homogeneity of $\rho$ deliver
    \begin{equation}\label{eq:simple}\rho(Y_{y,x})=-ry-(1-r)x.\end{equation}
    
    If $r=1$, we have for all $s,t\in v(I)$ with $s<t$ that $\rho(\vinverse(Y_{t,s}))=-t$, and  
    $\rho$ cannot be $v$-Meyer by Lemma~\ref{lem:spread}. 
    Else, if $r<1$, invoke Assumption~\ref{ass:Inada}, to choose $x,y\in I$ with $x<y$ such that 
    \[\frac{r}{v'(y)}>\frac{(1-r)p}{(1-p)v'(x)}.\]
    Set $s_0:=v(x)$, $t_0:=v(y)$, $a:=s_0(1-p)+t_0p$, and consider the set $J$ of all $t\in v(I)$ such that 
    $s(t):=(a-tp)/(1-p)<t$ and lies in $v(I)$. 
    In particular, $s_0=s(t_0)$ by construction. 
    By openness of $v(I)$, $t_0$ lies in the interior of $J$. 
    By \eqref{eq:simple}, the derivative of the function
    \[f(t):=\rho\big(\vinverse(Y_{t,s(t)})\big)=-rh(t)-(1-r)h(s(t))\]
    at $t=t_0$ is 
    \[f'(t_0)=\frac{(1-r)p}{(1-p)v'(\vinverse(s(t_0)))}-\frac{r}{v'(\vinverse(t_0))}=\frac{(1-r)p}{(1-p)v'(x)}-\frac{r}{v'(y)}<0.\]
    
    By Lemma~\ref{lem:spread}, $\rho$ cannot be $v$-Meyer in this case either.   
\end{proof}

\begin{proof}[Proof of Theorem~\ref{thm main 2}]
    Assume that $\rho^\infty\neq\worst$.  
    In view of Lemma~\ref{lem:essinf}, we can fix an event $A\in\CF$ with $p:=\P(A)\in(0,1)$ and $k:=|\rho^\infty(\ind_{A})|>0$. 
    For $s,t\in v(I)$ with $s<t$, 
    the estimate $\vinverse(s)\le \vinverse(Y_{t,s})\le \vinverse(t)$ implies together with antitonicity of $\rho$ that 
    $\rho\big(\vinverse(Y_{t,s})\big)\in[-\vinverse(t),-\vinverse(s)]$, i.e., there exists $\lambda\in[0,1]$ such that  
    \[\rho\big(\vinverse(Y_{t,s})\big)=-\lambda \vinverse(t)-(1-\lambda)\vinverse(s).\]   
    As   
    \[-\lambda \vinverse(t)-(1-\lambda)\vinverse(s)=\rho\big(\vinverse(Y_{t,s})\big)\le \rho^\infty\big(\vinverse(Y_{t,s})\big)=-k\vinverse(t)-(1-k)\vinverse(s),\]
    $\lambda\ge k$ must hold. 
    
    If $k=1$, it can be argued as in the proof of Theorem~\ref{thm main 1} that $\rho$ is not $v$-Meyer. 
    Else, choose  
$\gamma=\frac{1-k}{k}$ and $\delta=\frac{p}{1-p}$ 
in Assumption~\ref{ass:star}
and let $(t_n)\subseteq v(I)$ be the appropriate sequence. 
In particular, 
$\vinverse(t_n)\uparrow\infty$. 
Selecting the sequence $(\lambda_n)\subseteq[k,1]$ appropriately
\begin{align*}\rho\big(\vinverse(Y_{t_n,C-\delta t_n})\big)&=-\lambda_n\vinverse(t_n)-(1-\lambda_n)\vinverse(C-\delta t_n)\le-k\Big(\vinverse(t_n)+\gamma \vinverse(C-\delta t_n)\Big).
\end{align*}
As $n\to\infty$, this upper bound diverges to $-\infty$. 
Nevertheless, the sequence $(t_n)$ is increasing and $\E\big[Y_{t_n,C-\delta t_n}\big]=C(1-p)$
holds for all $n\in\N$. 
Thus Lemma~\ref{lem:spread} delivers that $\rho$ is not $v$-Meyer. 
\end{proof}

\end{document}